\documentclass[aps,prd,groupedaddress,preprint,eqsecnum,nofootinbib,showpacs]{revtex4}
\usepackage{graphicx,epsf,amssymb,amsbsy,amsfonts,amssymb,amsmath}
\usepackage[dvips]{color}

\usepackage{ifthen}

\newif\ifdraft
\drafttrue
\newif\ifpreprint
\preprinttrue

\def\sect#1{section~{\ref{#1}}}
\def\Sect#1{Section~{\ref{#1}}}
\def\sects#1#2{sections~{\ref{#1}} and~{\ref{#2}}}

\def\fig#1{fig.~{\ref{#1}}}
\def\Fig#1{Fig.~{\ref{#1}}}
\def\figs#1#2{figs.~{\ref{#1}} and~{\ref{#2}}}

\def\spa#1.#2{\left\langle#1\,#2\right\rangle}
\def\spb#1.#2{\left[#1\,#2\right]}
\def\spash#1.#2{\spa{\smash{#1}}.{\smash{#2}}}
\def\spbsh#1.#2{\spb{\smash{#1}}.{\smash{#2}}}
\def\sand#1.#2.#3{%
\left\langle\smash{#1}{\vphantom1}^{-}\right|{#2}%
\left|\smash{#3}{\vphantom1}^{-}\right\rangle}
\def\sandpp#1.#2.#3{%
\left\langle\smash{#1}{\vphantom1}^{+}\right|{#2}%
\left|\smash{#3}{\vphantom1}^{+}\right\rangle}
\def\sandpm#1.#2.#3{%
\left\langle\smash{#1}{\vphantom1}^{+}\right|{#2}%
\left|\smash{#3}{\vphantom1}^{-}\right\rangle}
\def\sandmp#1.#2.#3{%
\left\langle\smash{#1}{\vphantom1}^{-}\right|{#2}%
\left|\smash{#3}{\vphantom1}^{+}\right\rangle}

\def\tree{{\rm tree}}

\def\nn{\nonumber}

\def\eqn#1{eq.~(\ref{#1})}

\def\eqns#1#2{eqs.~(\ref{#1}) and~(\ref{#2})}

\def\Neqfour{{{\cal N}=4}}
\def\NeqFour{{{\cal N}=4}}
\def\Neqeight{{{\cal N}=8}}
\def\NeqEight{{{\cal N}=8}}

\def\be{\begin{equation}}
\def\ee{\end{equation}}
\def\bea{\begin{eqnarray}}
\def\eea{\end{eqnarray}}
\def\ba{\begin{eqnarray}}
\def\ea{\end{eqnarray}}

\def\Perm{{\cal P}}
\def\ve{\varepsilon}
\def\tlambda{{\tilde\lambda}}
\def\MHVbar{$\overline{\hbox{MHV}}$}

\newcommand{\Pflat}{P^{\flat}}

\def\tree{{\rm tree}}
\def\oneloop{{\rm 1\hbox{-}loop}}

\def\fourloop{{\rm 4\hbox{-}loop}}

\newbox\charbox
\newbox\slabox
\def\s#1{{      
        \setbox\charbox=\hbox{$#1$}
        \setbox\slabox=\hbox{$/$}
        \dimen\charbox=\ht\slabox
        \advance\dimen\charbox by -\dp\slabox
        \advance\dimen\charbox by -\ht\charbox
        \advance\dimen\charbox by \dp\charbox
        \divide\dimen\charbox by 2
        \raise-\dimen\charbox\hbox to \wd\charbox{\hss/\hss}
        \llap{$#1$} }}

\def\subtractfour#1{\ifthenelse{#1=5}{1}{\ifthenelse{#1=6}{2}
{\ifthenelse{#1=7}{3}{\ifthenelse{#1=8}{4}{\ifthenelse{#1=9}{5}
{\ifthenelse{#1=10}{6}{\ifthenelse{#1=11}{7}{\ifthenelse{#1=12}{8}
{\ifthenelse{#1=13}{9}{\ifthenelse{#1=14}{10}{}}}}}}}}}}}

\def\etab{\widetilde \eta}
\def\qbar{\tilde q}

\begin{document}

\ifpreprint
 UCLA/09/TEP/41
\fi

\title{On the Structure of Supersymmetric Sums in Multi-Loop Unitarity Cuts}

\author{Z.~Bern${}^a$, J.~J.~M.~Carrasco${}^a$, 
H.~Ita${}^a$,  H.~Johansson${}^a$ and  R.~Roiban${}^b$ }

\affiliation{
${}^a$Department of Physics and Astronomy, UCLA, Los Angeles, CA
90095-1547, USA  \\
${}^b$Department of Physics, Pennsylvania State University,
           University Park, PA 16802, USA
}

\date{March, 2009}

\begin{abstract}

In this paper we describe algebraic and diagrammatic methods, related
to the MHV generating function method, for evaluating and exposing the
structure of supersymmetric sums over the states crossing generalized
unitarity cuts of multi-loop amplitudes in four dimensions.  We focus
mainly on cuts of maximally supersymmetric Yang-Mills amplitudes.  We
provide various concrete examples, some of which are directly relevant
for the calculation of four-loop amplitudes.  Additionally, we discuss
some cases with less than maximal supersymmetry.  The results of these
constructions carry over to generalized cuts of multi-loop
supergravity amplitudes through use of the Kawai-Lewellen-Tye relations
between gravity and gauge-theory tree amplitudes.

\end{abstract}

\pacs{04.65.+e, 11.15.Bt, 11.30.Pb, 11.55.Bq \hspace{1cm}}

\maketitle


\section{Introduction}
\label{Introduction}

Multi-loop scattering amplitudes in maximally supersymmetric gauge and
gravity theories have received considerable attention in recent years
for their roles~\cite{BCDKS,ABDK,BDS,AM} in helping to confirm
and utilize Maldacena's AdS/CFT correspondence~\cite{Maldacena} and in
probing the ultraviolet structure of supergravity
theories~\cite{Finite,GravityThree,CompactThree}.

In particular, multi-loop calculations offer important insight into the
possibility that planar $\NeqFour$ super-Yang-Mills scattering
amplitudes can be resummed to {\it all} loop
orders~\cite{ABDK,BDS,AM}.  In ref.~\cite{ABDK} a loop iterative
structure was suggested, leading to the detailed BDS
conjecture~\cite{BDS} for planar
maximally-helicity-violating (MHV) amplitudes to all loop
orders.  Alday and Maldacena realized that certain planar scattering
amplitudes at strong coupling may be evaluated as the regularized area
of minimal surfaces in AdS$_5\times $S$^5$ with special boundary
conditions, and for four-point amplitudes they confirmed the BDS prediction.
Direct evidence suggests that the all-order resummation holds as well
for five-point amplitudes~\cite{TwoLoopFive}.  The structure of the
four- and five-point planar amplitudes is now understood as a
consequence~\cite{DHKS} of a new symmetry dubbed ``dual conformal
invariance''~\cite{DualConformal,BCDKS,FiveLoop}, with further
generalizations at tree level~\cite{DrummondYangian} and at infinite
't~Hooft coupling~\cite{BerkovitsMaldacena}.  However, beyond five
points, the BDS conjecture requires
modification~\cite{AM2,Lipatov,TwoLoopSixPt}.  High-loop calculations
in $\NeqFour$ super-Yang-Mills theory should also play a useful role
in clarifying the structure of subleading color contributions to the
soft anomalous dimension matrix of gauge
theories~\cite{SoftMatrixPapers}, once the evaluation of the required
nonplanar integrals becomes feasible at three loops and beyond.

In a parallel development, studies of multi-loop amplitudes in
$\NeqEight$ supergravity ~\cite{CremmerJuliaScherk} have suggested
that this theory may be ultraviolet finite in four
dimensions~\cite{Finite, GravityThree, CompactThree}, challenging the
conventional understanding of the ultraviolet properties of gravity
theories.  For a class of terms accessible by isolating one-loop
subamplitudes via generalized
unitarity~\cite{GeneralizedUnitarity,TwoLoopSplit,BCFGeneralized}, the
one-loop ``no-triangle'' property (\cite{OneloopMHVGravity,
NoTriangle,UnexpectedCancel, AHCKGravity}) shows that at least a
subset of these cancellations persist to all loop
orders~\cite{Finite}.  The direct calculation of the three-loop
four-point amplitude of $\NeqEight$ supergravity exposes cancellations
beyond those needed for ultraviolet finiteness in $D=4$ in all terms
contributing to the amplitude~\cite{GravityThree,CompactThree}.  Interestingly,
$M$ theory and string theory have also been used to argue either for
the finiteness of $\NeqEight$ supergravity~\cite{DualityArguments}, or
that divergences are delayed through at least nine
loops~\cite{Berkovits,GreenII}, though issues with decoupling towers
of massive states~\cite{GOS} may alter these conclusions.  A recent
direct field theory study proposes that a divergence may first appear at
the five loop order in $D=4$, though this can be softer if additional
unaccounted symmetries are present~\cite{HoweStelleRecent}.  If a
perturbatively ultraviolet-finite point-like theory of quantum gravity
could be constructed, the underlying mechanism responsible for the
required cancellations is expected to have a fundamental impact on our
understanding of gravity.

The recent studies of multi-loop amplitudes rely on the modern
unitarity method~\cite{UnitarityMethod, Fusing} as well as 
various refinements~\cite{GeneralizedUnitarity,TwoLoopSplit,
BCFGeneralized, FiveLoop, FreddyMaximal}.  
In this approach multi-loop
amplitudes are constructed directly from on-shell tree amplitudes.
This formalism takes advantage of the fact that tree-level amplitudes
are much simpler than individual Feynman diagrams, as well as makes use of
various properties that hold only on shell.  In particular, it
provides a means of using an on-shell superspace---which is much
simpler than its off-shell cousins---in the construction of
loops amplitudes.

Summing over the physical states of propagating fields is one
essential ingredient in higher-loop calculations.  In particular, the
modern unitarity method uses these sums over physical on-shell states
in the reconstruction of any loop amplitude in terms of covariant
integrals with internal off-shell lines.  In supersymmetric theories
the on-shell states can be organized in supermultiplets dictated by
the supersymmetry.  Systematic approaches to evaluate such
supersymmetric sums---or supersums---have recently been discussed in
refs.~\cite{FreedmanGenerating, FreedmanUnitarity, KorchemskyOneLoop,
AHCKGravity}.  As the calculations reach to ever higher loop orders
these sums become more intricate.  It is therefore helpful to expose
their structure
and simplify their evaluation as much as
possible.  In this paper we describe algebraic and diagrammatic methods
which are helpful in this direction.  These methods are the ones
used in the course of computing and confirming the four-loop
four-point amplitude of maximally supersymmetric Yang-Mills theory,
including nonplanar contributions.  The main aspects of the
construction of this amplitude, as well as the explicit results, will
be presented elsewhere~\cite{FourLoopNonPlanar}.  (The planar
contributions are given in ref.~\cite{BCDKS}.)

Supersymmetric cancellations were extensively discussed at one and two
loops in refs.~\cite{UnitarityMethod,Fusing, BRY, BDDPR} using a
component formalism that exploits supersymmetry Ward
identities~\cite{SWI}.  These supersums were relatively simple, making
it straightforward to sum over the contributions from the
supermultiplet in components.  The recent calculations of more
complicated amplitudes in refs.~\cite{TwoLoopFive, FiveLoop,
GravityThree,TwoLoopSixPt,LeadingApplications, CompactThree}, are
performed in ways obscuring the systematics of the supersums.  For
example, as explained in ref.~\cite{FiveLoop}, it is possible to avoid
evaluating (sometimes complicated) supersums in maximally
supersymmetric Yang-Mills theory via the method of maximal cuts, where
kinematics can be chosen to restrict scalars and fermions to a small
(even zero) number of loops.  Remarkably, this trick is sufficient to
construct ans\"atze for $\NeqFour$ super-Yang-Mills amplitudes.
However, any such ansatz needs to be confirmed by more direct
evaluations incorporating all particles in the supermultiplet, to
ensure that no terms are dropped.  It is therefore necessary to
compare the cuts of the ansatz with the cuts of the amplitude for more
general kinematic configurations, allowing all states to cross the
cuts.  The calculation of supersums is a crucial ingredient in carrying out
this comparison.  Moreover, formal studies of the ultraviolet behavior
of multi-loop amplitudes of supersymmetric theories, in particular of
$\NeqEight$ supergravity, are substantially aided by a formalism that
exposes the supersymmetric cancellations.

Nair's original construction of an on-shell superspace~\cite{Nair}
captured only MHV tree amplitudes in
$\NeqFour$ super-Yang-Mills theory; more recent developments extend
this to any helicity and particle configuration.  The approach
of~\cite{GGK, FreedmanGenerating, FreedmanUnitarity, FreedmanProof}
makes use of the MHV vertex expansion~\cite{CSW} to extend this
on-shell superspace to general amplitudes.  Another strategy,
discussed in refs.~\cite{AHCKGravity, RecentOnShellSuperSpace}, makes
use of the Britto, Cachazo, Feng, and Witten (BCFW) on-shell
recursion~\cite{BCFW} to extend the MHV on-shell superspace to general
helicity configurations.  A new key ingredient of this approach is a
shift involving anti-commuting parameters which may be thought of as
the supersymmetric extension of the BCFW shift of space-time momenta.
A recent paper uses shifts of anti-commuting parameters to construct a
new super-MHV expansion~\cite{Kiermaier}, which we do not use here.
With the unitarity method~\cite{UnitarityMethod, Fusing, BRY, BDDPR},
superspace expressions for tree amplitudes can be extended to loop
level.  One-loop constructions along these lines were discussed in
refs.~\cite{FreedmanGenerating, KorchemskyOneLoop, AHCKGravity}, while
various examples of supersums in higher-loop cuts, including four-loop
ones, have already been presented in ref.~\cite{FreedmanUnitarity}.

The MHV vertex expansion suggests an inductive structure for
supersymmetric cancellations.  Once these cancellations are 
exposed and understood for cuts with only MHV or \MHVbar{} tree amplitudes,
more general cuts with non-MHV amplitudes follow rather
simply~\cite{FreedmanUnitarity}.  Indeed, 
the prescription for evaluating these more general cuts involves 
summing over MHV contributions with shifts of certain on-shell
intermediate momenta.

To evaluate the supersymmetric sums that appear in unitarity cuts we
introduce complementary algebraic and diagrammatic approaches.  The
algebraic approach has the advantage of exposing supersymmetric
cancellations, in many cases leading to simple expressions.  It is
a natural approach for formal proofs.  In particular, it allows us to
systematically expose supersymmetric cancellations---within the
context of the unitarity method---sufficient for exhibiting the well
known~\cite{Mandelstam} all-loop ultraviolet finiteness of $\NeqFour$
super-Yang-Mills theory.  The diagrammatic approach gives us a means
of pictorially tracking contributions, allowing us to write down the
answer directly by drawing a set of simple diagrams.  It also leads to a
simple algorithms for writing down the results for any cut by sweeping
over all possible helicity labels.  Since it tracks contributions of
individual states, it can be easily applied to a variety of
cases with fewer supersymmetries.  To illustrate these techniques we
present various examples, including those relevant for
evaluating the four-loop four-point amplitude of $\NeqFour$
super-Yang-Mills theory~\cite{FourLoopNonPlanar}.  We will also show
that these techniques are not restricted to four-point amplitudes by
discussing some higher-point examples.

One potential difficulty with any four-dimensional approach is that
unitarity cuts are properly evaluated in $D$ dimensions~\cite{DDimUnitarity,
SelfDual}, since they rely on a form of dimensional
regularization~\cite{FDH} related to dimensional
reduction~\cite{Siegel}.  Moreover, a frequent goal in multi-loop
calculations is the determination of the critical dimension in which
ultraviolet divergences first appear.  Consequently, such calculations often 
need to be valid away from four dimensions.  This requirement
complicates the analysis significantly, because powerful
four-dimensional helicity methods~\cite{SpinorHelicity} can no longer
be used.  Any ansatz for an amplitude obtained with intrinsically
four-dimensional methods, such as the ones of the present paper, needs
to be confirmed through $D$-dimensional calculation.  Nevertheless,
the $D=4$ analysis offers crucial guidance for the construction of
$D$-dimensional amplitudes.  Additionally, 
$D=4$ methods appear to capture the  complete result for four-point 
$\NeqFour$ super-Yang-Mills  amplitudes with fewer than five 
loops~\cite{BRY,BDDPR,BCDKS,GravityThree}.

While difficulties appear to arise with extending the MHV diagram
expansion to general $\NeqEight$ supergravity tree
amplitudes~\cite{FreedmanGenerating}, they will not concern us here.
Instead we rely on the Kawai-Lewellen-Tye (KLT)
relations~\cite{KLT,GeneralKLT}, or their reorganization in terms of
diagram-by-diagram relations~\cite{TreeJacobi}, to obtain the sums
over supermultiplets in $\NeqEight$ supergravity cuts directly from
the cuts of corresponding $\NeqFour$ super-Yang-Mills theory
amplitudes.

This paper is organized as follows.  In \sect{TreeSuperspaceSection} we
review  on-shell superspace at tree level and introduce $SU(4)$
$R$-symmetry 
index diagrams.  In \sect{UnitarityMethodSection} we review the modern
unitarity method and present the general structure of supercuts.  In
\sect{LinearSystemSection} we explain how the supersums can be
evaluated in terms of the determinant of the matrix of coefficients of
a system of linear equations.  This section also contains various
examples of cuts of $\NeqFour$ super-Yang-Mills, including those of a
five-point amplitude at four loops.  \Sect{DiagramLoopSection}
describes supersums in terms of $R$-symmetry index diagrams, providing
pictorial means for tracking different contributions.  As discussed in
\sect{TrackingLoopSection}, these diagrams allow us to relate the cuts
of amplitudes with fewer supersymmetries to maximally supersymmetric
ones.  They also allow us construct a simple algorithm for obtaining all
contributions to cuts from purely gluonic ones.  Various three and
four-loop examples are presented in
\sects{DiagramLoopSection}{TrackingLoopSection}.  In
\sect{GravitySection} we outline the use of the KLT relations to carry
over the results for the sum over states in cuts of $\NeqFour$
super-Yang-Mills amplitudes to the corresponding ones of $\NeqEight$
supergravity theory.  Our conclusions are presented in
\sect{ConclusionSection}.

\section{On-shell superspace at tree level}
\label{TreeSuperspaceSection}

On-shell superspaces are useful tools for probing the properties
of supersymmetric field theories, providing information on their
structure without any complications due to unphysical degrees of
freedom.  
Here we review the construction of an on-shell superspace for
$\NeqFour$ super-Yang-Mills amplitudes.  In its original form,
devised by Nair~\cite{Nair}, it described maximally helicity violating
(MHV) gluon amplitudes and their supersymmetric partners.  While we will
depart at times from Nair's original construction, the main features will
persist.  This same superspace also captures general
amplitudes.  Indeed, there currently exists two methods for
constructing general amplitudes from MHV amplitudes: the MHV vertex
construction of Cachazo, Svr{\v c}ek and Witten~\cite{CSW} and the
on-shell recursion relation of Britto, Cachazo, Feng and
Witten (BCFW)~\cite{BCFW}.  The supersymmetric extension of the former
approach has been given in refs.~\cite{GGK, FreedmanGenerating,
FreedmanUnitarity, FreedmanProof}, while that of the latter approach
in refs.~\cite{AHCKGravity,RecentOnShellSuperSpace}.

To evaluate the supersum in unitarity cuts we will use an approach
based on MHV vertices, along the lines taken by Bianchi, Elvang,
Freedman and Kiermaier~\cite{FreedmanGenerating, FreedmanUnitarity}.
We will find that supersums involving only MHV and/or \MHVbar{} tree
amplitudes have a surprisingly simple structure.  We will also show
how the MHV vertex construction allows us to immediately carry over 
this simplicity, with only minor modifications, to more general cuts 
involving arbitrary non-MHV tree amplitudes.
 
The on-shell superspace of the type we will review here generalizes
easily to MHV and \MHVbar{} amplitudes in $\NeqEight$
supergravity.  Difficulties however, appear with the MHV vertex
construction of non-MHV gravity tree amplitudes because the on-shell
recursions used to obtain the expansion~\cite{GravityMHV} can fail to
capture all contributions~\cite{FreedmanGenerating}.  Such amplitudes
may nevertheless be found without difficulty through supersymmetric
extensions~\cite{AHCKGravity} of the on-shell BCFW recursion
relations~\cite{BCFW,CachazoLargez}, which do carry over to
$\NeqEight$ supergravity.  However, at present~\cite{BDDPR,
GravityThree, CompactThree} we find it advantageous to use the KLT
tree-level relations~\cite{KLT,GeneralKLT} or the recently discovered
diagram-by-diagram relations~\cite{TreeJacobi}, to obtain $\NeqEight$
supergravity unitarity cuts directly from those of $\NeqFour$
super-Yang-Mills theory.

\subsection{MHV amplitudes in $\NeqFour$ super-Yang-Mills}
\label{MHVAmplitudesSubSection}

The vector multiplet of the $\NeqFour$ supersymmetry algebra consists of
one gluon, four gluinos and three complex scalars, all in the adjoint
representation of the gauge group, which here we take to be $SU(N_c)$.
With all states in the adjoint representation, any complete tree-level
amplitude can be decomposed as
\begin{equation}
\mathbb{A}^\tree_n (1,2,3, \ldots, n)=g^{n-2} 
\sum_{\Perm (2,3,\ldots, n)} {\rm Tr}[T^{a_1}T^{a_2} T^{a_3}\cdots T^{a_n}] 
\, A^\tree_n (1,2,3, \ldots, n),
\label{TreeDecomposition}
\end{equation}
where $A_n^\tree$ are tree-level color-ordered $n$-leg partial
amplitudes.  The $T^{a_{i}}$'s are generators of the gauge group and
encode the color of each external leg $1,2,3 \ldots n$, with color
group indices $a_i$.  The sum runs over all noncyclic permutations of
legs, which is equivalent to all permutations keeping one leg fixed
(here leg~$1$).  Helicities and polarizations are suppressed.  We use
the all outgoing convention for the momenta to define the amplitudes.

All states transform in antisymmetric
tensor representations of the $SU(4)$ $R$-symmetry group such that
states with opposite helicities are in conjugate 
representations.  The $R$-symmetry and helicity quantum numbers
uniquely specify all on-shell states:
\def\hs{\hskip 1.5 cm}
\begin{equation}
g_+\,, \hs f_+^a\,, \hs s^{ab}, \hs f^{abc}_-\,, \hs g_-^{abcd} \,,
\label{MHVStates}
\end{equation}
where $g_\pm$ and $f_\pm$ are, respectively, the positive and negative
helicity gluons and gluinos while $s^{ab}$ are scalars.  (The scalars
are complex-valued and obey a self-duality condition which will not be
relevant here.)  These fields are completely antisymmetric in their
displayed $R$-symmetry indices---denoted by $a,b,c,d$---which
transform in the fundamental representation of
$SU(4)$, giving a total of 16 states in the on-shell multiplet.

Alternatively, we can use the dual assignment obtained by lowering the indices 
with a properly normalized Levi-Civita symbol 
$\varepsilon_{abcd}$, giving the fields,
\begin{equation}
g^+_{abcd}\,, \hs f^+_{abc}\,, \hs s_{ab}, \hs f^-_a\,, \hs g^- \,.
\label{MHVbarStates}
\end{equation}

We will use both representations to describe the amplitudes of
$\NeqFour$ super-Yang-Mills.  For MHV amplitudes we will mainly use
the states with upper indices in \eqn{MHVStates} whereas for \MHVbar{} 
we will use mainly the states with lower indices in \eqn{MHVbarStates}.  
This is a matter of convenience, and the two sets of 
states may be interchanged, as we will briefly discuss later 
in this section.

We begin by discussing the MHV amplitudes, which we define as an
amplitude with a total of eight (2 $\times$ 4 distinct)
upper $SU(4)$ indices.  (In order to respect $SU(4)$ invariance,
amplitudes of the fields in \eqn{MHVStates} must always come with $4m$
upper indices, where $m$ is an integer.  Furthermore amplitudes with
four or zero indices vanish as they are related by supersymmetry to
vanishing~\cite{SWI} amplitudes.)  Some simple examples of MHV
amplitudes, which we will use in \sect{DiagramSubSection}, are,
\begin{eqnarray}
&& {\rm (a)}:\;\; A_4^\tree(1^-_{g^{abcd}}, 2^-_{g^{abcd}}, 3^+_g, 4^+_g)  = 
         i {\spa1.2^4 \over \spa1.2 \spa2.3 \spa3.4 \spa4.1}\,, \nn \\
&& {\rm (b)}:\;\; A_4^\tree(1^-_{g^{abcd}}, 2^-_{f^{abc}}, 3^+_{f^d}, 4^+_g) = 
        i {\spa1.2^3 \spa1.3\over \spa1.2 \spa2.3 \spa3.4 \spa4.1}\,,
       \nn \\
&& {\rm (c)}:\;\; A_4^\tree(1^-_{f^{abc}}, 2^-_{f^{abd}}, 3_{s^{cd}}, 4^+_g) = 
       i {\spa1.2^2 \spa1.3 \spa2.3\over \spa1.2 \spa2.3 \spa3.4 \spa4.1}\,, 
\label{TreeExamples}
\end{eqnarray}
where $a,b,c,d$ are four distinct fundamental $SU(4)$ indices.  The
overall phases of these amplitudes depend on conventions.  We will fix
this ambiguity by demanding that the phases be consistent with the
supersymmetry algebra, which is automatically enforced when using
superspace.  The amplitudes are written in terms of the familiar
holomorphic and antiholomorphic spinor products,
\begin{eqnarray}
\spa{i}.{j}
&=& \langle i | j \rangle
= \bar{u}_-(p_i) u_+(p_j)
=\ve_{\alpha\beta} \lambda_i^{\alpha} \lambda_j^{\beta}\,,
\nn \\
\spb{i}.{j}
&=& [i|j]
= \bar{u}_+(p_i) u_-(p_j)
= \ve_{\dot\alpha\dot\beta} \tlambda_i^{\dot\alpha} \tlambda_j^{\dot\beta}\,,
\label{spinorproddef}
\end{eqnarray}
where the $\lambda_i^{\alpha}$ and $\tlambda_i^{\dot\alpha}$ are
commuting spinors which may be identified with the positive and
negative chirality solutions $|i\rangle=u_{+}(p_i)$ and
$|i]=u_{-}(p_i)$ of the massless Dirac equation and the spinor indices
are implicitly summed over.  These products are
antisymmetric, $\spa{i}.{j} = - \spa{j}.{i}$, $\spb{i}.{j} = -
\spb{j}.{i}$.

Momenta are related to these spinors via
\begin{equation}
p_{i}^\mu  \sigma_{\mu}^{\alpha{\dot\alpha}}=
\lambda_i^{\alpha}{\tilde\lambda}_i^{{\dot\alpha}}\, \hskip 5mm  
   {\rm or} \hskip 5mm p_{i}^\mu \sigma_\mu=|i\rangle[i|\,,
\end{equation}
and similar formul\ae ~hold for the expression of $p_{i}^\mu
{\overline\sigma}_{\,\mu}$.  We will often write 
simply $p_i=|i\rangle[i|$ or sometimes  $p=|p\rangle[p|$.  The
proper contractions of momenta
$p_{i}$  with spinorial objects will be implicitly assumed in the remainder
of the paper.  Typically, we will denote external momenta by $k_i$ 
and loop momenta by $l_i$.

A subtlety we must deal with is a slight inconsistency in the standard
spinor helicity formalism for massless particles when a state crosses
a cut.  In a given cut we will always have the situation that on one
side of a cut line the momentum is outgoing, but on the other side it
is incoming.  Thus across a cut we encounter expressions such as
$|$$-i\rangle [i|$, which is not properly defined in our all-outgoing
conventions and can lead to incorrect phases.  This is because the
spinor $|$$-i\rangle$ carries momentum $-p_i$, and thus it has an
energy component of opposite sign to that carried by the spinor $[i|$.
This problem is due to the fact that the spinor helicity formalism
does not distinguish between particle and antiparticle spinors, as has
been discussed and corrected in refs.~\cite{SignSubtlety} for the MHV
vertex expansion, and for BCFW recursion relations with fermions.  To
deal with this, we use the analytic continuation rule that the change of
of sign of the momentum is realized by the change in sign of the
holomorphic spinor~\cite{FreedmanUnitarity},
\begin{eqnarray}
p_i \mapsto -p_i
~~~&\leftrightarrow&~~~
\lambda_i^\alpha\mapsto -\lambda_i^\alpha \,,
~~~~~~
\tlambda_i^{\dot\alpha}\mapsto \tlambda_i^{\dot\alpha}\,,
 \nn \\
~~~&\leftrightarrow&~~
|-i\rangle \mapsto -|i\rangle \,,
~~~~ |-i] \mapsto |i] \,.  
\label{signrules}
\end{eqnarray}
%

\subsection{The MHV Superspace}

The supersymmetry relations between the different MHV amplitudes may
be encoded in an on-shell superspace, which conveniently packages all
amplitudes into a single object---the generating function or
superamplitude.  Each term in the superamplitude corresponds to a
regular component scattering amplitude.  
Depending upon the detailed formulation of the superspace, scattering
amplitudes of gluons, fermions and scalars are then formally extracted
either by the application of Grassmann-valued derivatives
\cite{FreedmanUnitarity}, or, equivalently, by multiplying with the 
appropriate wave functions and integrating over all Grassmann variables
\cite{Nair, WittenTopologicalString}.  
Effectively, these operations amount to selecting the component amplitude with
the desired external states.

The MHV generating function (or superamplitude) is defined as, 
\begin{equation}
{\cal A}^{\rm MHV}_n(1, 2, \ldots, n) \equiv
\frac{i}{\prod_{j=1}^n\langle j ~(j+1)\rangle}
\, \delta^{(8)} \Bigl(\sum_{j=1}^n\lambda_j^\alpha\eta_j^a \Bigr)\,,
\label{MHVSuperAmplitude}
\end{equation}
where the leg label $n+1$ is identified with the leg
label $1$, and $\eta_j^a$ are $4n$ Grassmann odd
variables labeled by leg $j$ and $SU(4)$ $R$-symmetry index $a$.  As
indicated by the cyclic denominator, this amplitude is color ordered
({\it i.e.,} it is the kinematic coefficient of a particular color trace in
\eqn{TreeDecomposition}), even though the numerator possesses full
crossing symmetry having encoded all possible MHV helicity and
particle assignments.  We suppress the delta-function factor
$(2\pi)^4\delta^{(4)}(\sum_i p_i)=
(2\pi)^4\delta^{(4)}(\sum_i\lambda_i{\tilde\lambda}_i)$ responsible 
for enforcing the
overall momentum conservation.

The eightfold Grassmann delta function in (\ref{MHVSuperAmplitude})
is a product of pairs of delta functions, each pair being associated
with one of the possible values of the $SU(4)$ $R$-symmetry index:
\begin{equation}
\delta^{(8)} \Bigl(\sum_{i=1}^n\lambda_i^\alpha\eta_i^a \Bigr)
= \prod_{a=1}^4 
\delta^{(2)} \Bigl(\sum_{i=1}^n\lambda_i^\alpha\eta_i^a \Bigr)\,.
\label{DeltaForm}
\end{equation}
This expression can be further expanded,
\begin{equation}
\delta^{(8)} \Bigl(\sum_{i=1}^n\lambda_i^\alpha \eta_i^a \Bigr)
 = \prod_{a=1}^4 \sum_{i<j}^n \spa{i}.{j} \eta_i^a  \eta_j^a \,,
\label{SpinorDeltaForm8}
\end{equation}
using the usual property of Grassmann delta functions that
$\delta(\eta) = \eta$.  Each monomial in $\eta$ in the superamplitude
corresponds to a different MHV amplitude.  In this form it is clear
that all terms indeed have eight upper $SU(4)$
indices, as expected for an MHV amplitude.

Similarly, one may define an on-shell \MHVbar{} superspace, whose
Grassmann parameters are $\etab$, in which the \MHVbar{}
superamplitude takes a form analogous to~(\ref{MHVSuperAmplitude}):
\begin{eqnarray}
\label{MHVBarConjSuperAmplitude0}
{\cal A}_n^{\rm \overline{MHV}}(1, 2, \ldots, n) & \equiv&
\frac{i (-1)^n}{\prod_{j=1}^n [ j ~(j+1)]}
\; \delta^{(8)} \Bigl(\sum_{j=1}^n\tilde 
                \lambda_j^{\dot\alpha}\etab_{j a}\Bigr) \nn \\
& = &
\frac{i (-1)^n}{\prod_{j=1}^n [ j ~(j+1)]}
\,  \prod_{a=1}^4 
\sum_{i<j }^n \spb{i}.{j} \etab_{ia}  \etab_{ja} \,.
\label{MHVBarConjSuperAmplitude}
\end{eqnarray}
The $SU(4)$ indices are now lowered, which implies that the component
\MHVbar{} amplitudes are built from the external states in
(\ref{MHVbarStates}) with a total of eight lower indices.

We note that the arguments of the MHV delta functions are the
super-momenta $Q^a$, and for \MHVbar{} are similarly the conjugate
super-momenta $\widetilde{Q}_a$,
\begin{eqnarray}
Q^{\alpha a}=\sum_i\lambda^\alpha_i\eta^a_i \,, ~~~~~~~~
\widetilde{Q}^{\dot \alpha}_a=\sum_i{\tilde\lambda}^{\dot
\alpha}_i\etab_{ia} \,,
\label{Supermomenta1}
\end{eqnarray}
where the index $i$ runs over all the external legs of the
amplitude.  Thus the purpose of the delta functions is to enforce
super-momentum conservation constraint in the respective superspaces.  For
later purposes we define the individual super-momenta of the external
legs,
\begin{eqnarray}
q^a_i=|i \rangle\eta_i^a\,,
~~~~~~~~
\qbar_{ia}= \etab_{ia}[ i|\, .  
\label{Supermomenta2}
\end{eqnarray}

The two superspaces can be related.  Following ref.~\cite{FreedmanUnitarity} 
we can rewrite the \MHVbar{} superamplitudes in the MHV
superspace (or $\eta$-superspace) via a Grassmann Fourier
transform.  For this purpose we define~\cite{FreedmanUnitarity} the
operator,
\begin{equation}
\widehat F \bullet\equiv \int \Bigl[\prod_{i,a} d \etab_{ia} \Bigr]
   \exp\Bigl(\sum_{b,j} \eta_j^b ~\etab_{jb}\Bigl)\,\bullet~~, 
\label{FourierOperation}
\end{equation}
which realizes this Fourier transform.  Then,
following~\cite{FreedmanUnitarity}, the \MHVbar{} superamplitude in the
$\eta$-superspace can be written as
\begin{equation}
\widehat F {\cal A}_n^{\rm \overline{MHV}}(1, 2, \ldots, n)=
     \frac{i (-1)^n}{\prod_{i=1}^n [ i ~(i+1)]} \prod_{a=1}^4 
     \sum_{i<j}^n \spb{i}.{ j} 
\partial_{\eta_i^a} \partial_{\eta_j^a} \eta_{1}^a \eta_{2}^a 
     \cdots \eta_{n}^a \,.
\label{MHVBarSuperAmplitude2}
\end{equation}
From this perspective, the Grassmann Fourier transform is then easily
expressed as the rule,
\begin{equation}
\spb{i}.{j} \etab_{ia}  \etab_{ja} \stackrel{\widehat F} 
\longrightarrow 
\eta_{1}^a\cdots \eta_{i-1}^a\, [i| \,\eta_{i+1}^a \cdots 
\eta_{j-1}^a \, |j]\, \eta_{j+1}^a \cdots \eta_{n}^a \,.
\label{FourierRule}
\end{equation}
Here the spinors $[i|$ and $ |j]$ are understood as being contracted
after they are brought next to each other by anticommuting them past
the various $\eta$ factors.  While the spinors
are generally taken as Grassmann-even, for the
purposes of this rule it is convenient to treat them as
Grassmann-odd.

However, in the above Fourier transformed \MHVbar{} amplitude the
notion of the numerator as a supermomentum conservation constraint
has been obscured.  This can be somewhat cured using a second
alternative presentation of the \MHVbar{} superamplitude in which we
consider an integral representation of the $\delta^{(8)}(\widetilde{Q})$,
\begin{equation}
{\cal A}_n^{\rm \overline{MHV}}(1, 2, \ldots, n)= 
\frac{i (-1)^n}{\prod_{j=1}^n [ j ~(j+1)]}
\,  
\int  \prod_{a=1}^4 d^2\omega^a 
 \prod_{i=1}^n
\exp{({\tilde\eta}_{ia}{\tilde\lambda}_i^{\dot\alpha}\omega_{\dot\alpha}^a)}\,,
\end{equation}
where $\omega_{\dot\alpha}^a$ are Grassmann odd integration
parameters, $d^2\omega^a =d\omega^a_{\dot 1} d\omega^a_{\dot 2}$.
The action of the Grassmann Fourier transform
(\ref{FourierOperation}) yields immediately \cite{KorchemskyOneLoop} a
product over one-dimensional Grassmann delta functions, one for each
external leg:
\be
\widehat F {\cal A}_n^{\rm \overline{MHV}}(1, 2, \ldots, n)=
\frac{i (-1)^n}{\prod_{j=1}^n [ j ~(j+1)]}
\,  \prod_{a=1}^4 \int d^2\omega^a 
\prod_{i=1}^n\delta (\eta^a_i-{\tilde\lambda}_i^{\dot\alpha}
\omega^a_{\dot\alpha})\,.
\label{aux_integral_MHVbar}
\ee
While somewhat obsfucated, for later purposes it is important to note the 
right-hand side of this equation is proportional to the
$\eta$-space supermomentum conservation constraint $\delta^{(8)}(Q)$
for $n>3$.
This relation may be exposed by taking an appropriate linear
combination \cite{KorchemskyOneLoop} of the arguments of the delta
functions:
\be
\sum_{i=1}^n
\lambda_i^\alpha(\eta^a_i-{\tilde\lambda}_i^{\dot\alpha}
\omega^a_{\dot\alpha}) =
\sum_{i=1}^n
(\lambda_i^\alpha\eta^a_i-(\lambda_i^\alpha{\tilde\lambda}_i^{\dot\alpha})
\omega^a_{\dot\alpha})=\sum_{i=1}^n\lambda_i^\alpha\eta^a_i \,,
\label{MHVbarSuperMomCons}
\ee
upon using the momentum conservation constraint
$\sum_i\lambda_i^\alpha{\tilde\lambda}_i^{\dot\alpha}=0$.  (For $n=3$
the Fourier transformed \MHVbar{} amplitude is not proportional to
$\delta^{(8)}(Q)$.  Even so, this amplitude still conserves
supermomentum and is invariant under
$Q$-supersymmetry~\cite{KorchemskyOneLoop}.)  While these
manipulations may be explicitly carried out at the expense of
introducing a Jacobian factor, it is frequently more convenient not to
do so.  Indeed, we will more often work directly with equation
(\ref{aux_integral_MHVbar}).

\subsection{Diagrammatic representation of MHV superamplitude}
\label{DiagramSubSection}

As mentioned, we are interested in simplifying the evaluation of sums
over the members of the $\NeqFour$ multiplet and uncovering their
structure.  For this purpose we introduce a diagrammatic approach for
capturing the superspace properties of MHV amplitudes.  These diagrams
will be in one-to-one correspondence with the contributions to any given 
cut amplitude, allowing us to map out the structure of its supersum.  
We will give rules for translating the diagrams into algebraic results,
including those for the Grassmann parameters needed to obtain the
correct relative signs.  While constructed for the maximally
supersymmetric Yang-Mills theory in four dimensions, the ideas behind
this method extend to theories with reduced supersymmetry (see
\sect{FewerSusySubsection}), being particularly well-suited for
studying deformations of $\NeqFour$ super-Yang-Mills theory.

\begin{figure}[th]
\centerline{\epsfxsize 4.5 truein \epsfbox{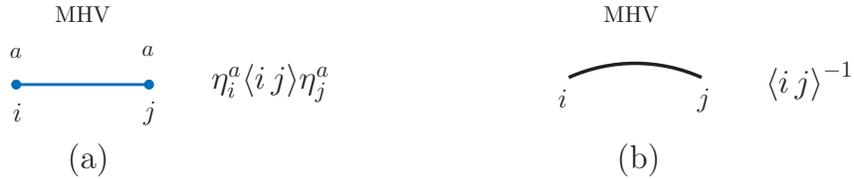}}
\caption[a]{\small  For an MHV amplitude the shaded (blue) ``index line'' 
(a) connecting leg $i$ to leg $j$ represents $\langle q_i^a \,
q_j^a\rangle$.  The two endpoints (and line) carry the same $SU(4)$
index.  A solid (black) line (b) without endpoint dots represents a
spinor product in the denominator.  }
\label{MHVruleFigure}
\end{figure}

\begin{figure}[th]
\centerline{\epsfxsize 5.6 truein \epsfbox{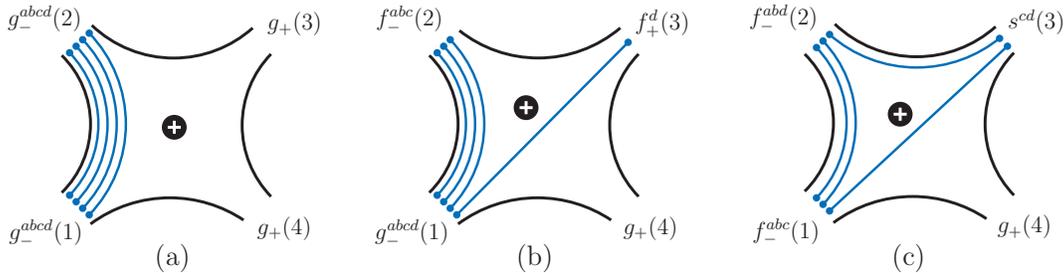}}
\caption[a]{\small Examples of $SU(4)$ index diagrams for specifying
numerator factors of MHV tree amplitudes.  The diagrams (a), (b) and
(c) correspond to the amplitudes in
\eqn{SuperTreeExamples}.  The shaded (blue) line connecting leg $i$ to
leg $j$ represents a factor of $\eta^a_i\spa{i}.{j}\eta^a_j$
respectively, and solid (black) lines represents
$\spa{i}.{j}^{-1}$.  The white ``$+$'' label on black background
indicates that the amplitude is holomorphic, or MHV.}
\label{SampleMHVTreesFigure}
\end{figure}

Inspecting the eightfold Grassmann delta function, 
as given in \eqn{SpinorDeltaForm8},
we recognize that the basic building block of the MHV amplitude numerators
is the spinor product of supermomenta,
\begin{equation}
 \spa{q_i^a}.{q_j^a} \equiv \eta_i^a \spa{i}.{j}  \eta_j^a \,.
\label{SpinorDeltaForm2}
\end{equation}
For each $SU(4)$ index, the delta function in \eqn{SpinorDeltaForm8}
is simply the sum over all such products.  We represent the
supermomentum product graphically by a shaded (blue) line connecting
point $i$ and $j$, as in \fig{MHVruleFigure}(a).  We will call this
object ``index line''.
%
%
In addition to the Grassmann delta function, color-ordered MHV
amplitudes also have another important structure, the cyclic spinor
string in the denominator,
\begin{equation}
( \spa{1}.{2} \spa{2}.{3} \spa{3}.{4} \spa{4}.{5} \ldots \spa{n}.{1} ) ^{-1} .
\label{SpinorString}
\end{equation}
This object has the same order as the trace of color-group generators,
and can be thought of as being in one-to-one correspondence with this
color structure.  The spinor products in the denominator of MHV
amplitudes will be represented by solid (black) lines without endpoint
dots shown in \fig{MHVruleFigure}(b).  The cyclicity of the MHV
denominator implies that these lines form closed loops, except for the
small gaps that we take to represent external states.  It is
convenient to draw the diagrams in a form reminiscent of string theory
world-sheets, as displayed in \fig{SampleMHVTreesFigure}.  The main
role of the solid (black) lines will be to span the background, or
canvas, on which the shaded (blue) $SU(4)$ index lines are drawn.  The
presentation of amplitudes in this world-sheet-like fashion provides
the necessary room to draw the index lines without cluttering the
figures.  These diagrams---which we will call ``index
diagrams''---capture the spinor structures of MHV tree amplitudes
along with the relative signs encoded by the superspace.

Given an MHV tree $n$-point amplitude with specified external states,
the rules for drawing the $SU(4)$ index diagram are simple: First draw
the $n$ solid (black) lines representing the cyclic spinor string of
the MHV amplitude denominator.  Leave $n$ gaps between these lines to
represent the external states, or legs.  Label these legs with the 
appropriate momentum, helicity and
$SU(4)$ indices.  If the same $SU(4)$ index appears on external legs
they should be connected by a shaded (blue) line with endpoint dots.  
This completes the diagram.

Consider, for example, the tree amplitudes in \eqn{TreeExamples}, whose
corresponding diagrams are shown in \fig{SampleMHVTreesFigure}.  The
``$+$'' and ``$-$'' labels on the external states indicate the
helicities, while the black-and-white-inverted ``$+$'' and ``$-$''
labels internal to the diagram indicates whether it is an MHV or
\MHVbar{} amplitude, respectively.  
We will refer to this property of being either MHV or \MHVbar{} as an amplitude's 
{\it holomorphicity}, as MHV amplitudes are built from
holomorphic $\lambda^\alpha$ spinors and \MHVbar{} amplitudes are
constructed from anti-holomorphic $\tilde \lambda^{\dot \alpha}$ spinors.  
From the above construction it follows that the index lines in the
diagrams of \fig{SampleMHVTreesFigure} are in one-to-one
correspondence to components in the MHV superamplitude, including the
Grassmann parameters.  Translating  from the figures to analytic expressions using the rules of \fig{MHVruleFigure}, we can easily write down these component amplitudes, 
\begin{eqnarray}
&&{\rm (a)}:\; \langle g^{1234}_{-}(1)g^{1234}_{-}(2)g_{+}(3)g_{+}(4)\rangle
 = i {\prod_{a=1}^4 \spa{q^a_1}.{q^a_2} \over 
\spa{1}.{2}\spa{2}.{3}\spa{3}.{4}\spa{4}.{1}} \,, \nn \\
&&{\rm (b)}:\; \langle g^{abcd}_{-}(1)f^{abc}_{-}(2)f^d_{+}(3)g_{+}(4)\rangle
 = i { \spa{q^a_1}.{q^a_2}  \spa{q^b_1}.{q^b_2}
    \spa{q^c_1}.{q^c_2}  \spa{q^d_1}.{q^d_3} 
 \over \spa{1}.{2}\spa{2}.{3}\spa{3}.{4}\spa{4}.{1}} \,,\nn \\
&&{\rm (c)}:\; \langle f^{abc}_{-}(1)f^{abd}_{-}(2)s^{cd}(3)g_{+}(4)\rangle 
= i {\spa{q^a_1}.{q^a_2}  \spa{q^b_1}.{q^b_2}  \spa{q^c_1}.{q^c_3} 
 \spa{q^d_2}.{q^d_3} \over \spa{1}.{2}\spa{2}.{3}\spa{3}.{4}\spa{4}.{1}} \,,
 \label{SuperTreeExamples}
\end{eqnarray}
where we have labeled the color ordered amplitudes (including
Grassmann parameters) using a ``correlator'' notation on 
the left hand side.
Repeated
indices are not summed over their values; rather, their values are
fixed and correspond to the particular choice of $SU(4)$ labels
identifying the external states.  For the amplitude to be nonvanishing,
the labels $a,b,c,d$ must be distinct.

\begin{figure}[th]
\centerline{\epsfxsize 5.6 truein \epsfbox{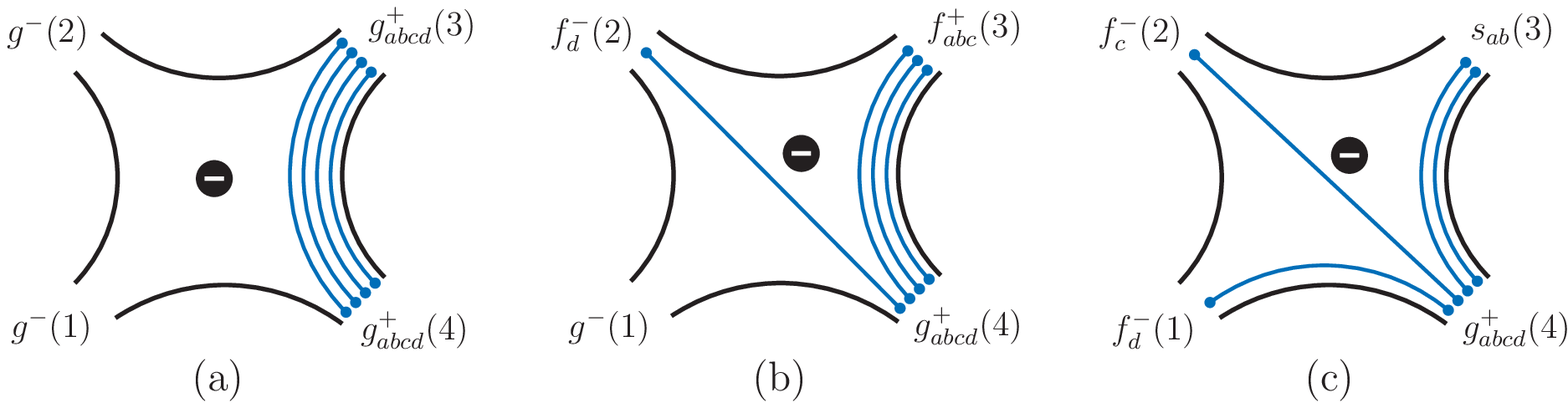}}
\caption[a]{\small The same amplitudes as in \fig{SampleMHVTreesFigure},
now in the \MHVbar{} representation.  The shaded (blue) line connecting
leg $i$ to leg $j$ represents a factor of
$\etab_{ia}\spb{i}.{j}\etab_{ja}$ respectively, and solid (black)
lines represents $\spb{i}.{j}^{-1}$.  The white ``$-$'' label on black
background indicates that the amplitude is anti-holomorphic, or
\MHVbar{}.}
\label{SampleMHVbarTreesFigure}
\end{figure}

Diagrams tracking the $SU(4)$ indices for \MHVbar{} amplitudes are
similar.  As a simple example, consider the same amplitudes as above,
but reinterpreted as \MHVbar{} amplitudes---for four-point amplitudes
(but no others) this is always possible.  In the \MHVbar{} form the
amplitudes are,
\begin{eqnarray}
&&{\rm (a)}:\; \langle g^{-}(1)g^{-}(2)g^{+}_{1234}(3)g^{+}_{1234}(4) \rangle=
 i {\prod_{a=1}^4 \spb{\qbar_{3a}}.{\qbar_{4a}} \over
 \spb{1}.{2}\spb{2}.{3}\spb{3}.{4}\spb{4}.{1}} \,, \nn \\
&&{\rm (b)}:\; \langle g^{-}(1)f_{d}^{-}(2)f^{+}_{abc}(3)g^{+}_{abcd}(4)\rangle
= i { \spb{\qbar_{3a}}.{\qbar_{4a}} \spb{\qbar_{3b}}.{\qbar_{4b}} 
\spb{\qbar_{3c}}.{\qbar_{4c}} \spb{\qbar_{2d}}.{\qbar_{4d}}
 \over \spb{1}.{2}\spb{2}.{3}\spb{3}.{4}\spb{4}.{1}} \,, \nn \\
&&{\rm (c)}:\; \langle f_{d}^{-}(1)f_{c}^{-}(2)s_{ab}(3)g^{+}_{abcd}(4)\rangle 
= i {\spb{\qbar_{3a}}.{\qbar_{4a}} \spb{\qbar_{3b}}.{\qbar_{4b}} 
\spb{\qbar_{2c}}.{\qbar_{4c}} \spb{\qbar_{1d}}.{\qbar_{4d}} 
  \over \spb{1}.{2}\spb{2}.{3}\spb{3}.{4}\spb{4}.{1}} \,,
\end{eqnarray}
where $\qbar_{ia}$ are the conjugate supermomenta defined in
\eqn{Supermomenta2}.  The index diagrams corresponding to these
expressions are displayed in \fig{SampleMHVbarTreesFigure}.  Now the
lines are interpreted in terms of conjugate or anti-holomorphic
spinors and Grassmann parameters.  As mentioned above, this is
indicated by the black-and-white-inverted ``$-$'' label on each
\MHVbar{} diagram.

If we wish to work entirely in the $\eta$-superspace for both MHV and
\MHVbar{} amplitudes, we must map the $\etab$ parameters to
$\eta$'s using the Grassmann Fourier transform $\widehat F$ in
\eqn{FourierOperation}.  This transformation is conveniently captured by
the rule in \eqn{FourierRule}, giving,
\begin{eqnarray}
&&{\rm (a)}:\; \widehat F \, 
\langle g^{-}(1)g^{-}(2)g^{+}_{1234}(3)g^{+}_{1234}(4) \rangle
= i {\prod_{a=1}^4 \eta_{1}^a \eta_{2}^a 
\spb{3}.{4} \over \spb{1}.{2}\spb{2}.{3}\spb{3}.{4}\spb{4}.{1}} \,,  \nn \\
&&{\rm (b)}:\; \widehat F \,
 \langle g^{-}(1)f_{d}^{-}(2)f^{+}_{abc}(3)g^{+}_{abcd}(4) \rangle 
= i { \eta_{1}^a\eta_{2}^a\spb{3}.{4} \eta_{1}^b\eta_{2}^b\spb{3}.{4} 
   \eta_{1}^c\eta_{2}^c\spb{3}.{4} \eta_{1}^d\eta_{3}^d\spb{4}.{2} 
   \over \spb{1}.{2}\spb{2}.{3}\spb{3}.{4}\spb{4}.{1}} \,,  \nn \\
&&{\rm (c)}:\; \widehat F \, 
\langle f_{d}^{-}(1)f_{c}^{-}(2)s_{ab}(3)g^{+}_{abcd}(4) \rangle
 = i { \eta_{1}^a\eta_{2}^a\spb{3}.{4} \eta_{1}^b\eta_{2}^b\spb{3}.{4}
   \eta_{1}^c\eta_{3}^c\spb{4}.{2} \eta_{2}^d\eta_{3}^d\spb{1}.{4}  
   \over \spb{1}.{2}\spb{2}.{3}\spb{3}.{4}\spb{4}.{1}} \,.
\end{eqnarray}

While perhaps less obvious for the time being, the utility of the
index diagrams will become apparent in \sect{DiagramLoopSection},
where they will allow a transparent bookkeeping of the helicity states
in unitarity cuts of multi-loop (super)amplitudes.

\subsection{MHV superrules for non-MHV superamplitudes}
\label{MHVSuperRuleSubsection}

\begin{figure}[t]
\centerline{\epsfxsize 5.  truein \epsfbox{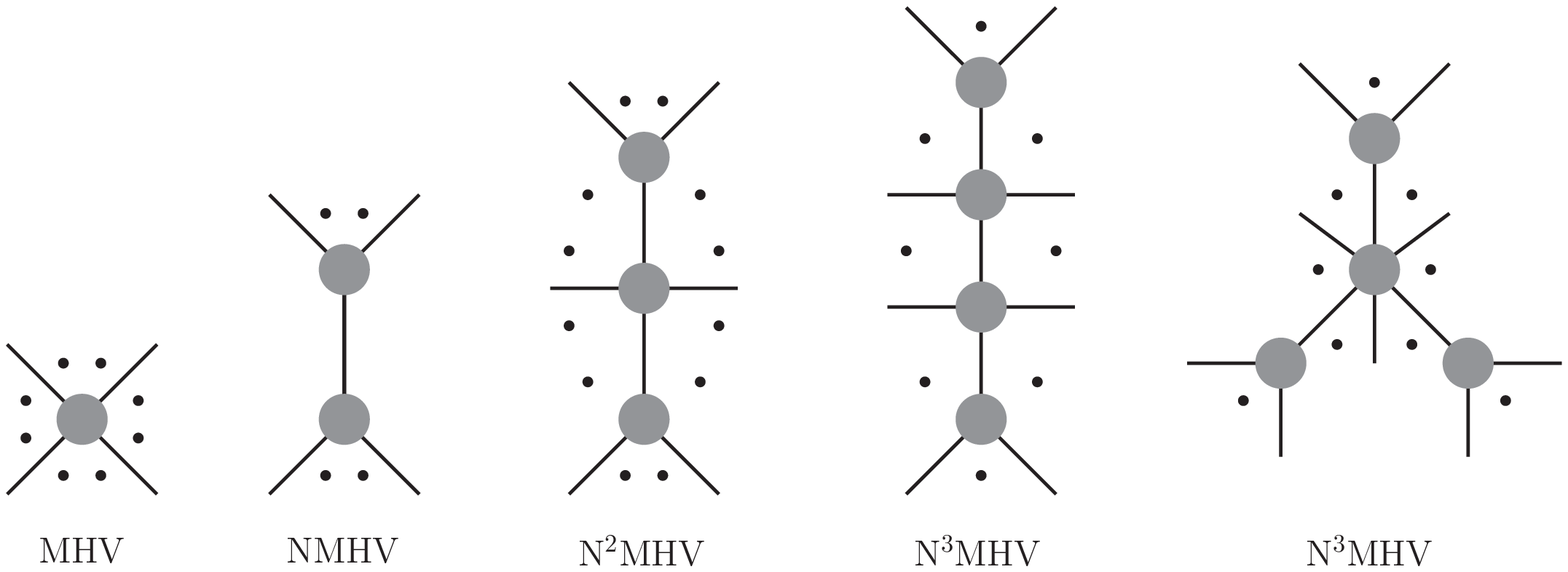}}
\caption[a]{\small The MHV vertex construction builds non-MHV
superamplitudes from MHV superamplitudes.  The blobs are MHV
superamplitudes, and the dots signify an arbitrary number of external
legs, of which a few are drawn explicitly.}
\label{CSWFigure}
\end{figure}

The MHV vertex construction generates non-MHV amplitudes from the MHV
ones via a set of simple diagrammatic rules.  Their validity has been proven 
in various ways, including the use of on-shell
recursion~\cite{Risager} and by realizing the MHV vertex rules as the
Feynman rules of a Lagrangian \cite{CSW_Lagrangian, CSW_QCD}.  The
former approach was recently shown to hold, with certain
modifications, for all amplitudes of $\NeqFour$ super-Yang-Mills
theory~\cite{FreedmanProof}, proving the validity of the MHV vertex
construction for the complete theory.  The latter approach was also
extended \cite{CSW_LagrangianSusy} to the complete $\NeqFour$
Lagrangian by carrying out an ${\cal N}=4$ supersymmetrization of the
MHV Lagrangian of refs.~\cite{CSW_Lagrangian}.  

The $n$-point N$^m$MHV gauge theory superamplitude (where the ``N''
stands for ``Next-to-'') contains gluon amplitudes with $(m+2)$
negative helicity gluons.  One begins its construction by drawing all
tree graphs with $(m+1)$ vertices, on which the external $n$ legs are
distributed in all possible inequivalent ways while maintaining the
color order.  Examples of these graph topologies are shown in
\fig{CSWFigure}.

 To each vertex one associates an MHV superamplitude
(\ref{MHVSuperAmplitude}).  As in the bosonic MHV rules, the
holomorphic spinor $\lambda_P$ associated to an internal leg is
constructed from the corresponding off-shell momentum $P$
using an arbitrary (but the same for all graphs)
null reference antiholomorphic spinor
$\zeta^{\dot\alpha}$,
\begin{equation} 
\lambda_{P\alpha} \equiv
P_{\alpha{\dot\alpha}}\zeta ^{\dot\alpha} \,.
\label{CSW_spinor}
\end{equation} 
Alternatively, the holomorphic spinor $\lambda_P=|\Pflat\rangle$ can be
defined in terms of a ``null projection'' of $P$, given 
by~\cite{K_proj, BBK},
\begin{equation}
P^\flat = P -\frac{P^2}{2\zeta\cdot P }\zeta\,,
\label{CSWMomShift}
\end{equation}
where $\zeta^\mu$ is a null reference vector.  In this form it is
clear that the momenta of every vertex are on shell, thus, at this stage,
the expression corresponding to each graph is a simple product of
$(m+1)$ well-defined on-shell tree superamplitudes.  (The analogous
construction for gravity amplitudes is more complicated due to the
fact that MHV supergravity amplitudes are not
holomorphic~\cite{GravityMHV}.)

To each internal line connecting two vertices one associates a
super-propagator which consists of the product between a standard
scalar Feynman propagator $i/P^2$ and a factor which equates the
fermionic coordinates $\eta$ of the internal line in the two vertices
connected by it.  The structure of the propagator depends on the
precise definition of the superspace, but such details are not
important for the following.  Upon application of the precise rules
for assembling the MHV vertex diagrams, the expression for the
N$^m$MHV superamplitude is given by
\begin{equation}
{\cal A}^{{\rm N}^m{\rm MHV}}_n = i^m
\sum_{\rm all~graphs}\int \Bigl[ \prod_{j=1}^{m} {d^4\eta_j\over P^2_j}\Bigr]
 {\cal A}^{\rm MHV}_{(1)} {\cal A}^{\rm MHV}_{(2)}\cdots 
  {\cal A}^{\rm MHV}_{(m)}{\cal A}^{\rm MHV}_{(m+1)}\,,
\label{CSWequation}
\end{equation}
where the integral is over the $4m$  internal Grassmann parameters 
($d^4\eta_j \equiv \prod_{a=1}^4d\eta_j^a$) associated with the internal legs, 
and each $P_j$ is the (off-shell) momentum of the $j$'th internal leg of the 
graph.  
The MHV superamplitudes appearing in the product
correspond to the $(m+1)$ vertices of the graph.  The
momentum and $\eta$ dependence of the MHV superamplitudes is
suppressed here.  We note, however, that the null projection of each 
internal momentum  $P^\flat_i$ and the Grassmann variable $\eta_i^a$ 
appear twice, in the form,
\begin{equation}
\cdots{\cal A}^{\rm MHV}_{(j)}(P^\flat_i,\eta_i^a)\cdots 
{\cal A}^{\rm MHV}_{(k)}(-P^\flat_i,\eta_i^a)\cdots
\label{CSWsewing}
\end{equation}

Each integration $\int d^4\eta_i$ in \eqn{CSWequation} selects the
configurations with exactly four distinct $\eta$-variables
$\eta^1_i\eta^2_i\eta^3_i\eta^4_i$ on each of the internal lines.
Since a particular $\eta^a_i$ can originate from either of two MHV
amplitudes, as per \eqn{CSWsewing}, there are $2^4$ possibilities that
may give non-vanishing contributions.  These contributions correspond
to the 16 states in the $\NeqFour$ multiplet, making it clear that the
application of $\int d^4\eta_i$ indeed yields the supersum.  However, for a given choice of external states, each term
corresponding to a distinct graph in (\ref{CSWequation}) receives
nonzero contributions from exactly one state for each internal leg.

Note that as far as sewing of amplitudes is concerned, it makes no
difference whether an intermediate state is put on-shell due to a cut
or due to the MHV vertex expansion.  This observation, implying that
sewing of general amplitudes proceeds by integrating over common
$\eta$ variables, will play an important role in our discussion of
cuts of loop amplitudes.

\begin{figure}[th]
\centerline{\epsfxsize 6.6 truein \epsfbox{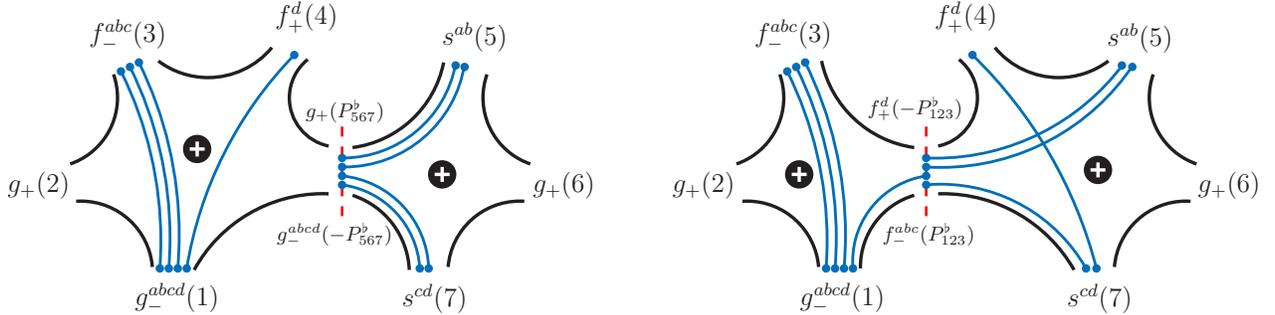}}
\caption[a]{\small The index lines for two out of the nine
diagrams of the MHV vertex expansion for the amplitude $\langle
g_-^{abcd}(1) g_+(2) f_-^{abc}(3)f_+^d(4) s^{ab}(5) g_+(6)
s^{cd}(7)\rangle$.  The dashed vertical (red) line signifies that the
intermediate state is on-shell.  The integration $\int d^4\eta$ will
force exactly four $SU(4)$ index lines to end (or start) on the
intermediate on-shell state.}
\label{CSWFlavorLinesFigure}
\end{figure}

We now illustrate the index diagrams, introduced in the previous
section, for the MHV-vertex expansion of an NMHV example.  Since the
index diagrams represent component amplitudes these diagrams clarify
the details of the $\NeqFour$ state sum.  First we note that according
to \eqn{CSWequation} an N$^m$MHV amplitude is a polynomial in $\eta$ of
degree $8(m+1)-4m=8+4m$, since there are $(m+1)$ MHV
amplitudes---which by definition contain eight $\eta$'s with upper
$SU(4)$ indices---and the Grassmann integration removes $4m$ of them.
Thus, an NMHV amplitude must have 12 ($3 \times 4$ distinct)
upper $SU(4)$ indices.

Let us consider the seven-point amplitude 
$\langle g_-^{abcd}(1) g_+(2) f_-^{abc}(3)
f_+^d(4)s^{ab}(5) g_+(6) s^{cd}(7)\rangle$ which is of this form.
There are a total of nine non-vanishing diagrams, of
which two are displayed as index diagrams in
\fig{CSWFlavorLinesFigure}, illustrating the sewing of gluonic and
fermionic states, respectively.  
Summing over the diagrams gives us the amplitude
\begin{eqnarray}
&& \hskip - .2 cm \langle g_-^{abcd}(1) g_+(2) f_-^{abc}(3) f_+^d(4) s^{ab}(5)
g_+(6) s^{cd}(7)\rangle \nn  \\
&& \hskip .2  cm = 
\int d^4 \eta_{P^\flat_{567}}
\langle g_-^{abcd}(1) g_+(2) f_-^{abc}(3) f_+^d(4) g_+(P^\flat_{567})\rangle
  \frac{i}{(P_{567})^2} 
\langle g_-^{abcd}(-P^\flat_{567}) s^{ab}(5) g_+(6) s^{cd}(7)\rangle \nn \\
&& \hskip .5  cm \null 
+ \int d^4 \eta_{P^\flat_{123}}
\langle f_+^d(4)  s^{ab}(5) g_+(6) s^{cd}(7)  f^{abc}_-(P^\flat_{123})\rangle
  \frac{i}{(P_{123})^2} 
\langle f_+^{d}(-P^\flat_{123}) g_-^{abcd}(1) g_+(2) f_-^{abc}(3) \rangle \nn\\
&& \hskip .5  cm \null
+ \cdots \nn 
\end{eqnarray}
\begin{eqnarray}
&&  = 
-i \frac{\spash{q_1^a}.{q_3^a} \spash{q_1^b}.{q_3^b} \spash{q_1^c}.{q_3^c} 
        \spash{q_1^d}.{q_4^d} }
{\spash1.2 \spash2.3 \spash3.4 \spash4.{\Pflat_{567}} \spash{\Pflat_{567}}.1} 
\frac{1}{(P_{567})^2} 
\frac{\spash{P^\flat_{567}}.{q_5^a} \spash{P^\flat_{567}}.{q_5^b} 
      \spash{P^\flat_{567}}.{q_7^c} \spash{P^\flat_{567}}.{q_7^d} }
 {\spash{P^\flat_{567}}.5 \spash5.6 \spash6.7 \spash7.{P^\flat_{567}}} 
\hskip 15.2 mm  \nn\\
&& \hskip 5 mm \null 
+ i \frac{\spash{P^\flat_{123}}.{q_5^a}
\spash{P^\flat_{123}}.{q_5^b} 
\spash{P^\flat_{123}}.{q_7^c}  \spash{q_4^d}.{q_7^d} }
{\spash4.5 \spash5.6 \spash6.7 \spash7.{P^\flat_{123}}
\spash{P^\flat_{123}}.4} 
\frac{1}{(P_{123})^2} 
\frac{\spash{q_1^a}.{q_3^a} \spash{q_1^b}.{q_3^b} 
      \spash{q_1^c}.{q_3^c} \spash{P^\flat_{123}}.{q_1^d} }
 {\spash{P^\flat_{123}}.1 \spash1.2 \spash2.3 \spash3.{P^\flat_{123}}} \nn\\
&& \hskip .5 cm \null +
 \cdots 
\end{eqnarray}
where,
\begin{equation}
P_{ijl} = k_{i}+ k_j + k_l \,, \hskip 2 cm 
\langle P^\flat \,q^a_i \rangle = \langle P^\flat \, i \rangle \eta^a_i \,,
\end{equation}
and we suppress all but the contributions of the two diagrams in
\fig{CSWFlavorLinesFigure}.  In the last equality we carried out the
Grassmann integration, which here only serves to convert the internal
four powers of $\eta$ to factors of $\pm 1$.  When using the MHV diagrams expansion in unitarity cuts of loop
amplitudes, as we will see in \sect{LinearSystemSection}, it is
generally convenient to delay carrying out the Grassmann integrations
until the complete cut is assembled.

We note that it is convenient to collect the various N$^m$MHV tree superamplitudes into a single generating function,
\begin{equation}
{\cal A}^\tree = {\cal A}^{\rm MHV} + {\cal  A}^{\rm NMHV} + 
{\cal  A}^{\rm N^2MHV}  + \cdots +  {\cal  A}^{{\rm N}^{(n-4)}{\rm MHV}} \,,
\label{GenSuperAmplitude}
\end{equation}
where $n$ is the number of external legs, and the sum terminates with
the \MHVbar{} amplitude, here written as an N$^{(n-4)}$MHV amplitude in
$\eta$ superspace.  The number of terms in this sum is $n-3$ for
$n\ge4$.  The three-point case should be treated separately since it
contains two terms, MHV and \MHVbar, which cannot be supported on the
same kinematics.

\section{Evaluation of Loop Amplitudes using the Unitarity Method}
\label{UnitarityMethodSection}

The direct evaluation of generalized unitarity cuts of ${\cal N}=4$
super-Yang-Mills scattering amplitudes requires summing over all
possible intermediate on-shell states of the theory.  
Various strategies for carrying out such sums over states have
recently been discussed in refs.~\cite{KorchemskyOneLoop, AHCKGravity,
FreedmanUnitarity}.  Here we review our current approach, which is
closely related to the generating function ideas of
ref.~\cite{FreedmanGenerating,FreedmanUnitarity}.  Additionally, we
present an analysis of the structure of the resulting factors and
expose various universal features.

\subsection{Modern unitarity method}

The modern unitarity method gives us a means for systematically
constructing multi-loop amplitudes for massless theories.  This method
and its various refinements have been described in some detail in
references~\cite{UnitarityMethod, Fusing,BRY,
GeneralizedUnitarity,TwoLoopSplit,BDDPR, BCFGeneralized, FiveLoop,
FreddyMaximal}, so here we will mainly review points salient to the
sums over all intermediate states appearing in maximally supersymmetric
theories.

\begin{figure}[th]
\centerline{\epsfxsize 6.5 truein \epsfbox{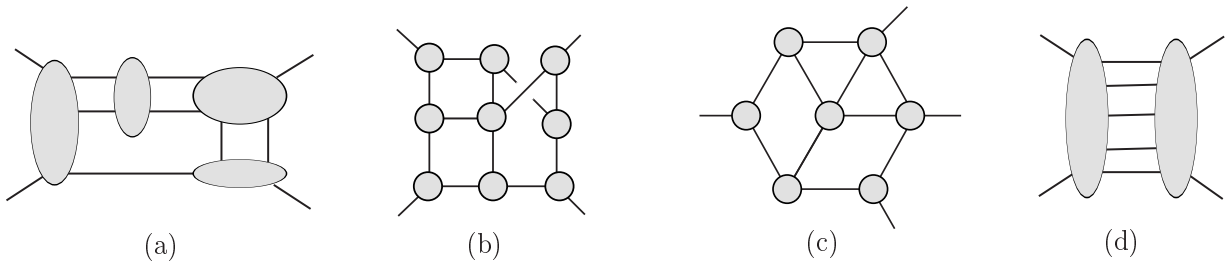}}
\caption[a]{\small Some examples of generalized cuts at four loops.
Every exposed line is cut and satisfies on-shell conditions.
Diagrams (b) and(c) are near maximal cuts.  In four
dimensions only MHV or \MHVbar{} tree amplitudes appear in cuts (a)-(c)
while in cut (d) non-MHV tree amplitudes appear.}
\label{SampleCutsFigure}
\end{figure}

The construction starts with an ansatz for the amplitude in terms of
loop momentum integrals.  We require that the numerator of each
integral is a polynomial in the loop and external momenta subject to
certain constraints, such as the maximum number of factors of loop
momenta that can appear.  The construction of such an ansatz is
simplest for the $\NeqFour$ super-Yang-Mills four-point amplitudes
where it turns out that the ratio between the loop integrand and the
tree amplitudes is a rational function solely of Lorentz invariant scalar 
products~\cite{BRY,BCDKS, FiveLoop}.  For higher-point amplitudes
similar ratios necessarily contain either spinor products or
Levi-Civita tensors, as is visible even at one
loop~\cite{UnitarityMethod}.

The arbitrary coefficients appearing in the ansatz are systematically
constrained by comparing generalized cuts of the ansatz to cuts of the
loop amplitude.  Particularly useful are cuts composed of $m$ tree
amplitudes of form,
\begin{equation}
\sum_{\rm states} A^\tree_{(1)} A^\tree_{(2)} A^\tree_{(3)} \cdots 
A^\tree_{(m)} \,,
\label{GeneralizedCut}
\end{equation}
evaluated using kinematic configurations that place all cut momenta on
shell, $l_i^2 = 0$.  Cuts which break up loop amplitudes into products
of tree amplitudes are generally the simplest to work with to
determine an amplitude, although one can also use lower-loop
amplitudes in the cuts as well.  In special cases, such as when there
is a four-point subamplitude, this can be
advantageous~\cite{FourLoopNonPlanar}.  In \fig{SampleCutsFigure}, we
display a few unitarity cuts relevant to four loops.  If cuts of the
ansatz cannot be made consistent with the cuts of the amplitude, then
it is, of course, necessary to enlarge the ansatz.

The reconstruction of an amplitude from a single cut configuration is
typically ambiguous as the numerator may be freely modified by adding
terms which vanish on the cut in question.  Consider, for example, a
particular two-particle cut with cut momenta labeled $l_1$ and $l_2$.
No expressions proportional to $l_1^2=0$ and $l_2^2=0$ are constrained
by this particular cut.  Such terms are instead constrained by other
cuts.  After information from all cuts is included, the only remaining
ambiguities are terms which are free of cuts in {\it every}
channel.
In the full amplitude these ambiguities add up to zero,
representing the freedom to re-express the amplitude into different 
algebraically equivalent forms. 
Using this freedom one
can find representations with different desirable properties, such as
manifest symmetries or explicit power
counting~\cite{GravityThree,CompactThree}.

For multi-loop calculations, generally it is best to organize the
evaluation of the cuts following to the method of maximal
cuts~\cite{FiveLoop}.  In this procedure we start from generalized
cuts \cite{GeneralizedUnitarity, TwoLoopSplit,BCFGeneralized} with the
maximum number of cut propagators and then systematically reduce the
number of cut propagators~\cite{FiveLoop}.  This allows us to isolate
the missing pieces of the amplitude, as well as reduce the
computational complexity of each cut.  A related procedure is the
``leading-singularity'' technique, valid for maximally supersymmetric
amplitudes~\cite{FreddyMaximal,LeadingApplications}.  These leading
singularities, which include additional hidden singularities, have
been suggested to determine any maximally supersymmetric
amplitude~\cite{AHCKGravity}.

At one loop, all singular and finite terms in amplitudes of massless
supersymmetric theories are determined completely by their
four-dimensional cuts~\cite{Fusing}.  Unfortunately, no such property
has been demonstrated at higher loops, although there is evidence that
it holds for four-point amplitudes in this theory through five
loops~\cite{BCDKS,GravityThree,FiveLoop}.  We do not expect that it
will continue for higher-point amplitudes.  Indeed, we know that for
two-loop six-point amplitudes terms which vanish in $D=4$ do
appear~\cite{TwoLoopSixPt}.  Even at four points, Gram determinants
which vanish in four dimensions, but not in $D$-dimensions, could
appear at higher-loop orders.

At present, $D$-dimensional evaluation of cuts is required to guarantee that
integrand contributions which vanish in four dimensions are not
dropped.  $D$-dimensional cuts~\cite{DDimUnitarity} make calculations
significantly more difficult, because powerful four-dimensional spinor
methods~\cite{SpinorHelicity} can no longer be used.  (Recently,
however, a helicity-like formalism in six dimensions has been
given~\cite{D6Helicity}.)  Some of this additional complexity is
avoided by performing internal-state sums using the (simpler) gauge
supermultiplet of $D=10,\, {\cal N}=1$ super-Yang-Mills theory instead
of the $D=4,\, \NeqFour$ multiplet.  In any case, it is usually much
simpler to verify an ansatz constructed using the simpler
four-dimensional analysis, than to construct the amplitude directly
from its $D$-dimensional cuts.

For simple four-dimensional cuts, the sum over states
in~\eqn{GeneralizedCut}, can easily be evaluated in components, making
use of supersymmetry Ward identities~\cite{SWI}, as discussed in
ref.~\cite{BDDPR}.  In some cases, when maximal or nearly maximal
number of propagators are cut, it is possible to choose ``singlet''
kinematics which force all or nearly all particles propagating in the
loops to be gluons in the $\NeqFour$ super-Yang-Mills theory~\cite{FiveLoop}.
However, for more general situations, we desire a systematic means for
evaluating supersymmetric cuts, such as the generating function
approach of ref.~\cite{FreedmanGenerating,FreedmanUnitarity}.

\subsection{General structure of a supercut}
\label{StructureSection}

Using superamplitudes, integration over the $\eta$ parameters of the
cut legs represents the sum over states crossing the cuts in
\eqn{GeneralizedCut}.  The generalized supercut is given by,
\begin{equation}
{\cal C} = 
\int \Bigl[ \prod_{i=1}^{k} {d^4\eta_i}\Bigr]
 {\cal A}^\tree_{(1)}
  {\cal A}^\tree_{(2)}
  {\cal A}^\tree_{(3)} \cdots 
   {\cal  A}^\tree_{(m)}\,,
\label{SuperCut}
\end{equation}
where ${\cal A}_{(j)}^{\rm tree}$ are generating functions
(\ref{GenSuperAmplitude}) connected by $k$ on-shell cut legs.  The
supercut incorporates all internal and external helicities and
particles of the $\NeqFour$ multiplet.  In most cases it is convenient
to restrict this cut by choosing external configurations, {\it e.g.}
external MHV or \MHVbar{} sectors (or even external helicities), {\it
etc}.  In many cases it is also convenient to expand out each ${\cal
A}^\tree$ into its N$^m$MHV components, and consider each term--consisting
of a product of such amplitudes---as a separate contribution.  We will
focus our analysis on such single terms, since as we will see they
form naturally distinct contributions, each being an $SU(4)$
invariant~\cite{KorchemskyOneLoop} expression.  As these contributions
correspond to internal quantities they must be summed over.  We note
that although in this discussion we restrict to cuts containing only
trees, it can sometimes be advantageous to consider cuts containing
also four and five-point loop amplitudes, since they
satisfy the same supersymmetry relations as the tree-level amplitudes.

If all tree amplitudes in the supercut have
fewer than six legs then each supercut contribution is of the form,
\begin{equation}
\int \Bigl[ \prod_{i=1}^{k} {d^4\eta_i}\Bigr]
 {\cal A}^{\rm MHV}_{(1)}\cdots 
  {\cal A}^{\rm MHV}_{(m')}
 \hat{\cal A}^{\rm \overline{MHV}}_{(m'+1)} \cdots 
 \hat {\cal  A}^{\rm \overline{MHV}}_{(m)}\,,
\label{SuperCutPart}
\end{equation}
where $\hat{\cal A}^{\rm \overline{MHV}} = \widehat F{\cal A}^{\rm
\overline{MHV}}$ uses the Grassmann Fourier transform $\widehat F$ in
\eqn{FourierOperation}.  For cuts where there are tree amplitudes with
more than five legs present, some cut contributions 
include non-MHV tree amplitudes.  For these we apply the MHV vertex
expansion (\ref{CSWequation}), which reduces these more complicated
cases down to a sum of similar expressions as \eqn{SuperCutPart} with
only MHV and \MHVbar{} amplitudes (and additional propagators).

Certain properties of the $\NeqFour$ super-Yang-Mills cuts can be
inferred from the structure of generalized cuts and the manifest
$R$-symmetry and supersymmetry of tree-level superamplitudes.  First
we note that a cut contribution that corresponds to a product of only
MHV tree amplitudes consists of a single term of the following
numerator structure,
\be
 \int  \Bigr[\prod_{i} d^4\eta_i  \Bigr] \prod_I
\biggl(\,\prod_{a=1}^4\delta^{(2)}(Q^a_I)\biggl)
= \prod_{a=1}^4  \biggl( \int  \Bigr[ \prod_{i} d\eta_i^a \Bigr] \prod_I\delta^{(2)}
(Q^a_I)  \biggr) \,,
\label{ProductCommutation}
\ee
where we have made it explicit that the product over the $SU(4)$
indices can be commuted past both the product over internal cut legs
$i$ and the product over tree amplitudes labeled by $I$.  Here
$Q_I^{a}=\sum_j\lambda_j\eta^a_j$ is the
total supermomentum of superamplitude ${\cal A}_I$, where $j$ runs
over all legs of ${\cal A}_I$.  For convenience we have also
suppressed the spinor index.  From the right-hand-side of
\eqn{ProductCommutation}, we conclude that the numerator factor
arising from the supersum of a cut contribution composed of only MHV
amplitudes is simply the fourth power of the numerator factor arising
from treating the index in $\eta^a$ as taking on only a single value.
 
A cut contribution constructed from only MHV and $\overline{\rm MHV}$
tree amplitudes has similar structure, though the details are slightly
different.  Using the fermionic Fourier transform operator
(\ref{FourierOperation}) any $n$-point $\overline{\rm MHV}$ tree
amplitude can manifestly be written as a product over four identical
factors, each depending on only one value of the $R$-symmetry index,
\begin{equation}
\prod_{a=1}^4 \; \int \Bigl[ \prod_{j}^{n} d\etab_{ja} e^{
\eta^a_j \etab_{ja}} \Bigr]\,
\delta^{(2)}\Bigl(\sum_{j=1}^n \tilde{\lambda}_j\etab^a_j\Bigr)\,.
\end{equation} 
Consequently, just as for cut contributions constructed solely from
MHV tree amplitudes, for the cases where only MHV and \MHVbar{} tree
amplitudes appear in a cut, the end result is that the fourth power
of some combination of spinor products appears in the numerator.  This
feature will play an important role in \sect{DiagramLoopSection}, 
simplifying the index diagrams that track the $R$-symmetry indices.

The super-MHV vertex expansion generalizes this structure to generic
cuts of $\NeqFour$ loop amplitudes.  As already mentioned, any non-MHV
tree superamplitude can be expanded as a sum of products of MHV
superamplitudes.  If we insert this expansion into a generalized cut,
we obtain a sum of terms where the structure of each term is the same
as a cut contribution composed purely of MHV amplitudes.  All that
changes is that the momenta carried by some spinors are shifted
according to \eqn{CSWMomShift}, and some internal propagators are made explicit.  We immediately deduce that the
numerator of each term is given by a fourth power of the numerator
factor arising when treating the index of $\eta^a$ as having a single
value.  This general observation is consistent with results found in
ref.~\cite{FreedmanUnitarity}.

The structure of the constraints due to supersymmetry may be further
disentangled.  It is not difficult to see that the cut of any
$\NeqFour$ super-Yang-Mills multi-loop amplitude is proportional to
the overall super-momentum conservation constraint on the
external supermomenta.
Similar observations have been used in a related context in
ref.~\cite{RSV_3, KorchemskyOneLoop, FreedmanProof}.  This property is
a consequence of supersymmetry being preserved by the sewing, which is
indeed manifest on the cut, as we now show.  Consider an arbitrary
generalized cut constructed entirely from tree-level amplitudes; using
the MHV-vertex super-rules, this cut may be further decomposed into a
sum of products of MHV tree amplitudes.
Each term in this sum contains a product of factors of the type
(\ref{DeltaForm}), one for each MHV amplitude in the product.  Using
the identity $\delta(A)\delta(B)=\delta(A+B)\delta(B)$ each such
product of delta functions may be reorganized by adding to the
argument of one of them the arguments of all the other
delta functions:
\be
\prod_{I=1}^m\delta^{(8)}(Q_I^a)=
\delta^{(8)}\Bigl(\sum_{I=1}^m Q_I^a \Bigr)
 \prod_{I=2}^m\delta^{(8)}(Q_I^a) \,,
\label{extract_tree}
\ee
where $m$ is the number of MHV trees amplitudes---including those from
a single graph of each MHV-vertex expansion.  In the conventions
(\ref{signrules}) in which a change of the sign of the four-momentum
$p_i$ translates to a change of sign of the holomorphic spinor
$\lambda_i$, and therefore also in $q^a_i=\lambda_i \eta^a_i$, we
immediately see that in the first delta function all $q^a_i$
corresponding to internal lines occurs pairwise with opposite sign,
and thus cancel, leaving only external variables,
\be
\delta^{(8)} \Bigl(\sum_{I=1}^m Q_I^a \Bigr) = 
\delta^{(8)} \Bigl(\sum_{i\in {\cal E}}\lambda_i \eta^a_i\Bigr) \,,
\label{OverallDelta}
\ee
where ${\cal E}$ denotes the set of external legs of the loop
amplitude whose cut one is computing.  Thus, this delta function
depends only on the external momentum configuration and is therefore
common to all terms appearing in this cut.  The generalized cuts
involving only tree amplitudes are sufficient for reconstructing the
complete loop amplitude~\cite{TwoLoopSplit}, therefore it is clear
that the superamplitude and all of its cuts are proportional to
$\delta^{(8)}(Q^a_{\cal E})$, assuming four-dimensional kinematics.

As can be seen from \eqns{aux_integral_MHVbar}{MHVbarSuperMomCons},
the discussion above, showing supermomentum conservation, goes through
unchanged for cuts containing $n$-point tree-level $\overline{\rm
MHV}$ amplitudes with $n\ge 4$.  This includes all cuts with real momenta. 
 For $n=3$, from
ref.~\cite{KorchemskyOneLoop}, we see that the supermomentum
conservation constraint of three-point amplitudes may be obtained from
their fermionic constraint upon multiplication by a spinor
corresponding to one of the external lines.  Using this observation,
it is then straightforward to show that for $n=3$
\eqns{extract_tree}{OverallDelta} continue to hold.

The explicit presence of the overall supermomentum conservation
constraint \eqn{OverallDelta} is sufficient to exhibit the
finiteness~\cite{Mandelstam} of $\NeqFour$ super-Yang-Mills theory.
Since, as we argued, the same overall delta function appears in {\it
all} cuts, it follows that the complete amplitude also has it as an
overall factor.  In fact, there is a strong similarity between the
superficial power counting that results from this and the
super-Feynman diagrams of an off-shell $N=2$ superspace.  Indeed, the
count corresponds to what we would obtain from the Feynman rules of a
superspace form of the MHV Lagrangian~\cite{CSW_LagrangianSusy} which
manifestly preserves half of the supersymmetries.

More concretely, for any renormalizable gauge theory with no more than
one power of loop momentum at each vertex, the superficial degree
of divergence is,
\begin{equation}
d_s =  4 - E + (D-4)L - p \,,
\label{SuperficialCount}
\end{equation}
where $L$ is the number of loops, $D$ the dimension, $E$ the number of
external legs and $p$ the number of powers of momentum that can be
algebraically extracted from the integrals as external momenta.  For
each power of numerator loop momentum that can be converted to an external
momentum, the superficial degree is reduced by one unit.  Taking $D=4$
and $p = 4$, corresponding to the four powers of external momentum
implicit in the overall delta function (\ref{OverallDelta}), we see
that $d_s <0$ for all loops and legs.  This also implies that $\Neqfour$
super-Yang-Mills amplitudes cannot contain any subdivergences as all
previous loop orders are finite.  It then follows inductively that the
negative superficial degree of divergence, for all loop amplitudes, is
sufficient to demonstrate the cancellations needed for all order finiteness.  
We note that although this displays the finiteness
of $\NeqFour$ super-Yang-Mills theory, not all cancellations are
manifest, and there are additional ones reducing the degree of
divergence beyond those needed for
finiteness~\cite{BDDPR,FiveLoop,HoweStelleNew}.

A similar analysis can be carried out for $\NeqEight$ supergravity; in
this case the two-derivative coupling leads to a superficial degree of
divergence which monotonically increases with the loop order.  Without
additional mechanisms for taming its ultraviolet behavior, this would lead to
the conclusion of that the theory is ultraviolet divergent.  As
discussed in refs.~\cite{Finite, GravityThree, CompactThree} 
direct evidence to all loop orders indeed points to the existence of much
stronger ultraviolet cancellations.


\section{The supersum as a system of linear equations}
\label{LinearSystemSection}

We now address the question of how to best carry out the evaluation of
multiple fermionic integrals, which can become tedious for
complicated multi-loop cuts.  An approach to organizing this
calculation, discussed in the following sections, is to devise
effective diagrammatic rules for carrying out these integrals.
Another complementary approach, discussed in this section, relies on 
the observation that the
fermionic delta functions may be interpreted as a system of
linear equations determining the integration variables (i.e.  the
variables $\eta$ corresponding to the cut lines) in terms of the
variables $\eta$ associated with the external lines of the amplitude.
From this standpoint, the integral over the internal $\eta$'s may be
carried out by directly solving an appropriately chosen system of
equations and evaluating the remaining supersymmetry 
constraints on the solutions of this system.  
While the relation between the
fermionic integrals and the sum over intermediate states in the cuts is
quite transparent, as we will see in later sections,
it is rather obscure to identify the contribution
of one particular particle configuration crossing the cut in the
solution of the linear system.

\subsection{Cuts involving MHV and MHV vertex expanded trees}
\label{ruleMHVSubSection}

Simple counting shows that after the overall supermomentum
conservation constraint is extracted, the number of equations
appearing in cuts of MHV amplitudes equals the number of integration
variables.  For such cuts the result of the Grassmann integration is
then just the determinant of the matrix of coefficients of that linear
system.  The same counting shows that the number of fermionic
constraints appearing in cuts of N$^k$MHV amplitudes is larger than
the number of integration variables.
One way to evaluate the integral is to determine the integration
variables by solving some judiciously chosen subset of the supermomentum 
constraints
and substitute the result into the remaining fermionic delta
functions.  Care must be taken in selecting the constraints being
solved, as an arbitrary choice may obscure the symmetries of the
amplitude.  One approach is to take the average over all possible
subsets of constraints determining all internal fermionic
variables.  Another general strategy is to select the fermionic
constraints with as few external momenta as possible.  Since the
integration variables are determined as ratio of determinants, all
identities based on over-antisymmetrization of Lorentz indices, such
as Schouten's identity, are accounted for automatically, generally
yielding simple expressions.

\begin{figure}[th]
\centerline{\epsfxsize 2.5 truein \epsfbox{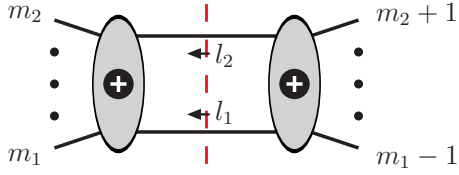}}
\caption{Supercut of a one-loop $n$-point MHV amplitude.  The white
``$+$'' labels on the black background signify that these blobs are
holomorphic vertices, or MHV superamplitudes.  The dashed (red) line
marks the cut, enforcing that momenta crossing the cut are on shell.
\label{OneLoopCutSuperSumFigure}}
\end{figure}

To illustrate this approach let us consider the example, shown in
\fig{OneLoopCutSuperSumFigure}, of the supercut of the one-loop $n$-point
MHV superamplitude
\be
{\cal C}^{{\rm fig.\ }\ref{OneLoopCutSuperSumFigure}}=
 \int d^4\eta_{l_1}\int d^4\eta_{l_2}  \,
{\cal A}^{\rm MHV}(-l_1,m_1,\dots, m_2, -l_2)
{\cal A}^{\rm MHV}(l_2, m_2+1,\ldots, m_1-1, l_1)\,.
\label{OneLoopCutSuperExample}
\ee
The only contribution to this cut is where both tree superamplitudes
are MHV; together they
contain the two delta functions,
\be
\delta^{(8)}\Bigl(-\lambda_{l_1}^\alpha\eta_{l_1}^a
             -\lambda_{l_2}^\alpha\eta_{l_2}^a
             +\sum_{i=m_1}^{m_2}\lambda_{i}^\alpha\eta_{i}^a \Bigr) \;
\delta^{(8)}\Bigl( \lambda_{l_1}^\alpha\eta_{l_1}^a
             +\lambda_{l_2}^\alpha\eta_{l_2}^a
             +\sum_{i=m_2+1}^{m_1-1}\lambda_{i}^\alpha\eta_{i}^a \Bigr) \,.
\ee
Adding the argument of the first delta function to the second one, as
discussed in (\ref{extract_tree}), exposes the overall supermomentum
conservation
\be
\delta^{(8)}\Bigl(-\lambda_{l_1}^\alpha\eta_{l_1}^a
             -\lambda_{l_2}^\alpha\eta_{l_2}^a
             +\sum_{i=m_1}^{m_2}\lambda_{i}^\alpha\eta_{i}^a\Bigr) \;
\delta^{(8)}\Bigl(\sum_{i=m_2+1}^{m_1-1}\lambda_{i}^\alpha\eta_{i}^a
             +\sum_{i=m_1}^{m_2}\lambda_{i}^\alpha\eta_{i}^a\Bigr)\,;
\ee
then, the value of the fermionic integral in
\eqn{OneLoopCutSuperExample} is the determinant of the matrix of
coefficients of the following system of linear equations,
 \be
\lambda_{l_1}^\alpha\eta_{l_1}^a+\lambda_{l_2}^\alpha\eta_{l_2}^a
=\sum_{i=m_1}^{m_2}\lambda_{i}^\alpha\eta_{i}^a \,, 
\ee 
interpreted as a system of equations for
$\eta_{l_1}^a$ and $\eta_{l_2}^a$; its determinant is
\begin{eqnarray}
J &=&  \det{}^{\! 4} \left|
\begin{array}{cc}
  \lambda_{l_1}^1 & \lambda_{l_2}^1 \\
  \lambda_{l_1}^2 & \lambda_{l_2}^2 
\end{array}
\right| = 
\langle l_1 l_2\rangle^4 \,.
\end{eqnarray}
Thus, the resulting cut superamplitude is just
\begin{eqnarray}
{\cal C}^{{\rm fig.\ }\ref{OneLoopCutSuperSumFigure}}&=&
-\delta^{(8)}\Bigl(\sum_{i=1}^n\lambda_i^\alpha\eta_i^a\Bigr)
\,\langle l_1 l_2\rangle^4\,\times
\frac{1}{ \langle m_2 l_2\rangle \, \langle l_2 l_1\rangle \,
\langle l_1 m_1\rangle \,
\prod_{i=m_1}^{m_2-1}\langle i(i+1)\rangle} \nn\\
&& \hskip 3 cm 
\times\frac{1}{\langle (m_1-1) l_1\rangle \,
\langle l_1 l_2\rangle \, \langle l_2 (m_2+1)\rangle \,
\prod_{i=m_2+1}^{m_1-2}\langle i(i+1)\rangle  }
\,.  \hskip 1 cm 
\end{eqnarray}
Extracting the gluon component we immediately recover the results of 
reference \cite{UnitarityMethod}, which had been obtained by using 
supersymmetry Ward identities~\cite{SWI} and explicitly summing over 
states crossing the cut.

\begin{figure}[t]
\centerline{\epsfxsize 5.1 truein \epsfbox{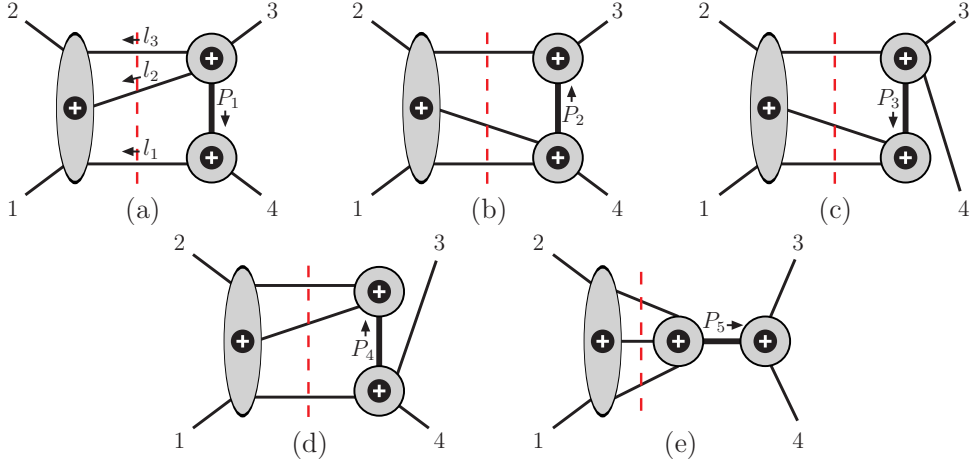}}
\caption{A three-particle supercut for the MHV four-gluon
amplitude.  This cut contribution contains one MHV and one NMHV
superamplitude.  The five-point NMHV amplitude is actually \MHVbar{}
but it is expanded using the MHV super-rules.  The thick line labeled
by $P$ marks the internal propagator.
Five additional contributions---not shown here---correspond to legs 1 
and 2 belonging to an ${\overline {\rm MHV}}$ amplitude and legs 3 and
4 to an MHV amplitude.
 \label{ThreeCutSuperSumFigure}}
\end{figure}

Let us now illustrate the interplay between supersum calculations and
the super-MHV vertex expansion.  The three-particle cut of the
two-loop four-gluon amplitude provides the simplest example in this
direction, as it contains an \MHVbar{} tree-level amplitude which may
be expanded in terms of MHV vertices, as shown in
\fig{ThreeCutSuperSumFigure}.  We will describe in detail the supercut
contribution in \fig{ThreeCutSuperSumFigure}(a) and quote the result
for the other ones in the figure.  Besides the contributions
shown in \fig{ThreeCutSuperSumFigure} there are additional
contributions which sum to the complex conjugate of these, ignoring
an overall four-point tree superamplitude factor.

The general strategy is to explicitly write down the constraints for a single value of the
$R$-symmetry index and then raise the final result to the fourth power, as
discussed in section~\ref{StructureSection}.  We find for \fig{ThreeCutSuperSumFigure}(a) the following three supermomentum constraints 
at each of the three MHV vertices,
\begin{eqnarray}
&&\delta^{(2)}(\lambda_1^\alpha\eta_1^a+\lambda_2^\alpha\eta_2^a
-\lambda_{l_1}^\alpha\eta_{l_1}^a-\lambda_{l_2}^\alpha\eta_{l_2}^a
-\lambda_{l_3}^\alpha\eta_{l_3}^a) \,
\delta^{(2)}(\lambda_3^\alpha\eta_3^a + \lambda_{P_1}^\alpha\eta_{P_1}^a
+\lambda_{l_2}^\alpha\eta_{l_2}^a +\lambda_{l_3}^\alpha\eta_{l_3}^a)\cr
&& \times \delta^{(2)}(\lambda_4^\alpha\eta_4^a
+\lambda_{l_1}^\alpha\eta_{l_1}^a - \lambda_{P_1}^\alpha \eta_{P_1}^a)\,.
\label{initial_const}
\end{eqnarray}
As before, we first isolate the
overall supermomentum conservation constraint by adding to the
argument of the first delta function the arguments of the second
and third ones\footnote{This is just one choice and the same result
can be obtained by adding the arguments of any two delta functions to
the argument of the third one.} and noticing the cancellation of all
spinors corresponding to the internal lines.
The remaining system of four equations involving the fermionic
variables for the internal lines are
the arguments of the second and third delta functions in equation
(\ref{initial_const}),
\begin{eqnarray}
-\lambda_{P_1}^\alpha \eta_{P_1}^a
\hphantom{+\lambda_{l_1}^\alpha\eta_{l_1}^a}
-\lambda_{l_2}^\alpha \eta_{l_2}^a -\lambda_{l_3}^\alpha\eta_{l_3}^a
 &=&\lambda_3^\alpha\eta_3^a \,,
\cr
+\lambda_{P_1}^\alpha\eta_{P_1}^a
-\lambda_{l_1}^\alpha\eta_{l_1}^a
\hphantom{-\lambda_{l_2}^\alpha \eta_{l_2}^a -\lambda_{l_3}^\alpha\eta_{l_3}^a}
&=&\lambda_4^\alpha\eta_4^a \,.
\end{eqnarray}
Its matrix of coefficients is
\be
\begin{pmatrix}
-\lambda_{P_1} ^\alpha& 0 & -\lambda_{l_2}^\alpha& -\lambda_{l_3}^\alpha \cr
+\lambda_{P_1} ^\alpha& -\lambda_{l_1}^\alpha& 0 & 0
\end{pmatrix} \,,
\ee
where each spinor $\lambda^\alpha_j$ should be thought of as a
submatrix with two rows and one column.  The determinant of this matrix
is just $(\langle l_1 \Pflat_1\rangle \langle l_2 l_3\rangle)$.
After restoring the four identical factors we thus find that the
supersum evaluates to 
\be 
(\langle l_1 \Pflat_1\rangle \langle l_2
l_3\rangle)^4 \delta^{(8)}(\lambda_1^\alpha \eta_1^a+\lambda_2^\alpha\eta_2^a
+\lambda_3^\alpha \eta_3^a+\lambda_4^\alpha \eta_4^a)\,.  
\ee
In obtaining this simple form, the explicit application of Schouten's identity was 
not required.%
\footnote{The same result may be obtained by explicitly solving the
system of constraints by expressing the equations in terms of spinor
inner products; however repeated use of Schouten's identity is
required in this case.}

Carrying out the same steps for the other four components (b), (c), (d) and (e)  in \fig{ThreeCutSuperSumFigure} gives us the complete
expression for this cut contribution,
\begin{eqnarray}
\hskip -.3 cm 
{\cal C}^{{\rm fig.\ }\ref{ThreeCutSuperSumFigure}}&=& \,
\delta^{(8)}(\lambda_1^\alpha\eta_1^a+\lambda_2^\alpha\eta_2^a
              +\lambda_3^\alpha \eta_3^a + \lambda_4^\alpha \eta_4^a) \nn\\
&\times & 
\frac{1}{\langle 12\rangle\langle 2 l_3 \rangle
\langle l_3 l_2 \rangle  \langle l_2 l_1 \rangle \langle l_1 1\rangle
}\biggl[
\frac{1}{\langle l_2 l_3 \rangle  \langle l_3 3 \rangle 
\langle 3 \Pflat_1\rangle\langle \Pflat_1 l_2\rangle}
\frac{1}{P_1^2}
\frac{1}{\langle 4 l_1\rangle\langle l_1 \Pflat_1\rangle
  \langle \Pflat_1 4\rangle}
       \left(\langle l_1 \Pflat_1\rangle \langle l_2 l_3\rangle \right)^4 
\nn\\
&& \null \hskip 2 cm 
+
\frac{1}{\langle \Pflat_2 l_3 \rangle  \langle l_3 3 \rangle 
\langle 3 \Pflat_2\rangle}
\frac{1}{P_2^2}
\frac{1}{\langle l_2 \Pflat_2\rangle \langle \Pflat_2 4 \rangle
              \langle 4 l_1 \rangle\langle l_1 l_2\rangle}
  \left(\langle l_3 \Pflat_2\rangle \langle l_1 l_2\rangle\right)^4 
\nn\\
&& \null \hskip 2 cm 
+
\frac{1}{\langle l_3 3\rangle  \langle  34 \rangle 
\langle 4 \Pflat_3\rangle\langle \Pflat_3 l_3 \rangle}
\frac{1}{P_3^2}
\frac{1}{\langle  \Pflat_3 l_1\rangle \langle l_1 l_2\rangle 
               \langle l_2 \Pflat_3\rangle}
   \left(\langle l_3 \Pflat_3\rangle \langle l_1 l_2\rangle\right)^4 
\nn\\
&& \null \hskip 2 cm 
+
\frac{1}{\langle l_2 l_3\rangle  \langle  l_3 \Pflat_4 \rangle 
\langle \Pflat_4 l_2\rangle}
\frac{1}{P_4^2}
\frac{1}{\langle  l_1 \Pflat_4\rangle \langle \Pflat_4 3\rangle 
               \langle 34\rangle\langle  4 l_1\rangle}
\left(\langle l_1 \Pflat_4\rangle \langle l_2 l_3\rangle \right)^4 
\biggr] \,,
\hskip 1 cm
\label{CSW_5pt} 
\end{eqnarray}
where, 
\begin{equation}
P_1 = k_4+ l_1 \,, \hskip .7 cm 
P_2 = k_3+ l_3 \,, \hskip .7 cm 
P_3 = l_1+ l_2 \,, \hskip .7 cm 
P_4 = l_2+ l_3 \,,
\label{Pdefs}
\end{equation}
and $\Pflat$ is defined in \eqn{CSWMomShift}.  Diagram (e) in
\fig{ThreeCutSuperSumFigure} gives a vanishing contribution.  The
dependence on the reference vector $\zeta$ cancels out in
\eqn{CSW_5pt}, as is simple to verify numerically.  This expression,
together with the five additional contribution (not shown in
\fig{ThreeCutSuperSumFigure}) arising from legs 1 and 2 belonging to
an \MHVbar{} amplitude and legs 3 and 4 to an MHV amplitude,
numerically agrees with the three-particle cut of the known planar
two-loop four-point amplitude~\cite{BRY,BDDPR}.

\subsection{Cuts with both MHV and \MHVbar{} trees}
\label{ruleMHVbarSubSection}

\begin{figure}[th]
\centerline{\epsfxsize 2 truein \epsfbox{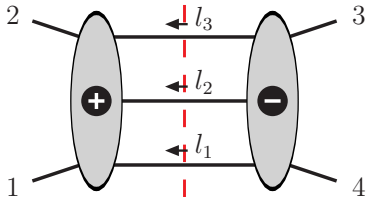}}
\caption{The same three-particle cut contribution as in
\fig{ThreeCutSuperSumFigure}, but where the right-hand-side \MHVbar{}
amplitude is not expanded using the MHV rules.  The ``$+$'' label
signifies a holomorphic vertex, or MHV superamplitude, and the ``$-$''
label signifies a anti-holomorphic vertex, or \MHVbar{}
superamplitude.
\label{ThreeCutSuperSumMHVbarFigure}}
\end{figure}

While the result obtained above is correct, the complexity of
\eqn{CSW_5pt} is somewhat unsettling.  This complexity comes from
expanding the \MHVbar{} amplitude in MHV diagrams.  For generic
non-MHV diagrams this strategy is useful, but for \MHVbar{} amplitudes
there is no need to do so.  Indeed previous evaluations of the above
cut~\cite{BRY, BDDPR, FreedmanUnitarity} without making use of the MHV
expansion give simpler forms.  In the same spirit, it is sometimes
convenient to use the \MHVbar{} representation of four-point
amplitudes.  As illustrated in
\fig{ThreeCutSuperSumMHVbarFigure}, we therefore reconsider the
previous example shown in \fig{ThreeCutSuperSumFigure}, but without
expanding the \MHVbar{} amplitude in MHV amplitudes.

The relevant fermionic integral (where we again keep explicitly a single
$R$-symmetry index and raise the result to the fourth power) is,
\begin{eqnarray}
{}&&\int d\eta^a_{l_1}d\eta^a_{l_2}d\eta^a_{l_3}d^2\omega^a\;
\delta^{(2)}(\lambda_1^\alpha \eta_1^a+\lambda_2^\alpha \eta_2^a
-\lambda_{l_1}^\alpha \eta_{l_1}^a-\lambda_{l_2}^\alpha\eta_{l_2}^a
-\lambda_{l_3}^\alpha \eta_{l_3}^a )\cr
{}&&~~~~~~~~\times
\delta(\eta^a_{l_1}-\tlambda^{\dot\alpha}_{l_1}\omega^a_{\dot\alpha}) \,
\delta(\eta^a_{l_2}-\tlambda^{\dot\alpha}_{l_2}\omega^a_{\dot\alpha}) \,
\delta(\eta^a_{l_3}-\tlambda^{\dot\alpha}_{l_3}\omega^a_{\dot\alpha}) \, 
\delta(\eta^a_3-\tlambda^{\dot\alpha}_3\omega^a_{\dot\alpha}) \,
\delta(\eta^a_4-\tlambda^{\dot\alpha}_4\omega^a_{\dot\alpha})\,.  \hskip 1 cm 
\end{eqnarray}
Here $\omega^a_{\dot \alpha}$ are the auxiliary integration variables
in equation (\ref{aux_integral_MHVbar}).  Adding the arguments of the
delta functions on the second line, with the appropriate weights, to
the argument of the delta function on the first line exposes the
overall supermomentum conservation.  We are then left with
\begin{eqnarray}
\hskip -.5 cm &&\delta^{(2)}(\lambda_1^\alpha\eta_1^a+\lambda_2^\alpha\eta_2^a
        +\lambda_3^\alpha\eta_3^a+\lambda_4^\alpha\eta_4^a)\\
\hskip -.5 cm 
&&\times\int d\eta^a_{l_1}d\eta^a_{l_2}d\eta^a_{l_3}d^2\omega^a
\delta(\eta^a_{l_1}-\tlambda^{\dot\alpha}_{l_1}\omega^a_{\dot\alpha}) \,
\delta(\eta^a_{l_2}-\tlambda^{\dot\alpha}_{l_2}\omega^a_{\dot\alpha}) \,
\delta(\eta^a_{l_3}-\tlambda^{\dot\alpha}_{l_3}\omega^a_{\dot\alpha}) \,
\delta(\eta^a_3-\tlambda^{\dot\alpha}_3\omega^a_{\dot\alpha}) \,
\delta(\eta^a_4-\tlambda^{\dot\alpha}_4\omega^a_{\dot\alpha}) \,.
\nn
\end{eqnarray}
The matrix of coefficients of the surviving system of constraints can be easily
read off,
\be
\begin{pmatrix}
1 & 0 & 0 & -\tlambda^{\dot 1}_{l_1} & -\tlambda^{\dot 2}_{l_1} \\
0 & 1 & 0 & -\tlambda^{\dot 1}_{l_2} & -\tlambda^{\dot 2}_{l_2} \\
0 & 0 & 1 & -\tlambda^{\dot 1}_{l_3} & -\tlambda^{\dot 2}_{l_3} \\
0 & 0 & 0 & -\tlambda^{\dot 1}_{3} & -\tlambda^{\dot 2}_{3} \\
0 & 0 & 0 & -\tlambda^{\dot 1}_{4} & -\tlambda^{\dot 2}_{4}
\end{pmatrix}\,.
\ee
Taking its determinant, raising it to the fourth power, and
restoring the remaining factors in the tree-level superamplitudes
gives the supercut contribution, 
\be
{\cal C}^{{\rm fig.} \ \ref{ThreeCutSuperSumMHVbarFigure}}=
\delta^{(8)}\Bigl(\sum_{i=1}^4\lambda_i^\alpha\eta_i^a \Bigr)\,[34]^4\,
\frac{1}{\langle 12\rangle\langle2 l_3\rangle\langle l_3 l_2 \rangle
         \langle l_2 l_1 \rangle\langle l_1 1\rangle}
\frac{1}{[34][4l_1][l_1l_2][l_2l_3][l_3 3]}\,.
\ee
This numerically matches \eqn{CSW_5pt}, again giving the proper
contribution to the cut four-gluon amplitude at two loops~\cite{BRY,BDDPR}.  

This calculation illustrates a general feature of supersums: if an
\MHVbar{} vertex appearing in a supercut has two external legs
attached to it, say $p$ and $k$, such as legs $3$ and $4$ in the
example above, then apart from the overall supermomentum conservation,
the supercut contribution is also proportional to the bracket product
of those two momenta, {\it i.e.} it contains a numerator factor,
\be
\delta^{(8)} \Bigl(\sum_{i\in {\cal E}}\lambda_i^\alpha \eta^a_i \Bigr)\,
 [pk]^4 \,.
\ee
As in \eqn{OverallDelta}, $\cal E$ denotes the set of external legs.  
This feature is related to the soft
ultraviolet properties of $\NeqFour$ super-Yang-Mills theory.

The \MHVbar{} superamplitudes  can also be used in the cuts
in conjunction with the MHV-vertex rules.  Indeed, any on-shell 
four-point amplitude may be
interpreted either as MHV or \MHVbar{} amplitudes as can be seen by directly
evaluating the $\omega$ integral in equation (\ref{aux_integral_MHVbar}) 
for $n=4$:
\be
\widehat F{\cal A}^{\overline{\rm MHV}}_4 = i
\delta^{(8)} \Bigl(\sum_{i=1}^4\lambda_i^\alpha\eta^a_i \Bigr)
\frac{[34]^4}{\langle 12\rangle^4[12][23][34][41]}= i
\delta^{(8)}  \Bigl(\sum_{i=1}^4\lambda_i^\alpha\eta^a_i \Bigr)
\frac{1}{\langle 12\rangle\langle 23\rangle\langle 34\rangle
\langle 41\rangle}\,.
\ee
Depending on context, choosing one interpretation of the four-point 
amplitude over the other can lead to more factors of loop momenta
in supersums being replaced by factors of external momenta thus 
making manifest more of the supersymmetric cancellations.  We will comment 
on an example in this direction at the end of section \ref{sec:higher_pts}.

\subsection{Cuts of higher-point superamplitudes \label{sec:higher_pts}}

\begin{figure}[th]
\centerline{\epsfxsize 6 truein \epsfbox{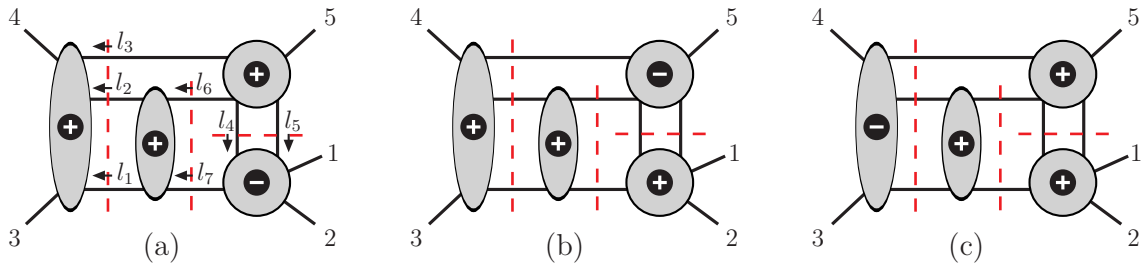}}
\caption{A generalized cut for a four-loop five-point MHV superamplitude
having three cut contributions (a), (b) and (c), corresponding to three
independent choices of holomophicity of the tree amplitudes comprising 
the cut.
\label{FourLoopFivePtCutSuperSumFigure}}
\end{figure}

The above techniques are by no means restricted to four-point
amplitudes.  To illustrate this, consider the supercut of the MHV
four-loop five-point amplitude shown in
\fig{FourLoopFivePtCutSuperSumFigure}.  For the displayed cut topology, these
are the three independent non-vanishing assignments of MHV or
\MHVbar{} configurations that
contribute to an external MHV configuration.  (Changing the MHV to an
\MHVbar{} label on the lone four-point amplitude is not an
independent choice, as the two cases are equivalent.)

For the cut contribution in figure \ref{FourLoopFivePtCutSuperSumFigure}(a)
the Jacobian of the system of constraints is
\be
J^{{\rm fig.\,}\ref{FourLoopFivePtCutSuperSumFigure}(a)}
=(\langle l_1 l_2\rangle\langle l_3 l_6\rangle [12])^4 \,,
\ee
leading to the following result for the supercut, 
\ba
\hskip -.2 cm 
{\cal C}^{{\rm fig.\,}\ref{FourLoopFivePtCutSuperSumFigure}(a)}\!
&=&-\delta^{(8)} \Bigl(\sum_{i=1}^5\lambda_i^\alpha\eta_i^a \Bigr)
(\langle l_1 l_2\rangle\langle l_3 l_6\rangle
[12])^4\frac{1}{\spa{3}.{4}\spa{4}.{l_3}
\spa{l_3}.{l_2}\spa{l_2}.{l_1}\spa{l_1}.{4}}\\
&\times&\frac{1}{\spa{l_1}.{l_2}\spa{l_2}.{l_6}
  \spa{l_6}.{l_5}\spa{l_5}.{l_1}}
    \frac{1}{\spa{5}.{l_5}\spa{l_5}.{l_4}\spa{l_4}.{l_6}
\spa{l_6}.{l_3}\spa{l_3}.{5}}
        \frac{1}{\spb{1}.{2}\spb{2}.{l_7}\spb{l_7}.{l_4} 
    \spb{l_4}.{l_5}\spb{l_5}.{1}} \,.
\nonumber
\ea

Similarly, the Jacobian for the system of constraints remaining, after
reconstructing the overall supermomentum conservation, for the cut
contributions in \fig{FourLoopFivePtCutSuperSumFigure}(b) and(c) are,
\begin{eqnarray}
J^{{\rm fig.\,}\ref{FourLoopFivePtCutSuperSumFigure}(b)}&=&
\spa{l_1}.{l_2}^4
(\spa{l_7}.{l_4} \spb{l_4}.{5} + \spa{l_7}.{l_5}\spb{l_5}.{5})^4 
= \spa{l_1}.{l_2}^4  (\spa{l_7}.1 \spb1.5 +\spa{l_7}.2 \spb2.5)^4
\,, \nn \\
J^{{\rm fig.\,}\ref{FourLoopFivePtCutSuperSumFigure}(c)}&=&
(\langle l_4 l_5\rangle\langle l_6 l_7\rangle [3 4] )^4 \,.
\end{eqnarray}
The complete contribution to the cut for these configurations is given
by multiplying these Jacobians by the appropriate spinor denominators
and the overall supermomentum delta function.  In the supersums
corresponding to \fig{FourLoopFivePtCutSuperSumFigure}(a) and (c) we
note the presence of bracket products of external momenta attached to
an \MHVbar{} tree amplitude; this illustrates a general property
described in \sect{ruleMHVbarSubSection}.  It is also worth pointing
out that if we reassign the four-point tree superamplitude ${\cal
A}^{\rm MHV}_4(l_1,l_2,-l_6,-l_7)$ in
\fig{FourLoopFivePtCutSuperSumFigure}(a) to be \MHVbar{}, then the
Jacobian becomes $J^{{\rm
fig.\,}\ref{FourLoopFivePtCutSuperSumFigure}(a)} =(\langle l_3|
k_3+k_4 |l_7] [12])^4$, so additional powers of external momenta come
out for this contribution.


\section{Supersums as $SU(4)$ index diagrams}
\label{DiagramLoopSection}

The algebraic approach of the previous section is quite effective for
the calculation of $\NeqFour$ supersums, as it elegantly avoids the
bookkeeping of individual states crossing the cuts.  However, it can be
advantageous to follow these contributions.  In this section will
discuss a complementary approach using a pictorial representation
of supercuts in terms of the index diagrams introduced in
\sect{DiagramSubSection}.

\subsection{Mixed superspace}
\label{MixedSuperspace}

As we have already seen, in the unitarity cuts it is convenient to use
both MHV and \MHVbar{} amplitudes.  However the need to Fourier
transform the amplitudes defined in $\etab$ superspace to $\eta$
superspace is sometimes inconvenient.  Therefore we will derive here
sewing rules for superamplitudes where $\eta$ and $\etab$ are on an
equal footing, which will then motivate the rules for the sewing of
index diagrams.

Consider an internal leg $i$ connecting two on-shell superamplitudes,
left ${\cal A}_L$ and right ${\cal A}_R$ in an arbitrary cut.  As
discussed \sect{StructureSection}, in the MHV $\eta$ superspace the
supersum over the states propagating through this leg is realized by 
the Grassmann integral,
\begin{equation}
\int \prod_{a=1}^4  d\eta_i^a \, {\cal A}_L {\cal A}_R \, .
\end{equation}
In \sect{StructureSection} we showed that each $SU(4)$ index
can be considered independently for tree amplitudes as well as in
supersums of cuts.  Therefore, it is sufficient to consider a single
index supersum of three cases: the internal leg $i$ connects
amplitudes of the type (a) MHV and MHV, (b) \MHVbar{} and \MHVbar{},
and (c) MHV and \MHVbar{} ,
\begin{eqnarray}
&&\hskip -.5 cm  {\rm (a)}:\;\; 
 \int d\eta_i^a {\cal A}^{\rm MHV}_L {\cal A}^{\rm MHV}_R \,,\nn \\
&& \hskip -.5 cm {\rm (b)}:\;\;
  \int d\eta_i^a  \left( \int d\etab_{ia} e^{\eta_i^a  \etab_{ia}} 
   {\cal A}^{\rm \overline{MHV}}_L\right) 
  \left( \int d\etab_{ia} e^{\eta_i^a  \etab_{ia}}
     {\cal A}^{\rm \overline{MHV}}_R\right)  
 = \int d\etab_{ia} {\cal A}^{\rm  \overline{MHV}}_L(\etab_{ia}) 
      {\cal A}^{\rm  \overline{MHV}}_R (-\etab_{ia}) \,,\nn \hskip .5 cm\\
&& \hskip -.5 cm {\rm (c)}:\;\;  \int d\eta_i^a {\cal A}^{\rm MHV}_L 
 \left( \int d\etab_{ia} e^{\eta_i^a  \etab_{ia}} 
  {\cal A}^{\rm \overline{MHV}}_R\right) 
  = \int d\eta_i^a d\etab_{ia} e^{\eta_i^a  \etab_{ia}}
    {\cal A}^{\rm MHV}_L  {\cal A}^{\rm \overline{MHV}}_R \,,
\label{IntegralCases}
\end{eqnarray}
where $a$ is taken to be a fixed $SU(4)$ $R$-symmetry index.  On the
left hand side of cases (b) and (c) we have applied the Grassmann
Fourier transform to the \MHVbar{} amplitudes, in order have a
well-defined supersum.  Note that case (b) can be interpreted as a
supersum in $\etab$ superspace, where the $\etab_{ia}$ has flipped
sign inside $ {\cal A}^{\rm \overline{MHV}}_R$ as is shown explicitly.
The sign flip happens because the
$\eta_i^a$ integral produces a delta function
$\delta(\etab_{ia}^{~L}+\etab_{ia}^{~R})$ enforcing this, where the
labels $L$ and $R$ are added to clarify which amplitude they originate
from.  Case (c) is more straightforward to simplify and it becomes a
mixed supersum correlating the $\eta$ and $\etab$ parameters.

Equation \ref{IntegralCases} motivates the definition of mixed
$\eta$-$\etab$ superspace operators for performing the supersum.  In
the three cases we have,
\begin{eqnarray}
&& \mbox{MHV-MHV}:\;\;\;\; \hat I^a_{i,\,++} \equiv \int d\eta_i^a \,,\nn \\
&& \mbox{\MHVbar{}-\MHVbar{}}:\;\;\;\; \hat I^a_{i,\, --}
    \equiv  \int d\etab_{ia} \,, \nn  \\
&& \mbox{MHV-\MHVbar{}}:\;\;\;\; \hat I^a_{i,\,+-} 
    \equiv  \int d\eta_i^a d\etab_{ia} e^{\eta_i^a  \etab_{ia}} \,,
\label{SumOperators}
\end{eqnarray}
where the $+$ and $-$ labels are shorthand for MHV and \MHVbar{}, respectively.

In terms of these operators, the sum over all members of the ${\cal
N}=4$ multiplet, in mixed superspace, is determined by the action of the
operator,
\begin{equation}
\hat I=\prod_{a=1}^4
 \left( \prod_{i \in {\rm internal}} \hat I_{i, \, {\rm case}_i}^a\right)\,,
\label{FullOperator}
\end{equation}
on the cut.  Here the label ``${\rm case}_i$'' labels the three cases
($++$,$--$,$+-$) given in \eqn{SumOperators}.  Although the individual
factors may be Grassmann odd, the ordering of the internal legs is
irrelevant after the $SU(4)$ index product is carried out. (Various
orderings differ only in an overall sign, which drops out in this
final product.)

In addition to the mixed supersum operator, a sign rule for sewing
$\etab_{ia}$ across a cut is required by the sign flip that appears in
case (b) in \eqn{IntegralCases}.  For incoming momenta, $p_{-i}=-p_i$,
we define the superamplitudes to be functions of $\eta^a_{-i}$ and
$\etab_{-ia}$, where
\begin{eqnarray}
&& \hskip 4 mm \eta_{-i}^a \rightarrow \eta_{i}^a, \hskip 1.8 cm  
\widetilde{\eta}_{-ia} \rightarrow -\widetilde{\eta}_{ia} \,.  
\label{etabsign}
\end{eqnarray}
This sign rule is also necessary in order to have conjugate
supermomenta $\etab_{ia} [i|$ transform correctly under sign flips of the
momentum direction.  Although case (c) in \eqn{IntegralCases} was
considered without this rule it can be shown to be consistent with the
mixed supersum operator \eqn{FullOperator} up to an overall sign which
drop out in the $SU(4)$ index product.

\begin{figure}[t]
\centerline{\epsfxsize 5.5 truein \epsfbox{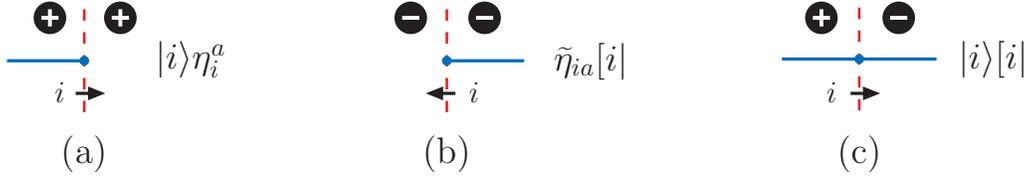}}
\caption[a]{\small Rules for index lines crossing a cut leg $i$
carrying momentum $p_i$.  If both sides of the cut are (a) MHV or both
are (b) \MHVbar{} then the index line ends at the cut.  This is
equivalent to the insertion of a super-momentum $|i\rangle \eta^a_i$
in the MHV case, or conjugate super-momentum $ \etab_{ia} [i|$ in the
\MHVbar{} case.  If one side is MHV and the other \MHVbar{} then the
index line is continuous across the cut and corresponds to the
insertion of $|i\rangle \eta^a_i \etab_{ia} [i|$, or as illustrated,
the insertion of the cut momentum $|i\rangle [i|$, as discussed in
\sect{Rules1and2}.  Dashed (red) lines mark cuts, solid (blue) denote
$SU(4)$ index lines, and plus or minus labels denote whether an
amplitude on a given side of a cut is MHV or \MHVbar{}.  The arrows
indicate the momentum direction.}
\label{BasicFlavorRulesFigure}
\end{figure}

Having defined the mixed superspace state sum, let us consider the
actions of the three types of sewing operators.  We note that the only
objects in the product ${\cal A}_L {\cal A}_R$ that survive the
supersum integrations of \eqn{SumOperators}, for leg $i$ and index
$a$, are those terms proportional to,
\begin{eqnarray}
&& {\rm (a)}:\;\; | i \rangle \eta_i^a=q^a_i \,, \nn \\
&& {\rm (b)}:\;\; \etab_{ia} [i| =\qbar_{ia} \,, \nn \\
&& {\rm (c)}:\;\; | i \rangle \eta_i^a  \etab_{ia} 
     [ i| \hskip 2mm {\rm or} \hskip 2mm 1 \,,
\label{transition}
\end{eqnarray}
where (a), (b) and (c) refers to the aforementioned cases, and where the ``1'' in case (c) denotes an absence of both $ \etab_{ia}$ and
$\eta_i^a$.  Furthermore, we note that since $\hat I^a_{i,\, +-}$ is a
Grassmann even operator we can immediately carry out the integration
of case (c),
\begin{equation}
{\rm (c)}:\;\;  | i \rangle  [ i| = p_i \hskip 2mm {\rm or} \hskip 2mm 1\,.
\label{CaseC}
\end{equation}
However this has to be done with some care, as will be discussed in
section \sect{Rule2Section}, where a precise rule will be given.

Interpreting the supermomenta of \eqn{transition} and momenta of
\eqn{CaseC} as parts of $SU(4)$ index lines, gives us the pictorial
rules displayed in \fig{BasicFlavorRulesFigure} for the transition
condition of an index line across a cut.  For an MHV-MHV transition
the index line ends (or starts) at the cut, corresponding to the
insertion of a supermomentum $| i \rangle \eta_i^a$. 
Similarly, for an \MHVbar{}-\MHVbar{} transition, 
the index line ends (or starts) at the cut, 
corresponding to the
insertion of a conjugate supermomentum $\etab_{ia} [i|$.
In contrast, for an MHV-\MHVbar{} transition,  the 
index lines are continuous across a cut."
This can happen in two ways, either the
two lines on each side meet at the cut, or there are no index line on
leg $i$ on either side of the cut.  The latter option corresponds to
the trivial insertion of a unit factor.  The former option can be
interpreted as either an insertion of a product between a
supermomentum and its conjugate $| i \rangle \eta_i^a \etab_{ia} [ i|$
as in \eqn{transition}(c), or it can be interpreted as an insertion of
momentum $| i \rangle [ i|$ according to \eqn{CaseC}, as is displayed
in \fig{BasicFlavorRulesFigure}(c).  These two interpretations will
give rise to two different sets of rules for carrying out the supersum
(see \sect{Rules1and2}).  In both cases the $SU(4)$ index-line
diagrams will be identical.

\subsection{One-loop warm-up}

\begin{figure}[t]
\centerline{\epsfxsize 6 truein \epsfbox{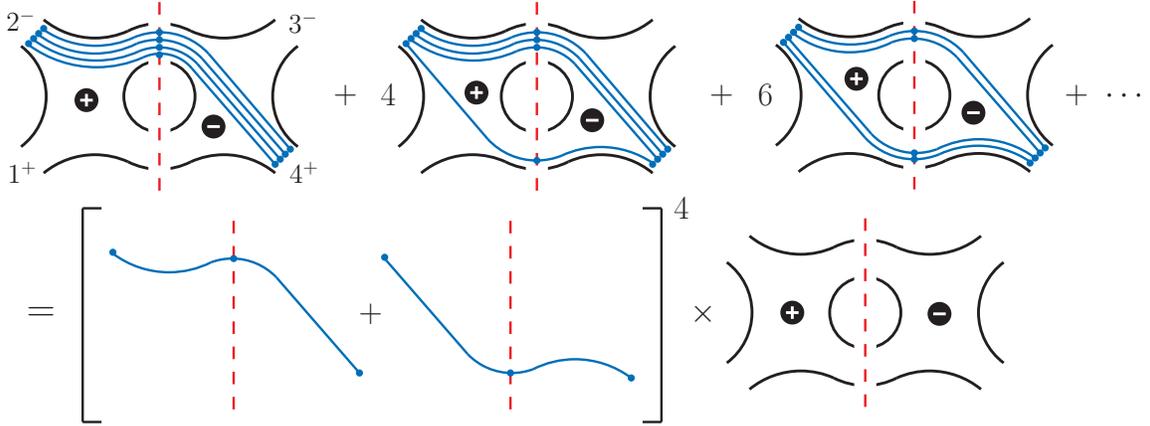}}
\caption[a]{\small A unitarity cut of the four-gluon amplitude
$A^\oneloop_4 (1^+,2^-,3^-,4^+)$, involving one MHV and one \MHVbar{}
superamplitude.  The top-left diagram represents a gluon loop, the
top-central diagram represents the four contributions in a fermion
loop, and the top-right diagram represents the six scalar state
contributions.  The ellipsis denote that four more fermion-loop and
one more gluon-loop contributions are suppressed.  
The bottom diagram illustrates that the 16 contributions may
be resummed, and that each index line may be treated independently.
The circle in each diagram is a one-loop ``hole'' and the dashed line
marks the cut.  The fourth power over the index lines should be
interpreted as a product over the four $SU(4)$ indices.}
\label{OneLoopMHVbarExampleFigure}
\end{figure}

We start with a simple one-loop example to pictorially illustrate the
state sum of a $\NeqFour$ cut.  We will postpone the analytic evaluation
of index diagrams to the following section.

Consider the one-loop cut of \fig{OneLoopMHVbarExampleFigure}.  Reading
off the index lines that end on external legs, this
cut corresponds to the purely gluonic amplitude
$A^\oneloop_4(1^+,2^-,3^-,4^+)$.  The left side of the cut is chosen to
be MHV and right side is \MHVbar{}, which means that the $SU(4)$ index
lines must be continuous through the cut.  The different diagrams in
the top of \fig{OneLoopMHVbarExampleFigure} correspond to the
different states in the $\NeqFour$ gauge supermultiplet.  There are
five such diagrams although only three are shown, the two hidden in
the ellipsis are horizontal flips of the first two shown.  The
combinatoric factors in front of each diagram are the distinct ways of
obtaining the same diagram, tracking of the $SU(4)$ labels.  As shown
in the figure, the sum over the diagrams can be interpreted as a
product over the four SU(4) indices, depicted as a fourth power.  This
is consistent with the general result discussed in
\sect{LinearSystemSection}: summing over the states crossing a cut
composed of a product of MHV and \MHVbar{} tree amplitudes is a sum of
terms raised to the fourth power.  In the diagrammatic language of
index lines this also leads to the simplification which allows us to
consider each of the four $SU(4)$ index-line factors
independently.  Thus in the remaining part of this paper all index
diagrams will be drawn for only a single $SU(4)$ index.

Interestingly, the index diagrams follow a ``sum over paths''
principle analogous to the one of quantum mechanics.  In our one-loop
example, a single continuous index line has two possible allowed
paths, crossing the cut through either the upper or lower internal
leg.  Thus, there are two terms for each index in the state sum or a
total of $2^4$ for the four index lines.  For cuts
which factorize into adjacent MHV amplitudes or adjacent \MHVbar{}
amplitudes, the index lines are discontinuous, or the ``paths'' are
broken into several pieces, as explained in the previous section.  See
the following sections for explicit examples of this.

More generally, for external gluon amplitudes the structure discussed
in the above one-loop example is quite generic for any configuration
of MHV and \MHVbar{} tree amplitudes appearing in a cut.  With
external gluons the four $SU(4)$ index lines all start on the same
legs, allowing us to treat each of the lines identically.  If some of
the external particles are scalars or fermions then the $SU(4)$ index
lines can start at different external legs, but in any case, each of
the four $SU(4)$ index lines can be treated independently.  As
discussed in \sect{TreeSuperspaceSection}, if a non-MHV tree amplitude
appears we simply insert its expansion in terms of MHV or \MHVbar{}
amplitudes into the cut, effectively reducing the evaluation of the
relevant supersums to the discussion above.

\subsection{Rules for converting diagrams to spinor expressions}
\label{Rules1and2}

As explained in \sect{DiagramSubSection}, each index line drawn for an MHV
tree amplitude (in a cut) corresponds to a factor $\spa{q^a_i}.{q^a_j}$,
and for an \MHVbar{} tree amplitude it corresponds to a factor
$\spb{\qbar_{ia}}.{\qbar_{ja}}$.  Since both $\spa{q_{i}^a}.{q_{j}^a}$
and $\spb{\qbar_{ia}}.{\qbar_{ja}}$ are Grassmann even as well as
symmetric under the exchange $i\leftrightarrow j$ it may seem to be a
straightforward task to convert the index diagrams to analytic
expressions.  However, in practice there are different strategies for
converting the Grassmann-valued numerators to spinor expressions, two
of which we describe here.  First we note that since the index
diagrams have pre-selected the terms that survive in the supersum, the
application of any supersum operator on an index diagram serves only
to convert the product of $\eta$'s and $\etab$'s to a $\pm1$ factor,
which can be achieved by simple replacements rules.  The two alternative
replacement rules are:

\subsubsection{Rule 1: Sign assignment in $\eta$-only superspace}
 One option, which will avoid the slightly more complicated
MHV-\MHVbar{} transition operator $\hat I^a_{i, \, +-}$, is to make
use of the Fourier transform and work only in $\eta$ superspace.  (It
also does not require the $\widetilde \eta$ sign flip for incoming
momenta given in \eqn{etabsign}.)  We Fourier transform all the
$\spb{\qbar_{ia}}.{\qbar_{ja}} $ factors according to the rule in
\eqn{FourierRule},
\begin{equation}
\spb{\qbar_{ia}}.{\qbar_{ja}} \stackrel{\widehat F} 
\longrightarrow \eta_{1}^a\cdots \eta_{i-1}^a\, [i| \,
\eta_{i+1}^a \cdots \eta_{j-1}^a \, |j]\, \eta_{j+1}^a \cdots \eta_{m}^a \, ,
\end{equation}
where $1,\ldots,m$ are the legs of the particular \MHVbar{}
amplitude that the $\spb{\qbar_{ia}}.{\qbar_{ja}}$ factor belongs
to.  Recall that in this rule the positions of $[i|$ and $|j]$ count giving
additional signs as they are pushed past the $\eta$'s.
Also note that for an odd number of legs $m$ the Fourier transform
maps the Grassmann even object $\spb{\qbar_{ia}}.{\qbar_{ja}}$ to a
Grassmann odd object, thus care has to be taken to not alter the
position of $\spb{\qbar_{ia}}.{\qbar_{ja}}$ relative to the position
of, say, $\spb{\qbar_{ib}}.{\qbar_{jb}} $ in the cut expression.

After the Fourier transform, every term in the cut will contain exactly
the same product of $\eta$'s, albeit in different orderings.  For each
term and each $SU(4)$ index this product can be converted to a $\pm$
sign by the replacement,
\begin{equation}
\eta^a_{i_1} \eta^a_{i_2}\cdots \eta^a_{i_{n-1}} \eta^a_{i_{n}} 
  \rightarrow {\rm signature}[i_{1}i_{2}\cdots i_{n-1}i_{n}] \,,
\end{equation}
where the signature function gives the signature of the permutation of
the legs relative to a canonical ordering, and here $n$ is the number
of internal legs plus the number of the external $\eta$'s.%
\footnote{The choice of canonical ordering is not
important since any two choices differ by an overall sign which drops out in
the product over the four $SU(4)$ indices.}  This rule is particularly easy to
automate.

\begin{figure}[t]
\centerline{\epsfxsize 5.1 truein \epsfbox{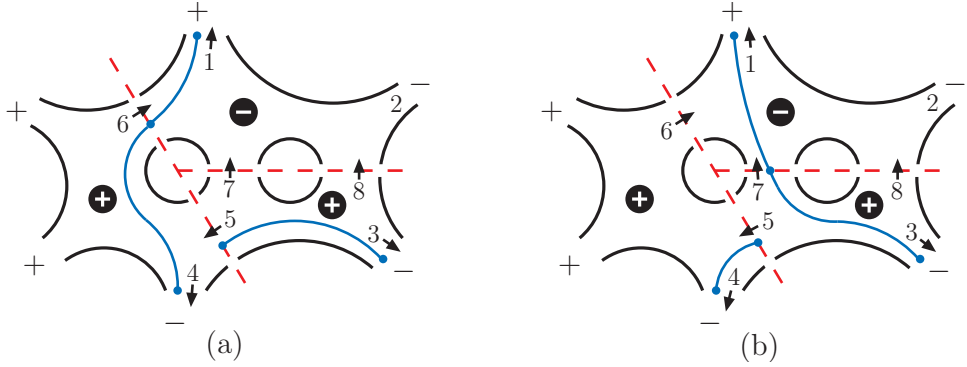}}
\caption[a]{\small Examples of contributions to a unitarity cut of a
six-point two-loop NMHV amplitude drawn as index-line diagrams for  
a single $SU(4)$ index.  Two
routings (a) and (b) of the index lines are shown; routing (c)
referred to in the main text is similar to (b), but where the longer 
index line, attached to legs 1 and 3, passes through cut leg 8
rather than cut leg 7.  Note that only legs
necessary for the subsequent discussion are labeled.}
\label{SixPtFlavorDiagramFigure}
\end{figure}

We will illustrate this rule by an example.  Consider the index
diagrams in \fig{SixPtFlavorDiagramFigure}, which correspond to a
particular contribution to a two-loop cut with gluonic
external states.  For a single $SU(4)$ index there are three
contributions (a), (b) and (c), two of which are shown.  Reading off
the numerator factors from the shaded (blue) index lines we have
\begin{eqnarray}
& {\rm (a)}:&\;\; (\eta_4 \spa{4}.{6} \eta_6)
(\eta_3 \spa{3}.{5} \eta_5)([1|\eta_2 |6] \eta_7 \eta_8)\;\,\rightarrow  
\spa{4}.{6} \spa{3}.{5} \spb{1}.{6}\,, \nn \\
& {\rm (b)}:&\;\; (\eta_5 \langle-5\, 4 \rangle \eta_4)
(\eta_7 \spa{7}.{3} \eta_3)([1|\eta_2 \eta_6 |7]  \eta_8) \rightarrow  
\spa{5}.{4} \spa{7}.{3} \spb{1}.{7}\,,\nn \\
& {\rm (c)}:&\;\;  (\eta_5 \langle-5\,4 \rangle \eta_4)
(\eta_8 \spa{8}.{3} \eta_3)([1|\eta_2 \eta_6\eta_7 |8] ) \rightarrow  
\spa{5}.{4} \spa{8}.{3} \spb{1}.{8}\,,
\end{eqnarray}
where we have suppressed the $SU(4)$ index since we consider only a
single component.  To get to the right-hand-side we first rearrange
the $\eta$'s using the rule~(\ref{FourierRule}) that the spinors
anticommute with the $\eta$'s, and then remove them after arranging
them into a chosen canonical order
$\eta_2\eta_3\eta_4\eta_5\eta_6\eta_7\eta_8$.  Leg 5 also carries a
negative sign since it is an incoming label in (b) and (c), this sign
must be properly extracted following \eqn{signrules}.  For external
gluons each of the four $SU(4)$ indices give identical results,
leading to the following numerator factor for the cut contribution:
\begin{equation}
\Bigl(  \spa{4}.{6} \spa{3}.{5} \spb{1}.{6} + 
\spa{5}.{4} \spa{7}.{3} \spb{1}.{7} + 
\spa{5}.{4} \spa{8}.{3} \spb{1}.{8} \Bigr)^4\, .
\end{equation}

\subsubsection{Rule 2: Sign assignment in a mixed $\eta$-$\etab$ superspace} 
\label{Rule2Section}
Alternatively, we can construct a rule that treats $\eta$ and $\etab$
on equal footing.   With this rule we must strictly impose the sign
rule \eqn{etabsign} that flips the sign of $\etab_i$ as well as
conjugate supermomenta $\qbar_i$ under momentum direction flips
$i\rightarrow -i$.  The mixed-superspace sign rules are based on the
observation in \sect{MixedSuperspace} that the MHV-\MHVbar{}
transition operator $\widehat I^a_{i, \, +-}$ can be immediately
applied to the cut to remove all Grassmann parameters associated with
internal lines on the border between MHV and \MHVbar{} amplitudes.
However, it must be done with some care, as is easily illustrated by
an example.  Consider the two ways of removing the $\eta_i^a\etab_{ia}$
factor,
\begin{eqnarray}
&& \spa{q^a_j}.{q^a_i} \spb{\qbar_{ia}}.{\qbar_{ka}} \rightarrow 
        \eta_j^a\spa{j}.{i} \spb{i}.{k}\etab_{ka}\,, \nn \\
&& \spb{\qbar_{ka}}.{\qbar_{ia}}\spa{q_i^a}.{q_j^a}  \rightarrow  
        \etab_{ka}\spb{k}.{i} \spa{i}.{j}\eta^a_j\,.
\label{SignProblem}
\end{eqnarray}
The two left-hand sides are clearly equal, but the two right-hand would
differ by signs since $\eta^a_j$ anticommute with
$\etab_{ka}$.  However, if we instead think of $\eta$'s and $\etab$'s as
living in two different mutually commuting Grassmann spaces then the
sign inconsistency in \eqn{SignProblem} is resolved.  Although
unconventional, this construction gives us a consistent treatment of the
sign of the index-line contributions.  We will not go further into the
details of proving that this assertion is valid.\footnote{A proof can
be constructed based on the observation that any term in the cut 
can be written so that the $\eta$ and $\tilde \eta$ parameters
are manifestly separated with the overall sign of the term 
unaffected.}
Instead we will state the final rules.

The rules that convert the index lines to spinor products, while
treating $\eta$ and $\etab$ on equal footing are: For each unbroken
index line, write down the corresponding spinor string (using momenta)
following either direction of the line.  Multiply with appropriate
Grassmann odd parameters at the endpoints of the line, as shown in
\fig{BasicFlavorRulesFigure}.  Use the sign rules of \eqn{signrules}
and \eqn{etabsign} to deal with the case of incoming momenta (or
supermomenta).  Now since each term in the cut has exactly the same
index-line endpoints (due to the spinor weight carried by these
points), every term will be multiplied by the same product of $\eta$'s
and $\etab$'s, albeit in different orderings.  The sign map for each
term is then,
\begin{equation}
\eta^a_{i_1} \eta^a_{i_2} \cdots \eta^a_{i_l}
 \; \etab_{j_1 a} \etab_{j_2 a} \cdots \etab_{j_m a} 
\rightarrow  {\rm signature}[i_{1}i_{2}\cdots i_{l}] \,
 {\rm signature}[j_{1}j_{2}\cdots j_{m}] \,,
\label{MixedSignature}
\end{equation} 
where the $\eta$'s commute with the $\etab$'s, $l$ is the number of
legs on an MHV-MHV border plus number of external $\eta$'s, and $m$ is
the number of legs at an \MHVbar{}- \MHVbar{} border plus the number
of external $\etab$'s.

\begin{figure}[t]
\centerline{\epsfxsize 3.6 truein \epsfbox{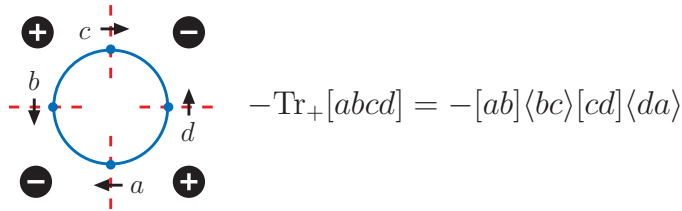}}
\caption[a]{\small According to  the mixed superspace sign rules (``rule 2'') a
closed loop of index lines corresponds to a (chiral) trace of only
momenta, no supermomenta, with an explicit insertion of a negative
sign, reflecting the fermionic nature of index lines.  (For clarity
the momenta are here directed so that no implicit sign comes out of
the spinors according to the sign rules \eqn{signrules}.)}
\label{FlavorLoopRuleFigure}
\end{figure}

An important special case is if the index lines form a closed
loop.  Then there are no Grassmann parameters present, only spinors
enter, or momenta in the form of a chiral trace, as shown in
\fig{FlavorLoopRuleFigure}.  The proper prescription for this case is to
insert an explicit factor $(-1)$ for each closed index loop.  This corresponds
to the standard prescription for fermion loops, and thus it reflects the
fermionic nature of the index lines.

To see how the mixed superspace works consider again the example in
\fig{SixPtFlavorDiagramFigure}.  We read off the diagrams, giving,
\begin{eqnarray}
& {\rm (a)}:&\;\;  (\etab_1 [1 | 6 |4\rangle \eta_4)
  (\eta_3 \spa{3}.{5} \eta_5)  ~~\rightarrow 
   -[1 | 6 |4\rangle \spa{3}.{5}\,, \nn \\
& {\rm (b)}:&\;\;  (\etab_1 [ 1 | 7 |3\rangle \eta_3) 
      (\eta_5 \langle -5\,4\rangle \eta_4)   \, \rightarrow
       ~\,~[1 | 7 |3\rangle \spa{5}.{4}\,, \nn \\
& {\rm (c)}:&\;\;  (\etab_1 [ 1 | 8 |3\rangle \eta_3) 
     (\eta_5 \langle -5\,4 \rangle \eta_4)  \, \rightarrow 
        ~\,~[1 | 8 |3\rangle \spa{5}.{4}\,.
\end{eqnarray}
where $[i|j|k\rangle\equiv \spa{i}.{j}\spb{j}.{k}$.  
As discussed above, we commute the $\etab$ past the $\eta$'s, and place
them in the canonical order $\eta_3\eta_4\eta_5\etab_1$, after which
they are removed.  The result is equivalent to the first rule, but
perhaps is simpler to carry out manually.

\subsection{Supersum simplifications}

\begin{figure}[t]
\centerline{\epsfxsize 6.1 truein \epsfbox{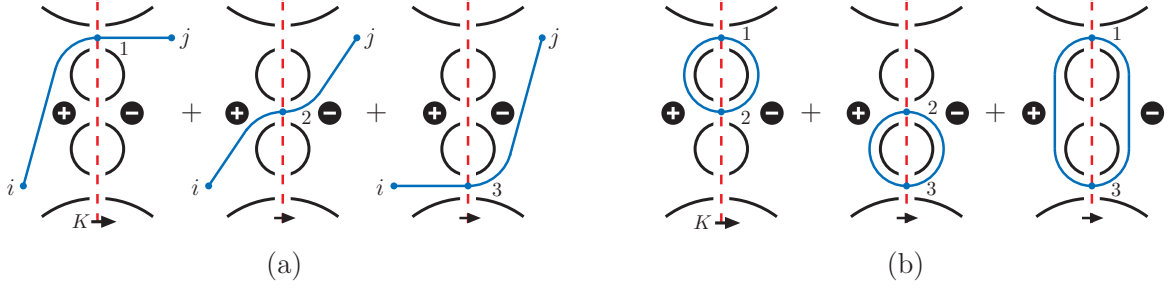}}
\caption[a]{\small Two examples of momentum conservation identities,
allowing us to convert loop momenta to external momenta.  In
(a) we have three index lines that can be summed up to $\eta_i\langle i | l_1 +  l_2 +  l_3 |j] \etab_j$.
Because the vertical cut cross the entire diagram the sum of loop 
momenta can be
re-expressed in terms of external momentum $K=l_1 + l_2 + l_3$ .  
Similarly, for (b) the sum
over index lines give $-2 l_1\cdot l_2 - 2 l_2 \cdot
l_3 - 2 l_1\cdot l_3=-K^2$.  }
\label{MomConsFigure}
\end{figure}

\begin{figure}[t]
\centerline{\epsfxsize 3 truein \epsfbox{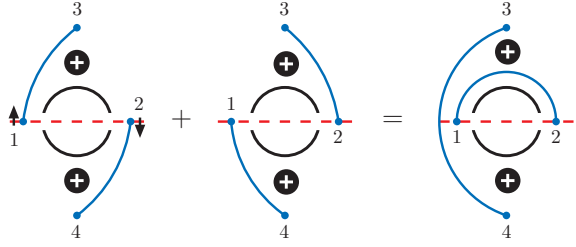}}
\caption[a]{\small A pictorial representation of Schouten's identity.
All index lines carry the same suppressed $SU(4)$ index.  (Note that
the supermomentum $q_i$ flips sign according to \eqn{signrules} depending on
which side of the dashed (red) cut line the shaded (blue) index line
extends.)}
\label{SchoutenFigure}
\end{figure}

\begin{figure}[t]
\centerline{\epsfxsize 5.7 truein \epsfbox{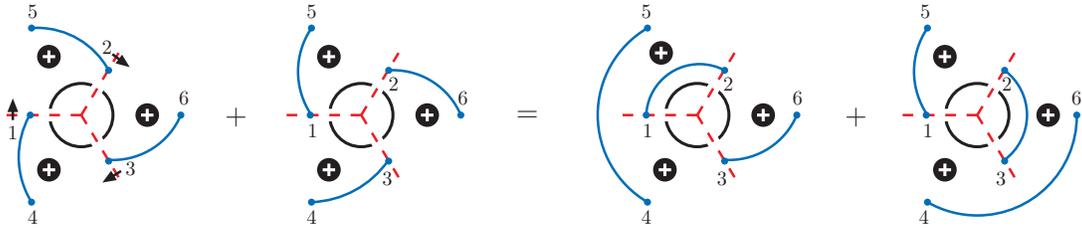}}
\caption[a]{\small Pictorial representation of more complicated
applications of Schouten's identity.}
\label{SchoutenGenFigure}
\end{figure}

In contrast to the algebraic approach of \sect{LinearSystemSection},
the index-diagram approach typically gives results that may be further
simplified.  In particular, in order to fully expose cancellations of
powers of loop momenta due to supersymmetry, rearrangements using
momentum conservation and Schouten's identity are generally
necessary.  Two typical situations where momentum conservation allows
us to pull out powers of loop momenta as external momenta are
displayed in \fig{MomConsFigure}.  Using the mixed-superspace rules (rule 2),
the index lines correspond to,
\begin{eqnarray}
&& {\rm (a)}:\;\; \eta_i\langle i | l_1 |j] \etab_j 
    + \eta_i\langle i | l_2 |j] \etab_j 
    + \eta_i\langle i | l_3 |j] \etab_j =  
  \eta_i\langle i | l_1 +  l_2 +  l_3 |j] \etab_j \,, \nn \\
&& {\rm (b)}:\;\; -\spa{l_1}.{l_2} \spb{l_2}.{l_1} 
    -\spa{l_2}.{l_3} \spb{l_3}.{l_2} 
    -\spa{l_1}.{l_3} \spb{l_3}.{l_1} = -(l_1 + l_2 + l_3)^2 \,,
\end{eqnarray}
where the $R$-symmetry indices have been suppressed, and where the
negative signs are due to the rule of \fig{FlavorLoopRuleFigure} for
closed index line loops.  In both cases we have a vertical cut
which runs from one side of a diagram to the other, therefore the loop
momentum sum corresponds to the external momentum $K=l_1 + l_2 + l_3$
crossing the cut, by momentum conservation.

Another important manipulation follows from Schouten's identity
displayed pictorially in \fig{SchoutenFigure}.  Reading off the index
diagrams we have,
\begin{equation}
\spa{-q_1}.{q_3} \spa{-q_2}.{q_4} + \spa{q_1}.{q_4} \spa{q_2}.{q_3} 
 = \spa{-q_1}.{q_2}\spa{q_3}.{q_4}\,,
\label{SchoutenEqn}
\end{equation}
in terms of supermomentum spinor products
(\ref{SpinorDeltaForm2}).  This can be written in a symmetric form.
Extracting the signs from the incoming supermomenta (\ref{signrules})
gives,
\begin{equation}
\spa{q_1}.{q_3} \spa{q_2}.{q_4} + \spa{q_1}.{q_4} \spa{q_2}.{q_3} 
+ \spa{q_1}.{q_2}\spa{q_3}.{q_4}=0\,,
\label{SchoutenEqn2}
\end{equation}
which expresses Schouten's identity as the statement that
$\spa{q_1}.{q_2}\spa{q_3}.{q_4}$ symmetrized over all legs vanishes.
(From this it also follows that all spinor strings involving $2n>2$
super-momenta vanish upon symmetrization.) In terms of regular bosonic
spinor products, this is equivalent to the usual Schouten's identity
where the anti-symmetrization of the spinor strings vanishes.  We note
that although the orginal index lines start and end on legs within a
single tree amplitudes, after an application of Schouten's identity in
\fig{SchoutenFigure}, they can begin and end on legs of different tree
amplitudes in the cuts.

Besides the basic
identity more complicated versions may be needed.  For
example, for the configuration in \fig{SchoutenGenFigure}, we have the
identity,
\begin{eqnarray}
&& \spa{q_1}.{q_4} \spa{q_2}.{q_5} \spa{q_3}.{q_6}
+ \spa{-q_1}.{q_5} \spa{-q_2}.{q_6} \spa{-q_3}.{q_4} \nn \\
&& \hskip 4 cm  =
\spa{q_4}.{q_5} \spa{-q_1}.{q_2} \spa{q_3}.{q_6}
+ \spa{-q_1}.{q_5} \spa{q_4}.{q_6} \spa{-q_2}.{q_3}\,, \hskip .5 cm 
\label{GenSchoutenEqn}
\end{eqnarray}
which is obtained by a composition of two applications of Schouten's identity.

We note that the identities presented in this section remains valid under
conjugation: $\langle \rangle \leftrightarrow []$, $q \leftrightarrow
\qbar$, $\eta \leftrightarrow \etab$, MHV $\leftrightarrow$ \MHVbar.

\subsection{Three-loop examples}
\label{ThreeLoopSubsection}

\begin{figure}[t]
\centerline{\epsfxsize 6.2 truein \epsfbox{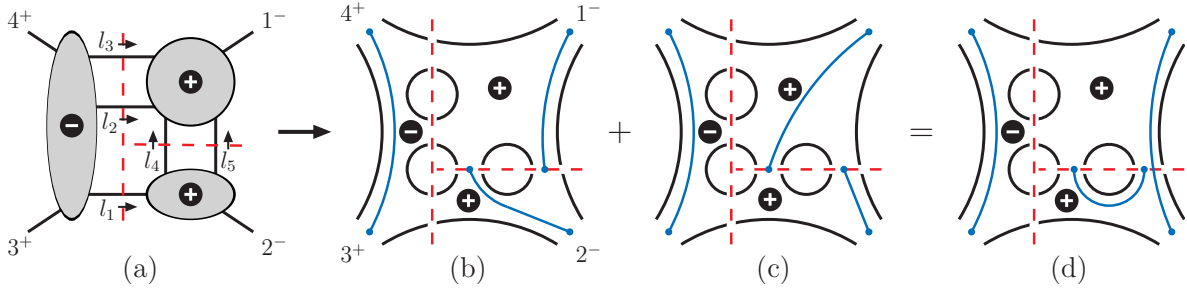}}
\caption[a]{\small A contribution of a three-loop cut (a) in terms of
index diagrams tracking only a single $SU(4)$ index (b) and (c).
Diagram (d) follows from applying Schouten's identity given in
\fig{SchoutenFigure} to (b) and (c). The index lines in  diagrams
(b) and (c) all
begin and end on legs of the same tree amplitude, but as in diagram 
(d), after application of Schouten's identity, an index line can
connect legs of different tree amplitudes.  The other independent
configuration of holomorphicity of the tree amplitudes for this cut is
given in \fig{ThreeLoopPlanarExample_bFigure}.
\label{ThreeLoopPlanarExample_aFigure}
}
\end{figure}

\begin{figure}[t]
\centerline{\epsfxsize 5 truein \epsfbox{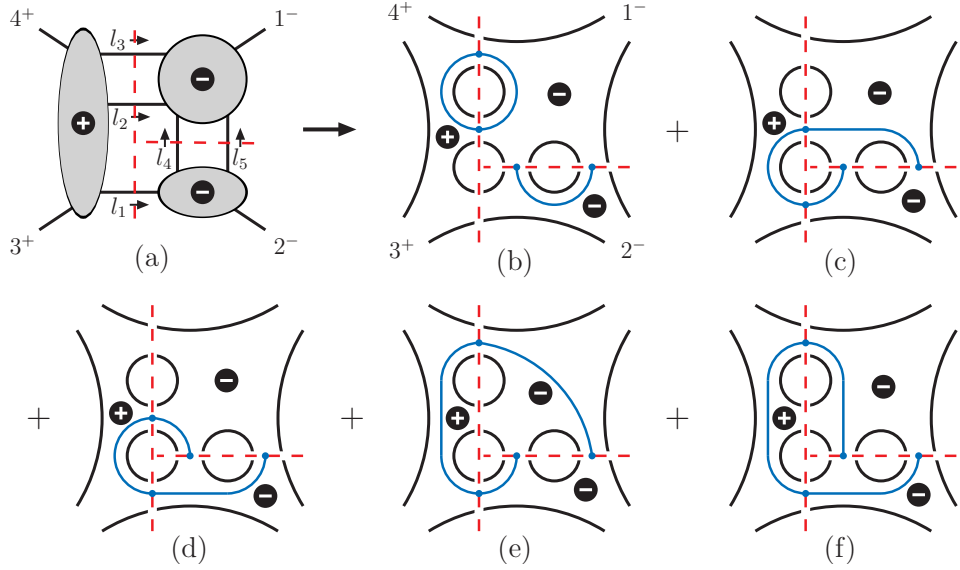}}
\caption[a]{\small The same cut topology as in
\fig{ThreeLoopPlanarExample_aFigure}, but for the other 
independent configuration of tree amplitudes.
\label{ThreeLoopPlanarExample_bFigure}
}
\end{figure}

To illustrate the use of the index diagrams in a non-trivial example
consider the cut of the three-loop four-point amplitude shown in
figs.  \ref{ThreeLoopPlanarExample_aFigure}(a) and
\ref{ThreeLoopPlanarExample_bFigure}(a) in terms of MHV and \MHVbar{}
tree amplitudes.  We have taken the external legs to be gluons with
helicity assignments $(1^-,2^-,3^+,4^+)$ allowing all possible states
of the $\NeqFour$ theory to circulate in the loops.  In this case there
are two distinct configurations of MHV and \MHVbar{} tree amplitudes
in the cuts separated into the two
figures.  As mentioned in \sect{DiagramSubSection}, the four-point
trees can be chosen to be either holomorphic or anti-holomorphic, so
flipping the identification of four-point trees from MHV to \MHVbar{}
does not lead to distinct contributions.

Consider first the configuration in
\fig{ThreeLoopPlanarExample_aFigure}, where two of the tree amplitudes
composing the cut are MHV and one is \MHVbar. 
We have,
\begin{eqnarray}
&& \hskip-40pt
C^{\rm fig.\; \ref{ThreeLoopPlanarExample_aFigure}}  = 
\sum_{\rm states}
A_5^{\rm \overline{MHV}}(3^+,4^+, l_3, l_2, l_1) \, 
A_5^{\rm MHV}(1^-, -l_5, -l_4, -l_2, -l_3) \,
A_4^{\rm MHV}(2^-,-l_1, l_4, l_5)
   \label{ThreeLoopPlanarExample3a} \hskip .3 cm \\
& =  & 
i \rho^{\rm fig.\;\ref{ThreeLoopPlanarExample_aFigure}}
{1 \over \spb3.4 \spb4.{l_3} \spb{l_3}.{l_2} \spb{l_2}.{l_1} \spb{l_1}.3} \, 
{1\over \spa1.{l_5} \spa{l_5}.{l_4} \spa{l_4}.{l_2}
      \spa{l_2}.{l_3} \spa{l_3}.1} \,
{1 \over \spa{2}.{l_1} \spa{l_1}.{l_4} \spa{l_4}.{l_5}
      \spa{l_5}.2} \,, \nn
\end{eqnarray} 
where the numerator result of the supersum contained in $\rho^{\rm fig.\;
\ref{ThreeLoopPlanarExample_aFigure}}$ can be obtained from the index
diagrams in \fig{ThreeLoopPlanarExample_aFigure}.  
The routings (b) and (c) are the only possibilities for a single $SU(4)$
index. This can be worked out following the rules that index lines,
corresponding to physical states, are discontinuous between two MHV
amplitudes and continuous between MHV and \MHVbar{} amplitudes.
Furthermore, every MHV and \MHVbar{} tree amplitude contains exactly one
index line per $SU(4)$ index.  Each line must properly attach to the external
assignment of $SU(4)$ indices (in this case the helicity of the
external gluons).  Using either set of
rules for reading the diagrams in \sect{Rules1and2} gives the
single-index-line numerator,
\begin{equation}
-\spb{\qbar_3}.{\qbar_4} \spa{q_{l_4}}.{q_2} \spa{q_{1}}.{q_{l_5}}
-\spb{\qbar_3}.{\qbar_4}\spa{q_{l_5}}.{q_2}\spa{q_1}.{q_{l_4}}
=\spb{\qbar_3}.{\qbar_4}\spa{q_1}.{q_2}\spa{q_{l_4}}.{q_{l_5}} \,.
\end{equation}
The right-hand side corresponds to
\fig{ThreeLoopPlanarExample_aFigure}(d) which is obtained from
\fig{ThreeLoopPlanarExample_aFigure}(b) and (c) after applying the
pictorial Schouten's identity in \fig{SchoutenFigure}.  Dropping the
Grassmann parameters and raising the result to fourth power
immediately yields,
\begin{equation}
\rho^{\rm fig.\; \ref{ThreeLoopPlanarExample_aFigure}}
 =\spb3.4^4 \spa1.2^4 \spa{l_4}.{l_5}^4 \,.
\label{ThreeLoopPlanarExampleSuperSum3a}
\end{equation}

\begin{figure}[t]
\centerline{\epsfxsize 4.8 truein \epsfbox{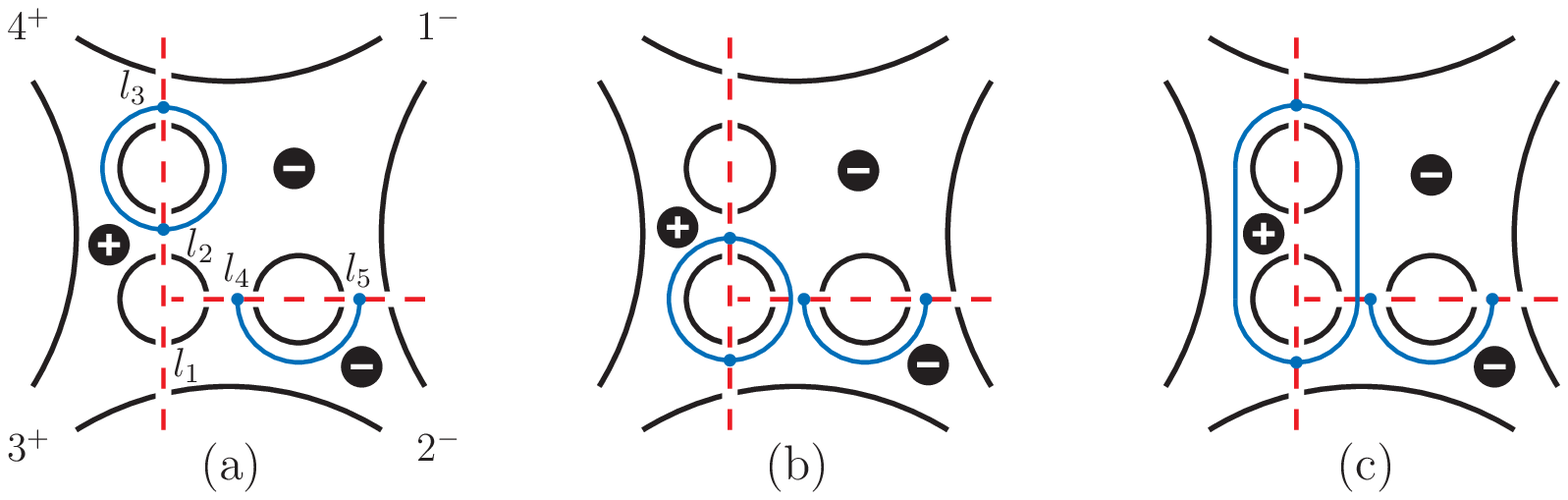}}
\caption[a]{\small Simplified results after applying Schouten's
identity.  Diagram (a) is just diagram (b) of
\fig{ThreeLoopPlanarExample_bFigure}, while diagram (b) is obtained by
combining diagrams (c) and (d) of \fig{ThreeLoopPlanarExample_bFigure}
via the pictorial Schouten's identity in
\fig{SchoutenFigure}.  Similarly, diagram (c) comes from combining
diagrams (e) and (f) of \fig{ThreeLoopPlanarExample_bFigure}.  This
form exposes supersymmetric cancellation, allowing us to extract
factors depending only on external momenta from each cut numerator.}
\label{ThreeloopPlanarSchoutenFigure}
\end{figure}

The other distinct contribution of holomorphicity of tree amplitudes in
\fig{ThreeLoopPlanarExample_bFigure}, while somewhat more complicated,
is quite similar.  For this contribution we have,
\begin{eqnarray}
C^{\rm fig.\; \ref{ThreeLoopPlanarExample_bFigure}} &\! =\! & \sum_{\rm states}
A_5^{\rm MHV}(3^+,4^+, l_3, l_2, l_1) \, 
A_5^{\rm \overline{MHV}}(1^-, -l_5, -l_4, -l_2, -l_3) \,
A_4^{\rm \overline{MHV}}(2^-,-l_1, l_4, l_5) \label{MHVCut3LoopExample3b} \\
&=& \!
i \rho^{\rm fig.\; \ref{ThreeLoopPlanarExample_bFigure}}
{1 \over \spa3.4 \spa{4}.{l_3} \spa{l_3}.{l_2} 
         \spa{l_2}.{l_1} \spa{l_1}.3} \, 
{1 \over \spb1.{l_5} \spb{l_5}.{l_4} \spb{l_4}.{l_2} 
        \spb{l_2}.{l_3} \spb{l_3}.1} \,
{1 \over \spb{2}.{l_1} \spb{l_1}.{l_4} \spb{l_4}.{l_5} 
   \spb{l_5}.{2} } \,.  \nn
\end{eqnarray}
The result of the state sum is contained in the 
factor $\rho^{\rm fig.\; \ref{ThreeLoopPlanarExample_bFigure}}$ and 
can be read off from the index lines in \fig{ThreeLoopPlanarExample_bFigure}.
Using the mixed superspace rules (rule 2), 
the five diagrams in this figure (b)-(f) yield a numerator factor,
\begin{eqnarray}
&&-\spa{l_2}.{l_3}\spb{l_3}.{l_2} \spb{\qbar_{l_4}}.{\qbar_{l_5}}- \etab_{l_4}[l_4| l_1 l_2 | l_5]  \etab_{l_5}- \etab_{l_4}[l_4| l_2 l_1 | l_5]  \etab_{l_5}- \etab_{l_4}[l_4| l_1 l_3 | l_5]  \etab_{l_5}- \etab_{l_4}[l_4| l_3 l_1 | l_5]  \etab_{l_5}\nn \\
&& \hskip 2 cm 
=-(\spa{l_2}.{l_3}\spb{l_3}.{l_2}+\spa{l_1}.{l_2}\spb{l_2}.{l_1}+\spa{l_1}.{l_3}\spb{l_3}.{l_1}) \spb{\qbar_{l_4}}.{\qbar_{l_5}} \nn \\
&&\hskip 2 cm 
 =-(l_1+l_2+l_3)^2 \spb{\qbar_{l_4}}.{\qbar_{l_5}} \nn \\
&&\hskip 2 cm 
 =-s \spb{\qbar_{l_4}}.{\qbar_{l_5}} \,.
\label{ThreeLoopPlanarExampleSuperSum3b}
\end{eqnarray}
The second line in this equation is obtained by applying the pictorial
Schouten's identity displayed in \fig{SchoutenFigure} to the second
and third contributions in \fig{ThreeLoopPlanarExample_bFigure}, as
well as to the fourth and fifth.  This gives the second line
corresponding to the diagrams displayed in
\fig{ThreeloopPlanarSchoutenFigure}.  The result on the last line of
\eqn{ThreeLoopPlanarExampleSuperSum3b} follows from
momentum conservation $(l_1+l_2+l_3)^2 = (k_1 + k_2)^2 = s$.
Stripping the anticommuting parameters and raising the
result to the fourth power gives us the desired numerator,
\begin{equation}
\rho^{\rm fig.\; \ref{ThreeLoopPlanarExample_bFigure}}= s^4
\spb{l_4}.{l_5}^4\,.
\end{equation}
This displays a cancellation of a total of eight powers of loop
momenta from the numerator of the cut.

Rather remarkably, we see that after dividing out the four-point tree
amplitude, the two contributions
(\ref{ThreeLoopPlanarExampleSuperSum3a}) and
(\ref{ThreeLoopPlanarExampleSuperSum3b}) corresponding to
\figs{ThreeLoopPlanarExample_aFigure}{ThreeLoopPlanarExample_bFigure}
are complex conjugates of each other
\begin{equation}
\frac{C^{\rm fig.\; 
 \ref{ThreeLoopPlanarExample_bFigure}\rm (b)}}{A^\tree_4} = \Biggl(
\frac{C^{\rm fig.\; 
\ref{ThreeLoopPlanarExample_aFigure}\rm (a)}}{A^\tree_4} \Biggr)^* \,.
\label{DaggerRelation3}
\end{equation}
The tree amplitude in this equation is given (see \eqn{TreeExamples}) by, 
\begin{equation}
A_4^\tree(1^-,2^-, 3^+, 4^+) =  
i \frac{\spa1.2^4}{\spa1.2 \spa2.3 \spa3.4\spa4.1} = 
i \frac{\spb3.4^4}{\spb1.2 \spb2.3 \spb3.4\spb4.1} \,.
\end{equation}
which are the MHV and \MHVbar{} forms of the four-gluon amplitude.  To
make the relation (\ref{DaggerRelation3}) manifest, on the left side
we use the MHV form of the tree amplitude while on the right side we
use the \MHVbar{} form.  Thus, after removing an overall factor of the
tree amplitude the two contributions add up to a real expression.  A
consequence of this observation is that the cut can be expressed
entirely in terms of scalar products of momenta multiplied by an
overall factor of the tree amplitude.  (Terms containing the
Levi-Civita tensor cancel.)  For amplitudes other than four-point
ones, this property no longer holds~\cite{Fusing}.

\section{Tracking Contributions}
\label{TrackingLoopSection}

The $R$-symmetry index-diagram approach allows us to track the
contributions of individual states in four-dimensional cuts.  This
observation has some interesting consequences.  In particular, as we
outline below, we can give rules for constructing cuts of amplitudes
in various theories with fewer supersymmetries.  We also use this
observation to obtain rules for finding the contributions of the
complete $\NeqFour$ supermultiplet starting from the easily enumerated
purely gluonic contributions.  We illustrate this with some
non-trivial four-loop examples, relevant to the construction of the
complete four-loop four-point amplitude of the $\NeqFour$
theory~\cite{FourLoopNonPlanar}.


\subsection{Cases with fewer supersymmetries}
\label{FewerSusySubsection}

Certain theories with reduced supersymmetry may be constructed 
simply by truncating the spectrum of the $\NeqFour$ theory.
As discussed in \sect{DiagramLoopSection} the supersums contributing
to cuts of amplitudes with external gluons are always the fourth power
of a sum of terms
\begin{eqnarray}
	(A+B+C+\dots)^{\cal N}\,,\hskip 1 cm \NeqFour\,,
	\label{NeqFourSupersum}
\end{eqnarray}
where the summands $A, B, C,\dots$ represent the possible spinorial
numerator factors encoded by the $SU(4)$ index diagrams.  They
correspond to the possible paths, or routings, of an index line, after
all $\eta$'s and $\etab$'s have been removed.  After expanding
(\ref{NeqFourSupersum}), the terms are in one-to-one correspondence to
individual particles and helicity configurations.  In particular,
index lines routed in groups of four correspond to purely-gluonic
states and give the numerator terms $A^4,B^4,C^4\dots$---a fact which
we exploit below in \sect{SimpleRulesSubsection}.  Combinations where
the four index lines follow different routings give rise to the cross
terms in the product (\ref{NeqFourSupersum}).  These terms correspond
to cases where scalar and fermion fields of the supermultiplet
propagate, {\it e.g.} terms such as  $A^3 B$ or $A^2 B^2$
arise from fermion and scalar states in
the loops (see \fig{OneLoopMHVbarExampleFigure} for explicit
examples).  Precise tracking of the matter fields through the cuts is
dictated by the index-line diagrams.

For a few theories which are closely related to $\NeqFour$
super-Yang-Mills theory it is possible to write down closed form
expressions for the cuts of their scattering amplitudes in terms of
the $\NeqFour$ cuts.
This is a consequence of the fact that the ${\cal N}=4$ vector multiplet
decomposes in a direct sum of representations of ${\cal N}<4$
supersymmetry algebras.  
By systematically dropping contributions following their $R$-charges, 
we obtain cuts of amplitudes in theories with reduces
supersymmetry and a field content which is a subset of that of
$\NeqFour$ SYM.

Starting from the $\NeqFour$ spectrum we may eliminate one ${\cal N}=2$
hypermultiplet to obtain the spectrum of the pure ${\cal N}=2$
super-Yang-Mills theory.  This can be done by expressing the
representations of the $SU(4)$ $R$-symmetry in representations of an
$SU(2)\times SU(2)\times U(1)$ subgroup and restricting to states
transforming trivially under one $SU(2)$ factor.  Without loss of
generality, this may be taken to act on indices 3 and 4; this
truncation breaks $SU(4)$ down to $SU(2)$, giving the following
states:
\def\hs{\hskip 1.5 cm}
\begin{equation}
 g_+\,, \hs f_+^a\,, \hs s^{ab}\,, \hs s^{34}\,, \hs f^{b34}_-\,, \hs
g_-^{ab34} \,;
\label{NeqTwoStates}
\end{equation}
here $a,b=1$ or $2$ are the $SU(2)$ $R$-symmetry indices.  Although 
indices 3 and 4 no longer plays the role of group indices, we keep
them as labels to distinguish the states and to keep notation 
uniform with the $\NeqFour$ case.  As expected, there are two
fermions which, on-shell, correspond to four states $f_+^a$ and
$f^{b34}_-$.  The two scalar fields are complex conjugates
$(s^{34})^*=s^{12}$, thus the counting of on-shell states is
consistent with the ${\cal N}=2$ gauge multiplet.

In terms of the index diagrams this truncation implies that we should
keep only those diagrams where indices 3 and 4 are grouped together.  
For external gluons, this gives the following cut numerator,
\begin{equation}
(A+B+C+\dots)^2(A^2+B^2+C^2+\dots)\,,
\label{NeqTwoSupersum}
\end{equation}
where $A, B, C$ represent the same terms as in \eqn{NeqFourSupersum},
and the squares $A^2, B^2, C^2$ are a consequence of the above requirement
that two indices are always grouped together in the diagrams.

In the same spirit,  by dropping one chiral multiplet from the  ${\cal
N}=2$ spectrum we obtain the on-shell ${\cal N}=1$ gauge supermultiplet, 
\def\hs{\hskip 1.5 cm}
\begin{equation}
	g_+\,, \hs f_+^a\,, \hs f^{234}_-\,, \hs g_-^{a234} \, .
\label{NeqOneStates}
\end{equation}
By requiring that the fields transform trivially in the $2,3,4$
directions we remove all scalars and all but one fermion.  Although
this also fixes the index $a=1$ we keep the label $a$ ``covariant'' as
a reminder that it should treated differently from the others.  

This truncation is reflected at the level of index diagrams as 
three lines, corresponding to three indices taking the values $2,3,4$,
always being grouped together, while the remaining line being allowed to
have an independent routing.
For external gluons, the resulting cut numerator factor is then,
\begin{equation}
(A+B+C+\dots)(A^3+B^3+C^3+\dots)\, .
\label{NeqOneSupersum}
\end{equation}

By truncating away all fields carrying $R$-charges, the $\NeqFour$
theory is reduced to pure (${\cal N}=0$) Yang-Mills theory.  The cut
numerators may then be identified with those index diagrams in
which all four index lines follow the same path.  This eliminates all 
contributions from ``matter'' fields and yields the numerator,
\begin{eqnarray} 
	(A^4+B^4+C^4+\dots) \,.
	\label{NeqZeroSupersum}
\end{eqnarray}

The above formul\ae\ for cut numerators can be summarized in a single
closed form,
\begin{eqnarray} 
(A+B+C+\dots)^{\cal N}\times (A^{4-{\cal N}}+B^{4-{\cal N}}+
C^{4-{\cal N}}+\dots)\,,\quad {\cal N}<4\,,
	\label{GenSupersum}
\end{eqnarray}
which holds for ${\cal N}=0,1,2,3$, where the ${\cal N}=3$ case is
identical to the $\NeqFour$ super-Yang-Mills case
\eqn{NeqFourSupersum}.  This is in line with the well-known on-shell
equivalence of the ${\cal N}=3$ and ${\cal N}=4$ super-Yang-Mills
theories~\cite{N3N4Equiv}.  In equation (\ref{GenSupersum}) the first
factor represents the supersymmetric summation over index lines with
${\cal N}$ independent $R$-symmetry indices, the second factor
corresponds to the controlled truncation of index diagrams, so that
$4-{\cal N}$ indices are always grouped together.  This formula is
consistent with one-loop expressions for cuts found in, {\it e.g.,}
refs.~\cite{Fusing,UnexpectedCancel}

In fact, the above closed form for the cut numerator implies that the
amplitudes of these theories can be assembled into generating
functions.   
We illustrate this by introducing such generating functions for
the MHV tree amplitudes for the minimal gauge multiplets of 
${\cal N}<4$ super-Yang-Mills theory,
\begin{equation}
{\cal A}^{\rm MHV}_n(1, 2, \ldots, n)=
\frac{i}{\prod_{j=1}^n\langle j ~(j+1)\rangle}
\, \Big( \prod_{a=1}^{\cal N}\delta^{(2)}( Q^a)\Big)\,
\Bigl(\sum_{i<j}^n \spa{i}.{j}^{4-{\cal N}} 
\prod_{a={\cal N}+1}^{4} \eta_i^a  \eta_j^a\Bigr) \,,
\label{MHVLessSuperAmplitude}
\end{equation}
with ${\cal N}$ counting the number of supersymmetries,
$Q^a=\sum_{i=1}^n \lambda_i \eta_i^a$, and $n\ge 3$.  Each monomial in
the super-amplitude corresponds to an MHV amplitude, where the
external states match the spectra of the respective supersymmetric
theory.  By keeping all four $\eta^a$ for each leg, we have a uniform
bookkeeping device for amplitudes in any theory obtainable by
truncating the spectrum of the $\NeqFour$ theory.  Through the MHV
expansion, this generalizes as well to the non-MHV amplitudes of these
theories.

As a consistency check we have confirmed that the amplitudes grouped
in the generating functions, for each value of ${\cal N}$, form a
closed set under factorization, thus ensuring that internal states in these
amplitudes are in the spectrum of external states.  Equipped with the
generating functions we may follow ref.~\cite{FreedmanGenerating} and
use supersymmetry to validate the interactions.  The explicit
super-momentum constraints in eq.~(\ref{MHVLessSuperAmplitude})
ensures that superamplitudes are annihilated by the super-charges,
$Q^a$ with $a=1,\dots {\cal N}$.  This property is sufficient to
link all MHV amplitudes in the generating function
(\ref{MHVLessSuperAmplitude}) to the gluonic Park-Taylor amplitudes by
supersymmetry, and ensures correct couplings.

Interestingly, following the discussion in \sect{StructureSection},
from super-momentum conservation, the cut of any ${\cal N}<4$
super-Yang-Mills multi-loop super amplitude ${\cal A}_n$ is
proportional to the overall super-momentum conservation constraint,
\be
\prod_{a=1}^{\cal N}\delta^{(2)}(Q^a)\,.
 \label{SuperMomentumConstraint}
\ee
As for the $\NeqFour$ theory, the fact that this structure factors out
in all cuts implies that complete on-shell loop amplitudes also
contain a factor of the overall supermomentum conservation constraint.

The considerations outlined here can be generalized to other theories
and particle spectra.  An example in this direction are orbifolds of
$\NeqFour$ SYM.  While the spectra of such theories are still obtained
by truncation of the $\NeqFour$ spectrum, the fact that the gauge
group is intertwined nontrivially with the truncation makes this
generalization nontrivial.  It has been shown \cite{orbifolds} that
planar scattering amplitudes in the orbifolded theory are, up to
trivial numerical factors, the same as those of the parent theory, to
all orders in perturbation theory.  Nonplanar amplitudes are, however,
different.  The fact that supersum calculations do not depend on
whether amplitudes are planar or not, hints that a closer relation
might exist between the amplitudes of the orbifolded and parent theory
even at the nonplanar level.

Considerations similar to those discussed above also hold for
supergravity, where one can write down generating functions for the
MHV and \MHVbar{} sectors for ${\cal N}<8$ starting from the
$\NeqEight$ generating function given in
refs.~\cite{FreedmanGenerating,FreedmanUnitarity}.  Furthermore, one
can write down generating functions for more general
non-supersymmetric matter content.  One interesting example is dictated
by the set of index diagrams with even numbers of index lines routed
identically, giving a bosonic state-sum,
\begin{eqnarray} 
	(A^2+B^2+C^2+\dots)^{2}\,,
	\label{NonSupersum}
\end{eqnarray}
corresponding to a theory of gluons and scalars 
arising from the dimensional reduction of pure Yang-Mills theory 
from six to four dimensions.  The amplitudes of this theory thus also 
possess a generating function description.  It should also be possible
to extend these considerations to theories 
not obtainable from $\NeqFour$ super-Yang-Mills theory by
truncation.


\subsection{A simple algorithm for evaluating $\NeqFour$ supersums}
\label{SimpleRulesSubsection}

Consider now a generalized cut which breaks an $n$-gluon amplitude of
$\NeqFour$ super-Yang-Mills theory at $L$ loops into a product of tree
amplitudes.  As discussed above, the purely gluonic contributions are
represented in index diagrammatic language by grouping all index lines
into sets of four following identical paths through the
diagrams. The key observation is that the purely gluonic diagrams
cover all possible paths.  This allows us to use the enumeration of only
gluonic helicity configurations in the cuts to obtain the contributions of all
other states.  The relative signs between terms are then determined by
dressing with anticommuting parameters as discussed in 
\sect{DiagramLoopSection}.  

The simplified rules for obtaining the $\NeqFour$
super-Yang-Mills numerators of the $n$-gluon amplitudes from the
purely gluonic cases are:
\begin{itemize}

\item Identify all non-vanishing purely gluonic helicity choices.  If
the cut contains a tree amplitude which is neither MHV nor \MHVbar{},
expand it in MHV vertices as discussed in \sect{ruleMHVSubSection}.
Each helicity choice then belongs to one independent configuration of
holomorphicity of MHV and \MHVbar{} tree amplitudes.  (Recall that at
four points, the MHV and \MHVbar{} tree amplitudes are equivalent and
should be treated as as dependent.)  Each independent configuration 
of holomorphicity will form a distinct contribution,
which are summed over at the end.

\item For each independent choice of holomorphicity, form the 
sum over all gluonic helicity configurations, assigning one 
power of $\eta_i \spa{i}.{j} \eta_j$ for
MHV amplitudes with negative helicity legs $i$ and $j$ and one power
of $\tilde \eta_i \spb{i}.{j} \tilde \eta_j$ for \MHVbar{} amplitudes
with positive helicity legs $i$ and $j$.

\item Apply the Fourier transform rule (\ref{FourierRule}) and
anticommute the $\eta_i$ and $[i|$ to a standard ordering, picking up
relative signs between terms in the sum.

\item After removing the common factor of the anticommuting parameters
ordered in a standard form, raise the sum to the fourth power.

\item The denominator for a given configuration of MHV and \MHVbar{} 
tree amplitudes in the cuts is the product of denominators
for each tree amplitude, as well as any propagators from MHV 
expansions.

\item Sum over the contributions of the independent choices
of holomorphicity.
\end{itemize}


\subsection{Four-loop examples}

\begin{figure}[t]
\centerline{\epsfxsize 5.2 truein \epsfbox{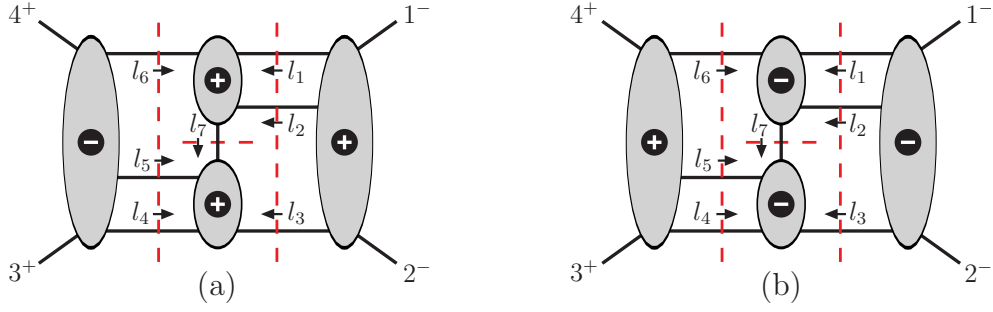}}
\caption[a]{\small An example of a four-loop planar cut.
Case (a) gives the singlet helicity 
configuration, where only a single gluonic helicity configuration 
contributes.  
Case (b) gives non-singlet helicity configurations where all particles in 
the $\Neqfour$ multiplet contribute.
}
\label{FourLoopPlanarExampleFigure}
\end{figure}

To give an illustration of the above rules, we consider supersums in
the evaluation of some non-trivial cuts of four-loop amplitudes.
First consider the planar generalized cut of the four-loop amplitude
$A_4^{\fourloop}(1^-, 2^-, 3^+, 4^+)$ shown in
\fig{FourLoopPlanarExampleFigure}.  There are two distinct
configurations of MHV and \MHVbar{} tree amplitudes.
\Fig{FourLoopPlanarExampleFigure}(a) is a singlet helicity
configuration. The helicity configuration of the internal lines is
uniquely fixed once the external lines are specified. Thus, according
to our rules only a single term appears in the sum raised to the
fourth power.  The value of contribution (a) is then,
\begin{eqnarray}
C^{\rm fig.\; \ref{FourLoopPlanarExampleFigure}\rm (a)} &=& 
A_5^{\rm MHV}(1^-, 2^-, l_3^+, l_2^+, l_1^+) \, 
A_4^{\rm MHV}(-l_3^-, -l_4^+, -l_5^+, -l_7^-) \, 
A_4^{\rm MHV}(-l_1^-, -l_2^-, l_7^+, -l_6^+) \,  \nn \\
\null &&\hskip 2 cm \times 
A_5^{\rm \overline{MHV}}(3^+, 4^+, l_6^-, l_5^-, l_4^-) \nn \\
&=&
-{\spa1.2^4 \over \spa1.2 \spa2.{l_3} \spa{l_3}.{l_2} 
                  \spa{l_2}.{l_1} \spa{l_1}.1 } \,
{\spa{l_3}.{l_7}^4  \over \spa{l_3}.{l_4} \spa{l_4}.{l_5} \spa{l_5}.{l_7} 
                           \spa{l_7}.{l_3} } \nn \\
&& \hskip 1 cm \null \times 
 {\spa{l_2}.{l_1}^4 \over \spa{l_1}.{l_2} \spa{l_2}.{l_7} \spa{l_7}.{l_6}
       \spa{l_6}.{l_1} } \, 
 {\spb3.4^4 \over \spb3.4 \spb4.{l_6} \spb{l_6}.{l_5} \spb{l_5}.{l_4}
                  \spb{l_4}.{3} } 
\,.
\label{FourloopPlanara}
\end{eqnarray}
This result is valid for all the gauge multiplet of
${\cal N} \le 4$ supersymmetric theories, since only gluons contribute
here.

Now consider the more complicated case in
\fig{FourLoopPlanarExampleFigure}(b) involving non-singlet
contributions.  We have,
\begin{eqnarray}
C^{\rm fig.\; \ref{FourLoopPlanarExampleFigure}\rm (b)} &=& 
\sum_{\rm states} A_5^{\rm \overline{MHV}}(1^-, 2^-, l_3, l_2, l_1) \, 
A_4^{\rm \overline{MHV}}(-l_3, -l_4, -l_5, -l_7) \, 
A_4^{\rm \overline{MHV}}(-l_1, -l_2, l_7, -l_6) \, \hskip .7 cm  \nn \\
\null &&\hskip 2 cm \times  
A_5^{\rm MHV}(3^+, 4^+, l_6, l_5, l_4) \nn \\
&=&
-\rho^{\rm fig.\; \ref{FourLoopPlanarExampleFigure}\rm (b)} 
{1 \over \spb1.2 \spb2.{l_3} \spb{l_3}.{l_2} 
                           \spb{l_2}.{l_1} \spb{l_1}.1 } \,
{1  \over \spb{l_3}.{l_4} \spb{l_4}.{l_5} \spb{l_5}.{l_7} 
                           \spb{l_7}.{l_3} } \nn \\
&& \hskip 1 cm \null \times 
 {1 \over \spb{l_1}.{l_2} \spb{l_2}.{l_7} \spb{l_7}.{l_6}
       \spb{l_6}.{l_1} } \, 
 {1  \over \spa3.4 \spa4.{l_6} \spa{l_6}.{l_5} \spa{l_5}.{l_4}
                  \spa{l_4}.{3} } 
\,,
\end{eqnarray}
where $\rho^{\rm fig.\; \ref{FourLoopPlanarExampleFigure}\rm (b)}$
accounts for the sum over multiplet.  There are a total of eight
distinct purely gluonic helicity configurations, obtained by listing
out the non-vanishing possibilities which maintain the holomorphicity
of \fig{FourLoopPlanarExampleFigure}(b).  Using the rules in the
previous section, the gluonic numerator factors can be converted to eight
primitive contributions,
\begin{eqnarray}
&&A=
\spa{l_4}.{l_5} \spb{l_4}.{l_5} \spb{l_2}.{l_7}\spb{l_1}.{l_3} 
\,, \hskip 3mm 
B=
\spa{l_4}.{l_5} \spb{l_4}.{l_5} \spb{l_7}.{l_1}\spb{l_2}.{l_3}
\,, \hskip 3mm 
C=
\spa{l_4}.{l_6} \spb{l_4}.{l_7} \spb{l_2}.{l_6}\spb{l_1}.{l_3}
\,,\nn \\&&
D=
\spa{l_4}.{l_6} \spb{l_4}.{l_7}  \spb{l_6}.{l_1}\spb{l_2}.{l_3}
\,, 
\hskip 3mm 
E=
\spa{l_5}.{l_6}\spb{l_5}.{l_7} \spb{l_2}.{l_6}\spb{l_1}.{l_3}
 \,, 
\hskip 3mm
F= 
\spa{l_5}.{l_6}\spb{l_5}.{l_7} \spb{l_6}.{l_1}\spb{l_2}.{l_3}
\,,\nn \\&&
G=
\spa{l_4}.{l_6}\spb{l_2}.{l_1}\spb{l_3}.{l_4}\spb{l_6}.{l_7} 
\,, 
\hskip 3mm H=
\spa{l_5}.{l_6}\spb{l_2}.{l_1}\spb{l_3}.{l_5}\spb{l_6}.{l_7} 
\,.
\end{eqnarray}
The sum over these eight terms exhibits the supersymmetric
cancellations after using Schouten's identity and momentum 
conservation,
\begin{equation}
A+B+C+D+E+F+G+H=
s \spb{l_1}.{l_2} \spb{l_7}.{l_3}
\,,
\end{equation}
where $s=(k_1+k_2)^2$.  We may then assemble the supersum for ${\cal
N}<4$ following the discussion in \sect{FewerSusySubsection}; using
\eqn{GenSupersum} we obtain,
\begin{equation}
\rho^{\rm fig.\; \ref{FourLoopPlanarExampleFigure}\rm (b)}= 
(s \spb{l_1}.{l_2} \spb{l_7}.{l_3})^{\cal N}(A^{4-{\cal N}}+B^{4-{\cal N}}+C^{4-{\cal N}}+D^{4-{\cal N}}+E^{4-{\cal N}}+F^{4-{\cal N}}+G^{4-{\cal N}}+H^{4-{\cal N}}),
\label{FourloopPlanarNumerb}
\end{equation}
which is valid for the minimal ${\cal N}=0,1,2,3$ supersymmetric gauge
multiplets.  The case ${\cal N}=3$ is equivalent to ${\cal N}=4$,
\begin{equation}
\rho^{\rm fig.\; \ref{FourLoopPlanarExampleFigure}\rm (b)}_{{\cal N}=4} = 
s^4 \spb{l_1}.{l_2}^4 \spb{l_7}.{l_3}^4 
\,.
\end{equation}

As for the ${\cal N}=4$ three-loop example in
\sect{ThreeLoopSubsection}, the ${\cal N}=4$ case (but not ${\cal N}
\le 2$) exhibits the property that the two contributions in
\fig{FourLoopPlanarExampleFigure}, are complex conjugates after
dividing by an overall factor of the tree amplitude,
\begin{equation}
\frac{i C^{\rm fig.\; 
   \ref{FourLoopPlanarExampleFigure}\rm (b)}_{{\cal N}=4}}{A^\tree_4} = \Biggl(
\frac{i C^{\rm fig.\; 
   \ref{FourLoopPlanarExampleFigure}\rm (a)}}{A^\tree_4}
\Biggr)^* \,,
\label{DaggerRelationFourLoop}
\end{equation}
where the $i$ is inserted to correct for an overall phase that depends
on the loop order.

\begin{figure}[t]
\centerline{\epsfxsize 5 truein \epsfbox{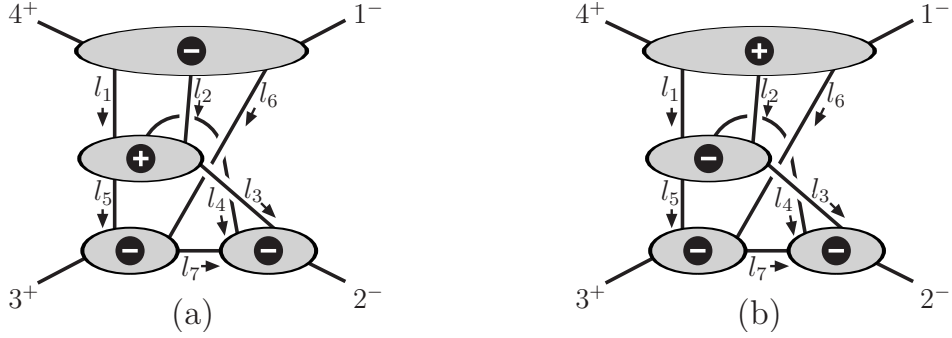}}
\caption[a]{\small A nontrivial nonplanar cut at four loops.  The cuts
(a) and (b) represent the two distinct contributions to the cuts.  As
discussed in the text it using the rules developed here it is
straightforward to write down the expression corresponding to these
diagrams.  All visible legs are on-shell.}
\label{FourLoopNonPlanarExampleFigure}
\end{figure}

As a nonplanar example, consider the cut depicted in
\fig{FourLoopNonPlanarExampleFigure}.  As far as the supersums are
concerned the planarity or nonplanarity of the cut, is of little
consequence, with the only difference appearing in the spinor
denominators which are identical for all terms in the sum.  This is an
especially useful cut because it checks a large number of
contributions to the four-loop amplitude, including the most
complicated nonplanar integrals.  As for previous examples, it turns
out that there are two distinct choices of holomorphicity,
corresponding to (a) and (b) in \fig{FourLoopNonPlanarExampleFigure},
since any helicity configuration falls into one of these two classes.
In the first class (a) we have seven distinct gluonic helicity
choices.  In the second class (b) we have eight distinct gluonic
helicity choices.

We write down the target expression from the cuts
using the above rules.  For (a) we have the cut contribution
\begin{eqnarray}
C^{\rm fig.\; \ref{FourLoopNonPlanarExampleFigure}\rm (a)} 
&=& \sum_{\rm states}
A_5^{\rm \overline{MHV}}(1^-,l_6,l_2,l_1,4^+)\, 
A_5^{\rm MHV}(-l_1,l_4,-l_2,l_3,l_5) \, \nn \\
&&\null \times 
A_4^{\rm \overline{MHV}}(-l_4,-l_3,2^-,-l_7) \, 
A_4^{\rm \overline{MHV}}(-l_5,-l_6,l_7,3^+) \,,
\label{FourLoopNonplanarCuta}
\end{eqnarray}
where the ${\cal N}=4$ supersum factor is
\begin{eqnarray}
\rho^{\rm fig.\; \ref{FourLoopNonPlanarExampleFigure}\rm (a)}_{{\cal N}=4} &=& 
   \biggl[\spa{l_1}.{l_4} \spb{l_1}.4 \spb{l_4}.{l_7} \spb3.{l_6} + 
       \spa{l_1}.{l_3} \spb{l_1}.4 \spb{l_3}.{l_7} \spb3.{l_6}
        + \spa{l_2}.{l_3} \spb{l_2}.4 \spb{l_3}.{l_7} \spb3.{l_6} \nn \\ 
&& \null 
        + \spa{l_2}.{l_4} \spb{l_2}.4 \spb{l_4}.{l_7} \spb3.{l_6} 
        + \spa{l_3}.{l_5} \spb{l_3}.{l_7} \spb{l_5}.3 \spb4.{l_6}  \nn \\ 
&& \null 
        + \spa{l_4}.{l_5} \spb{l_4}.{l_7} \spb{l_5}.3 \spb4.{l_6} + 
          \spa{l_4}.{l_3} \spb{l_3}.{l_4} \spb4.{l_6} \spb3.{l_7}
\biggr]^4 
\,.
\label{FourloopNPNumera}
\end{eqnarray}
By making repeated use of Schouten's identity this 
simplifies to 
\begin{equation}
\rho^{\rm fig.\; \ref{FourLoopNonPlanarExampleFigure}\rm (a)}_{{\cal N}=4}=
 \big(\spa1.2 \spb{l_7}.2 \spb{l_6}.3 \spb1.4 \big)^4 
\,.
\end{equation}

For configuration (b) in \fig{FourLoopNonPlanarExampleFigure} we have the 
contribution, 
\begin{eqnarray}
C^{\rm fig.\; \ref{FourLoopNonPlanarExampleFigure}\rm (b)} &=& 
 \sum_{\rm states}
A_5^{\rm MHV}(1^-,l_6,l_2,l_1,4^+)\, 
A_5^{\rm \overline{MHV}}(-l_1,l_4,-l_2,l_3,l_5) \, \nn \\
&&\null \times 
A_4^{\rm \overline{MHV}}(-l_4,-l_3,2^-,-l_7) \, 
 A_4^{\rm \overline{MHV}}(-l_5,-l_6,l_7,3^+) \,.
\label{FourLoopNonplanarCutb}
\end{eqnarray}
In this case the ${\cal N}=4$ factor from summing over the states
crossing the cuts is
\begin{eqnarray}
 \rho^{\rm fig.\; \ref{FourLoopNonPlanarExampleFigure}\rm (b)}_{{\cal N}=4}
 &=& 
  \biggl[
      \spa1.{l_6} \spb{l_5}.{l_3}  \spb{l_4}.{l_7} \spb3.{l_6} + 
      \spa1.{l_1} \spb{l_1}.{l_4}  \spb{l_3}.{l_7} \spb{l_5}.3 
\label{FourloopNPNumerb}\\ 
&& \null 
    + \spa1.{l_2} \spb{l_3}.{l_7} \spb{l_4}.{l_2} \spb3.{l_5} + 
      \spa1.{l_1} \spb{l_1}.{l_3}  \spb{l_4}.{l_7} \spb3.{l_5} + 
      \spa1.{l_2} \spb{l_2}.{l_3}  \spb{l_4}.{l_7} \spb3.{l_5} \nn \\ 
&& \null 
    + \spa1.{l_6} \spb{l_3}.{l_7} \spb{l_4}.{l_5} \spb3.{l_6} + 
      \spa1.{l_1} \spb{l_1}.{l_5}  \spb{l_3}.{l_4} \spb3.{l_7} + 
      \spa1.{l_2} \spb{l_2}.{l_5}  \spb{l_3}.{l_4} \spb3.{l_7} \biggr]^4 
\nn \,.
\end{eqnarray}
After repeatedly applying Schouten's identity we obtain, 
\begin{equation}
 \rho^{\rm fig.\; \ref{FourLoopNonPlanarExampleFigure}\rm (b)}_{{\cal N}=4}
=
\big( \spa1.2 \spb2.3 \spb{l_3}.{l_4}\spb{l_5}.{l_7} \big)^4 
\,.
\end{equation}
Although not manifest in the form we present here, the two
contributions to the cut in
\eqns{FourLoopNonplanarCuta}{FourLoopNonplanarCutb}, satisfy a complex
conjugation relation similar to the one in
\eqn{DaggerRelationFourLoop}.  To obtain the nonplanar cuts for the
${\cal N} < 4$ supersymmetric theories, we match to the numerator
forms in \eqns{FourloopNPNumera}{FourloopNPNumerb} to
\eqn{NeqFourSupersum} and use the form in \eqn{GenSupersum} to replace
the numerators with the appropriate ones.


\section{From $\Neqfour$ super-Yang-Mills theory to $\NeqEight$ supergravity}
\label{GravitySection}

Many of the tools presented in previous sections, which were derived
from the on-shell superspace of $\NeqFour$ super-Yang-Mills, carry
directly over to $\NeqEight$ supergravity.  For cuts that factorize
loop amplitudes into only MHV and \MHVbar{} tree amplitudes, the
methods of the previous sections can be generalized to $\NeqEight$
supergravity by replacing $\delta^{(8)}(Q^a) \rightarrow
\delta^{(16)}(Q^a)$, and by suitably replacing the other factors in
the amplitudes with the crossing symmetric gravity expressions.  In
this case the $R$-symmetry index runs up to eight.  However, at
present the existence of a complete set of MHV expansion rules for
gravity has not been fully established~\cite{FreedmanGenerating}.  As
such, there are many gravity cuts that cannot be handled directly by
relying on an MHV expansion.  One may use the BCFW recursion form of
the tree-level superamplitudes in the unitarity cuts, but this has not
been studied systematically beyond one
loop~\cite{AHCKGravity,N8TreeSuperAmplitude}.  Furthermore, the issue
of four-dimensional cuts being insufficient for reconstructing the
$D$-dimensional amplitude is more pressing in the case of gravity.  The
presence of twice as many powers of momenta in the numerators of
gravity diagrams, compared to gauge theory, offers more possibilities
for expressions that vanish in four dimensions, but not in $D$
dimensions, to appear in the cuts.  An example of such an object is the
Gram determinant $\det(p_i\cdot p_j)$, with at least five independent
momenta (including loop momenta).

A method that effectively tackles both of these problem is described in
refs.~\cite{BDDPR,GravityThree}: the tree level Kawai-Lewellen-Tye
(KLT ) relations can be used to relate cuts of $\NeqEight$
supergravity to sums of products of cuts $\NeqFour$ super-Yang-Mills
theory, with additional kinematic factors.  Since the KLT relations are
valid in $D$ dimensions, the gravity cuts determined through their use
will automatically be valid in arbitrary dimensions if the 
corresponding Yang-Mills cuts 
are.

Schematically, the KLT relations are of the form
\begin{equation}
M_n^\tree  = \sum_{i,j} g_{ij} A_n^{(i)}  A_n^{(j)} \,,
\end{equation}
where $M^\tree_n$ is an $n$-point $\NeqEight$ supergravity amplitude, the
$A_n^{(i)}$ are color-stripped
$n$-point tree amplitudes in $\NeqFour$ super-Yang-Mills theory labeled
by an index $i$, implicitly incorporating all labels appearing
in the amplitudes.  The 
$g_{ij}$ are polynomials in kinematic invariants $s_{lm}=(k_l+k_m)^2$
of degree $(n-3)$.  The precise form of the relations
for any number of external legs may be found in
ref.~\cite{OneloopMHVGravity}.  While their derivation from the
(super)gravity Lagrangian remains obscure, it was recently shown that
the KLT relations are equivalent to relations between numerator
factors of individual tree diagrams~\cite{TreeJacobi}.

Generalized unitarity cuts in $\NeqEight$ supergravity are
constructed, in much the same way as in $\NeqFour$ SYM, as products of
tree-level amplitudes.  Because the $\NeqEight$ supergravity multiplet
is the tensor product of two $\NeqFour$ super-Yang-Mills vector
multiplets, when applying the KLT relations, the supersymmetric
sums appearing in the cuts for supergravity amplitudes can be
re-expressed as two copies of supersymmetric sums for Yang-Mills
amplitudes.  For example, for a cut that breaks the amplitude
into two tree amplitudes we have~\cite{BDDPR},
\bea
M_n^{L\rm-loop}  \Bigr|_{\rm cut}=\sum_{\NeqEight}
M_{n_1}^{\rm tree}M_{n_2}^{\rm tree}
&=&\sum_{\NeqEight}
\left(\sum_{i,j} g_{ij} A_{n_1}^{(i)}  A_{n_1}^{(j)}\right)
\left(\sum_{k,l} g_{kl} A_{n_2}^{(k)} A_{n_2}^{(l)}\right)
\cr
&=&\sum_{i,j,k,l} g_{ij}g_{kl}
\left(\sum_{\NeqFour}A_{n_1}^{(i)}A_{n_2}^{(k)} \right)
\left(\sum_{\NeqFour}A_{n_1}^{(j)}A_{n_2}^{(l)} \right)\,,
\eea
where the $\sum_{\NeqFour}A_{n_1}^{(i)}A_{n_2}^{(k)}$ are
color-stripped cuts of $\NeqFour$ super-Yang-Mills amplitudes.  Any
cut which decomposes a loop amplitude into a product of trees works
similarly.  Thus, instead of evaluating the supergravity cuts starting
from $D$-dimensional supergravity tree amplitudes, it is generally
more efficient to assemble them from simpler cuts of the Yang-Mills
amplitude via the KLT relations ~\cite{CompactThree}.  

Since the KLT relations also hold for gravity theories with fewer
supersymmetries than the maximal number, the gauge theory discussion
in \sect{FewerSusySubsection} can be carried over to gravity as well.
Whenever a gravity theory is the low-energy limit of a string theory,
we are guaranteed that the KLT relations will hold; this includes the
vast number of heterotic string constructions~\cite{Heterotic}.  The
relations appear to apply even more generally than dictated by the
heterotic string constructions~\cite{GeneralKLT}.  
In general, the KLT construction may give undesirable
states in the tensor product, such as an dilaton and antisymmetric
tensor in the ${\cal N}=0$ case; to remove their contributions 
additional projections would be required.


\section{Conclusions}
\label{ConclusionSection}

In this paper we described techniques for evaluating
sums over the multiplet of states appearing in the
four-dimensional generalized unitarity cuts of multi-loop 
super-Yang-Mills amplitudes.  We used these techniques to expose
general features of the cut amplitudes.

Our approach for evaluating the supersums in cuts is inspired by the
one of Bianchi, Elvang, Kiermaier and
Freedman~\cite{FreedmanGenerating, FreedmanUnitarity} and based on the
MHV expansion of tree amplitudes~\cite{CSW,GGK,FreedmanProof}.  Here we
reorganized the contributions in two ways: first, as a linear system
of algebraic equations, and, in the second, in terms of diagrams
tracking the flow of $SU(4)$ $R$-symmetry indices.  An important
advantage of the algebraic approach is that simplifications based on
Schouten's identity are obtained automatically.  This is a
natural approach for carrying out formal derivations of properties of
amplitudes.  On the other hand, the diagrammatic approach makes it
straightforward to construct results by drawing simple diagrams and leads
to an easily programmable algorithm for evaluating supersums by
sweeping over possible purely gluonic configurations.  The expressions
obtained this way can be further simplified through use of 
Schouten's identity and momentum conservation; we described graphical
rules for carrying out such manipulations, whose effect is to improve
the power count by replacing some of the numerator 
loop momenta of cuts with external momenta.

We also used the index-diagram approach to construct a generating
function for certain theories with less-than-maximal supersymmetry.
This is straightforward because the index-diagram approach tracks the
contributions of individual configurations of states in the cuts.  This
allowed us to give simple rules determining the contributions of
various gauge multiplets to cuts.  It should be possible to further
generalize these considerations to supersymmetric theories with arbitrary
matter content.

In general, completely determining the integrand of amplitudes
requires the evaluation of unitarity cuts in an arbitrary number of
dimensions.  In particular, use of dimensional regularization to
control infrared or ultraviolet singularities implies that the
amplitudes cannot be evaluated in strictly four dimensions.
Nevertheless, in practical calculations, four-dimensional cuts provide
invaluable guidance for constructing an ansatz, whose cuts can be
verified through the more complicated $D$ dimensional cuts.
The efficient and systematic evaluation of supermultiplet sums in
arbitrary dimensions remains an important open problem.  One 
obstacle arises from the strong dependence of on-shell superspaces on
the specific dimensionality of space-time, making it difficult 
to treat all dimensions in a unified way.  Another difficulty 
is the absence of a formalism as efficient as four-dimensional 
spinor helicity in general dimensions.  Recent
progress towards solving the latter problem is given in
ref.~\cite{D6Helicity}, where a six-dimensional helicity-like formalism
is constructed.

The KLT~\cite{KLT,GeneralKLT,TreeJacobi} relations allow us to
rewrite any product of tree-level amplitudes in $\NeqEight$
supergravity representing the generalized cut of some multi-loop
amplitude directly in terms of double products of cuts of $\NeqFour$
super-Yang-Mills amplitudes~\cite{BDDPR}.  This allows us to
immediately carry over to $\NeqEight$ supergravity $\NeqFour$
super-Yang-Mills evaluations of supersums.  
 Higher-loop studies of $\Neqeight$ supergravity
should help shed further light on the recent proposal that $\NeqEight$
supergravity may be a perturbatively finite theory of quantum
gravity~\cite{Finite, DualityArguments, GravityThree}.

In summary, the techniques presented here clarify the structure of
unitarity cuts in supersymmetric theories.  These should be helpful in
future studies of the properties of multi-loop amplitudes via the
unitarity method.  In particular these methods are important parts of
the construction of the four-loop four-point nonplanar amplitudes of
$\NeqFour$ super-Yang-Mills theory~\cite{FourLoopNonPlanar}, which
will probe the multiloop infrared and ultraviolet structures of gauge
theories, and aid in the construction of the corresponding $
\NeqEight$ supergravity amplitudes.  These amplitudes will allow for
a definitive determination of the four-loop ultraviolet behavior
of the two maximally supersymmetric theories in various dimensions.


\section*{Acknowledgments}
\vskip -.3 cm 

We thank Lance Dixon for many helpful discussions and collaboration on
this and related topics.  We also thank Nima Arkani-Hamed, and David
Kosower for a number of stimulating discussions.  We thank Henriette
Elvang and Dan Freedman for helpful comments on our manuscript.  We
also thank them for sending us their results for non-trivial cuts of
four-loop amplitudes, obtained via their generating function
method~\cite{FreedmanUnitarity}, providing strong consistency checks.
We thank Academic Technology Services at UCLA for computer support.
This research was supported by the US Department of Energy under
contracts DE--FG03--91ER40662, DE--FG02--90ER40577 (OJI) and the US
National Science Foundation under grant PHY-0608114.  R.R.
acknowledges support by the A.~P.  Sloan Foundation.  J.~J.~M.~C.  and
H.~J.  gratefully acknowledge the financial support of Guy Weyl Physics
and Astronomy Alumni Grants.

\end{document}